\documentclass[
aps,
prd,            
reprint,      
letterpaper,           
preprintnumbers,
 superscriptaddress,
amsmath,        
amssymb,        
floatfix]{revtex4-2}
\usepackage{silence}
\usepackage[T1]{fontenc}
\usepackage{graphicx}    
\usepackage[normalem]{ulem}
\usepackage{xcolor}     
\usepackage{bm}         
\usepackage{braket}
\usepackage[export]{adjustbox}
\usepackage{placeins}
\usepackage[colorlinks=true, linkcolor=blue, citecolor=blue, urlcolor=blue]{hyperref}
\usepackage{booktabs}

\begin{document}
	\title{\textbf{Gaussian ergotropy and spatial correlations: accelerated vacuum versus a thermal field}} 
\author{Siwei Li}
\affiliation{School of Mathematics and Statistics, Hainan Normal University, Haikou 571158, China}

\author{Xiaofen Huang} 
\email{huangxf1206@163.com}
\affiliation{School of Mathematics and Statistics, Hainan Normal University, Haikou 571158, China}

\author{Shao-Ming Fei}
\affiliation{School of Mathematical Sciences, Capital Normal University, Beijing 100048, China}
\begin{abstract}
	Can the spatial correlations of a quantum environment be revealed through work extraction? 
	We address this question using two harmonic-oscillator detectors coupled to either an accelerated vacuum or a matched inertial thermal field.
	 We establish a quantitative connection between the spatial correlations of the environment and the equilibrium Gaussian ergotropy, showing that the leading extractable work is governed by a response-weighted spatial correlation.
	 Although the two environments have identical local spectra, their different spatial correlations give rise to distinct work responses.
	 We further identify a finite-distance blind point at which a single measurement configuration fails to distinguish the two environments.
	 By reversing the direct coupling between the detectors, we develop a two-setting scheme that removes this ambiguity and reconstructs the underlying spatial correlations from work measurements.
	 This reconstruction remains effective beyond the weak-coupling regime. We also show that spatial correlations can provide additional global work even when the detector pair is separable, demonstrating that entanglement is not a necessary resource for this advantage. Finally, we investigate the transient evolution of Gaussian ergotropy and reveal its oscillatory decay and dependence on acceleration and detector parameters. 
	 Our results establish Gaussian ergotropy as an operational probe of spatial correlations in bosonic environments.
\end{abstract}
\maketitle
\section{Introduction}

The ergotropy quantifies the maximum work extractable from a quantum state through cyclic unitary transformations and provides a central quantity in quantum thermodynamics \cite{Allahverdyan2004}. For composite systems, the extractable work depends on both the correlations among subsystems and the class of operations allowed for its extraction. These aspects have been investigated in connection with work stored in correlations, collective work extraction through entangling operations, and the difference between local and global ergotropy \cite{AlickiFannes2013,PerarnauLlobet2015,Mukherjee2016,Francica2017,Touil2022,Francica2022}. Related studies have further examined the role of correlations in quantum batteries, including their effects on extractable work and ergotropy transport \cite{Andolina2019,Shi2022,Simon2025}. The contribution of quantum coherence to work extraction has also been studied in both closed and open quantum systems \cite{Francica2020,Cakmak2020}. For bosonic systems described by quadratic Hamiltonians, restricting the allowed transformations to Gaussian unitaries leads to Gaussian ergotropy, for which analytical expressions have been derived in terms of the covariance matrix and its symplectic eigenvalues \cite{Hovhannisyan2020,Rodriguez2025}. More recently, ergotropic properties have also been employed to investigate entanglement in continuous-variable Gaussian systems \cite{PoloRodriguez2026}. These developments provide a thermodynamic framework for studying the role of correlations in the work extraction of continuous-variable quantum systems.

The Unruh effect establishes a fundamental connection between quantum field fluctuations and thermal physics, whereby a uniformly accelerated detector in the Minkowski vacuum responds as if immersed in a thermal bath at temperature $T_U=a/(2\pi)$ \cite{unruh1976,takagi1986,crispino2008}. Together with Hawking radiation and the Gibbons--Hawking effect \cite{Hawking1975,Gibbons1977}, it has motivated extensive studies of particle-detector response and thermalization under acceleration and in curved spacetimes, including finite-time interactions and spatially extended detectors \cite{Lima2019,Moustos2018,Garay2016,Moustos2022,Perche2021,Conroy2022}. In parallel, relativistic quantum-information research has explored observer-dependent entanglement and quantum communication \cite{AlsingMilburn2003,FuentesSchuller2005,Alsing2006}, continuous-variable and multipartite entanglement, and entanglement harvesting \cite{Reznik2003,LinHu2010,pozas2015,Salton2015,NgMannMartinMartinez2018,GallockYoshimura2021,Barman2022,LinMondal2024Alpha,liu2023fermionic,wang2020genuine,Li2023Quantum,zhang2025entanglement,liu2025entanglement}, as well as quantum coherence \cite{HuangSitu2018Coherence,wu2021quantum,li2025multiqubit,liao2025quantum,Kaczmarek2025Coherence,Du2025BasisIndependentCoherence}, Bell nonlocality \cite{zhang2023hawking,Kaczmarek2024Signatures,mi2025genuine,Kaczmarek2026Nonlocal,Ye2025BellMermin,LinMondal2026}, entropic uncertainty relations \cite{feng2013,Wang2024Entropic,Du2026QMAEUR}, quantum steering \cite{Wang2016GaussianSteering,Gong2022GaussianSteering,liu2018influence,wu2025fermionic,wu2025gaussian}, and quantum discord \cite{Datta2009,Kanno2016,WangWang2021,Wu2023Monogamy,LinMondal2024Alpha,huang2026,li2026geometric}. More recently, quantum work extraction has emerged as a complementary perspective in relativistic quantum thermodynamics, with investigations of accelerated Unruh--DeWitt quantum batteries \cite{Hao2023}, coherence-assisted ergotropy as a probe of Unruh thermality \cite{Zhang2025CoherentErgotropy}, relativistic quantum Otto cycles \cite{GallockYoshimura2024}, and ergotropy in de Sitter and anti-de Sitter spacetimes \cite{Hao2026}. Collectively, these investigations encompass discrete- and continuous-variable systems, diverse quantum correlations, and a variety of accelerated and curved-spacetime backgrounds, broadening our understanding of quantum information and thermodynamic processes in relativistic settings.

For spatially separated detectors, the field correlations sampled along distinct worldlines introduce an additional dependence on their relative geometry. While the local response is determined by the field correlation function evaluated along an individual trajectory, the collective dynamics also involves the cross Wightman function connecting the two detector trajectories. Field-mediated entanglement generation and harvesting, acceleration-dependent inter-detector interactions, and spatially dependent response spectra have been investigated extensively in this context \cite{LinHu2010,pozas2015,Salton2015,WangWang2021,ChengZhouYu2022,ChenHuYu2022,ZhouChengYu2025,Huang2026Spectra}. Related investigations have also examined the influence of spacetime curvature and horizon structure on correlations extracted by multiple detectors \cite{NgMannMartinMartinez2018,GallockYoshimura2021,Barman2022,LinMondal2024Alpha}. These studies highlight the importance of inter-detector field correlations in determining the collective behavior of spatially separated quantum systems.

Harmonic-oscillator detectors offer a natural framework for investigating how field correlations affect the collective dynamics of spatially separated quantum systems. Their linear coupling to a free scalar field leads to a quadratic system--environment Hamiltonian, allowing the detector dynamics to be treated within the covariance-matrix formalism beyond perturbation theory \cite{Brown2013}. This approach is closely related to quantum Brownian motion, where the environmental degrees of freedom can be integrated out to obtain the effective dissipation and fluctuation kernels governing the system dynamics \cite{feynman1963,caldeira1983,grabert1988,hu1992,FlemingRouraHu2011}. Extensions to multiple coupled oscillators interacting with a common environment have further established the role of bath-mediated correlations in their dynamics and decoherence \cite{ChouYuHu2008,paz2008,lin2009,hsiang2015}. The resulting covariance-matrix description naturally incorporates both local fluctuations and inter-detector correlations, providing a suitable basis for studying their thermodynamic consequences.

In this work, we investigate whether Gaussian ergotropy can serve as a quantitative probe of spatial field correlations by distinguishing environments with identical local spectra but different spatial cross spectra. We consider two transversely separated harmonic-oscillator detectors with equal proper acceleration, linearly coupled to a common massless real scalar field and directly coupled through a two-mode-squeezing interaction. We compare the Minkowski vacuum sampled by uniformly accelerated detectors with an inertial thermal field at the same Unruh temperature, and also consider independent local thermal baths as an uncorrelated reference. Starting from the corresponding field correlation functions, we construct the oscillator-bath models and derive the equilibrium covariance matrices using the Green-function formalism. In the weak-coupling regime, we establish an analytical relation between the equilibrium Gaussian ergotropy and a response-weighted spatial correlation of the field. We identify a finite-distance blind point at which the single-setting work responses of the two environments coincide despite their different spatial correlations. To resolve this ambiguity, we propose a two-setting protocol with opposite direct squeezing couplings, $\lambda=\pm g$, and derive an explicit inversion relation that reconstructs the response-weighted spatial correlation from the corresponding ergotropies, eliminating the common local work scale. We further show that the additional work accessible through joint Gaussian operations is proportional to the square of this correlation when the direct coupling is switched off, and can remain nonzero even for separable detector states. We also investigate finite-coupling effects using the full, unexpanded Green functions, quantifying their corrections to the analytical predictions and the reconstructed correlations. Finally, we study the finite-time evolution of Gaussian ergotropy and characterize its oscillatory decay under the accelerated-vacuum environment.

The paper is organized as follows. Sections~\ref{hmd}--\ref{tau} introduce the oscillator-bath model and develop the covariance-matrix dynamics and equilibrium Gaussian ergotropy. Sections~\ref{cr:main}--\ref{sec:readout} establish the relation between spatial correlations and work extraction and present the two-setting reconstruction. Sections~\ref{st:global} and \ref{sec:dynamics} discuss the additional global work and finite-time dynamics, respectively. Finally, Sec.~\ref{c} presents our conclusions.
	\section{Oscillator-bath representation of the accelerated vacuum and the matched thermal field \label{hmd}}
	We consider a massless real scalar field in Minkowski spacetime, where the metric line element is given by
	\begin{equation}
		\mathrm{d}s^2 = -\mathrm{d}t^2 + \mathrm{d}x^2 + \mathrm{d}y^2 + \mathrm{d}z^2.
		\label{dugui}
	\end{equation}
	Suppose two detectors, Alice and Bob, moving with uniform proper acceleration along the $x$ direction. They are separated by a distance $d$ in the transverse $y$ direction, with the $z$ coordinate fixed at zero.
	
	The worldlines of the two detectors are given by
	\begin{equation}
		x_A = \left( \frac{\sinh(a\tau)}{a},\, \frac{\cosh(a\tau)}{a},\, 0,\, 0 \right),
		\label{xa}
	\end{equation}
	\begin{equation}
		x_B = \left( \frac{\sinh(a\tau)}{a},\, \frac{\cosh(a\tau)}{a},\, d,\, 0 \right),
		\label{xb}
	\end{equation}
	where both detectors possess the same proper acceleration, $a_A = a_B = a$, and share a common Rindler time parameter $\tau$.
	
	The massless real scalar field obeys the massless Klein--Gordon equation:
	\begin{equation}
		\Box \hat{\phi}(x) = 0,
	\end{equation}
	where $\Box = -\partial_t^2 + \partial_x^2 + \partial_y^2 + \partial_z^2$ is the d'Alembertian operator.
	
	The field operator $\hat{\phi}(x)$ is expanded in plane waves \cite{peskin1995}:
	\begin{equation}
		\hat{\phi}(x) = \int \frac{d^3\mathbf{k}}{(2\pi)^3} \frac{1}{\sqrt{2|\mathbf{k}|}}
		\left[ \hat{a}_{\mathbf{k}} e^{-\mathrm{i}|\mathbf{k}|t + \mathrm{i}\mathbf{k}\cdot\mathbf{x}} + \hat{a}_{\mathbf{k}}^\dagger e^{+\mathrm{i}|\mathbf{k}|t - \mathrm{i}\mathbf{k}\cdot\mathbf{x}} \right],
		\label{changfuction}
	\end{equation}
	where $\mathbf{k}$ is the wave vector, $|\mathbf{k}|$ is the wave number, and $\hat{a}_{\mathbf{k}}$ and $\hat{a}_{\mathbf{k}}^\dagger$ are the annihilation and creation operators for mode $\mathbf{k}$, respectively.
	
	The Minkowski vacuum satisfies
	\begin{equation}
		\hat{a}_{\mathbf{k}} \ket{0}_M = 0, \qquad \forall \mathbf{k},
	\end{equation}
	where
	$\left[ \hat{a}_{\mathbf{k}},\, \hat{a}_{\mathbf{k}'}^\dagger \right] = (2\pi)^3 \delta^3(\mathbf{k} - \mathbf{k}'), \left[ \hat{a}_{\mathbf{k}},\, \hat{a}_{\mathbf{k}'} \right] = 0.$
	
	The two-point correlation function of the quantum field, i.e., the Wightman function, is defined as \cite{wightman1956,birrell1982,wald1994,takagi1986}
	\begin{equation}
		W_{ij}(x,x') = \langle 0_M | \hat{\phi}(x) \hat{\phi}(x') | 0_M \rangle, \quad i,j \in \{A,B\},
		\label{wfoction}
	\end{equation}
	where the diagonal elements describe the autocorrelations of individual detectors, while the off-diagonal elements describe the cross-correlations between the two detectors.
	
	Equation~(\ref{wfoction}) is given explicitly by
	\begin{equation}
		W_{ij}(x,x')
		= \int \frac{\mathrm{d}^3 k}{(2\pi)^3} \frac{1}{2|\mathbf{k}|}
		e^{-\mathrm{i}|\mathbf{k}|\Delta t + \mathrm{i}\mathbf{k}\cdot\Delta\mathbf{x}}.
	\end{equation}
	
	We first evaluate the solid-angle integral by exploiting the spatial spherical symmetry. Let \( r = |\Delta\mathbf{x}| \) denote the spatial separation between the two spacetime points, \( k = |\mathbf{k}| \) the magnitude of the wave vector, and \( \theta \) the angle between \( \mathbf{k} \) and \( \Delta\mathbf{x} \), so that \( \mathbf{k}\cdot\Delta\mathbf{x} = kr\cos\theta \). The angular integral yields
	\begin{equation}
		\int \mathrm{d}\Omega \, e^{\mathrm{i}kr\cos\theta} = \frac{4\pi \sin(kr)}{kr}.
	\end{equation}
	
	Substituting this result into the Wightman function of the Minkowski vacuum and decomposing the three-dimensional wave-vector integral into radial and angular parts, we obtain after successive simplifications
	\begin{equation}
		\begin{aligned}
			W_{ij}(x,x')
			= \frac{1}{8\pi^2 \mathrm{i} r} \int_0^\infty \mathrm{d}k \left[ e^{-\mathrm{i}k(\Delta t - r)} - e^{-\mathrm{i}k(\Delta t + r)} \right].
		\end{aligned}
	\end{equation}
	
	The two integrals derived above are oscillatory integrals over the positive half-line, and they do not converge in the ordinary Riemann sense.
	We adopt the standard $i\epsilon$ prescription by analytically continuing the time interval as $\Delta t \to \Delta t - \mathrm{i}\epsilon$ with $\epsilon \to 0^+$, which defines the integrals in the distributional sense. The integral identity reads
	\begin{equation}
		\int_0^\infty e^{-\mathrm{i} k u} \mathrm{d}k = \pi\delta(u)-\mathrm{i}\,\operatorname{PV}\frac1u,
	\end{equation}
	where $\operatorname{PV}$ denotes the Cauchy
	principal value.
	
	Substituting this expression into the one-dimensional Wightman function, we obtain
	\begin{equation}
		W_{ij}(x,x') = -\frac{1}{4\pi^2} \frac{1}{(\Delta t - \mathrm{i}\epsilon)^2 - |\Delta \mathbf{x}|^2}. 
		\label{wXX}
	\end{equation}
	
	We let $\Delta\tau = \tau - \tau'$. Using Eqs.~\eqref{xa} and~\eqref{xb}, we obtain
	\begin{equation}
		\begin{split}
			&\Delta t = \frac{2}{a}\cosh\frac{a(\tau+\tau')}{2}\sinh\frac{a\Delta\tau}{2}, \\
			&\Delta x = \frac{2}{a}\sinh\frac{a(\tau+\tau')}{2}\sinh\frac{a\Delta\tau}{2}.
		\end{split}
	\end{equation}
	
	Therefore,
	\begin{equation}
		\Delta t^2 - \Delta x^2
		= \frac{4}{a^2}\sinh^2\frac{a\Delta\tau}{2}.
		\label{dt2dx2}
	\end{equation}
	
	Substituting Eq.~\eqref{dt2dx2} into Eq.~\eqref{wXX}, we arrive at the single-detector Wightman function:
	\begin{equation}
		W_{AA}(\Delta\tau) = 	W_{BB}(\Delta\tau) = -\frac{a^2}{16\pi^2} \frac{1}{\sinh^2\left[ \frac{a}{2}\left( \Delta\tau - \mathrm{i}0^+ \right) \right]}.
		\label{waa}
	\end{equation}
	It is worth noting that the resulting Wightman function depends exclusively on the proper time interval $\Delta\tau$.

	We define the Fourier transform of the two-point function as
	\begin{equation}
		\widetilde{W}_{ij}(\Omega) = \int_{-\infty}^{\infty} \mathrm{d}\Delta\tau \, e^{i\Omega\Delta\tau} W_{ij}(\Delta\tau),
		\label{fly}
	\end{equation}
	where \(\Omega\) is the frequency conjugate to the detector's proper time.

	Taking the Fourier transform of Eq.~\eqref{waa}, we obtain (For details, see Appendix~\ref{C}):
	\begin{equation}
		\widetilde{W}_{AA}(\Omega) = \widetilde{W}_{BB}(\Omega) = \frac{\Omega}{2\pi} \frac{1}{1 - e^{-2\pi\Omega/a}}, \qquad \Omega > 0.
		\label{fwaa}
	\end{equation}
	
	The mean occupation number of quanta with frequency $\Omega$ at temperature $T$ obeys the Bose--Einstein distribution:
	\begin{equation}
		n(\Omega) = \frac{1}{e^{\Omega/T} - 1}.
	\end{equation}
	
	A simple algebraic identity for the Bose distribution can be obtained by multiplying its numerator and denominator by $e^{-\Omega/T}$:
	\begin{equation}
		n(\Omega) + 1 = \frac{1}{1 - e^{-\Omega/T}}.
	\end{equation}
	
	By directly comparing Eq.~\eqref{fwaa} with the above identity, we find that the factor \(1/(1-e^{-2\pi\Omega/a})\) in the Wightman spectrum takes precisely the form \(n(\Omega)+1\). The Unruh temperature can thus be directly read off \cite{unruh1976,crispino2008}
	\begin{equation}
		T_U = \frac{a}{2\pi},
	\end{equation}
	where $T_U$ is the standard Unruh temperature for a uniformly accelerated observer with proper acceleration $a$.
	
	Introducing the thermal occupation number evaluated at the Unruh temperature
	\begin{equation}
		n_U(\Omega) = \frac{1}{e^{2\pi\Omega/a} - 1},
	\end{equation}
	the positive-frequency Wightman function can be recast into the physically transparent form
	\begin{equation}
		\widetilde{W}_{AA}(\Omega) = \widetilde{W}_{BB}(\Omega) = \frac{\Omega}{2\pi} \left[ n_U(\Omega) + 1 \right], \qquad \Omega > 0.
	\end{equation}
	
	For $\Omega>0$, we readily obtain:
	\begin{equation}
		\widetilde{W}_{AA}(-\Omega) = \widetilde{W}_{BB}(-\Omega) = \frac{\Omega}{2\pi} n_U(\Omega).
	\end{equation}
	This satisfies the detailed balance condition for thermal equilibrium, i.e., the Kubo--Martin--Schwinger (KMS) \cite{kubo1957,martin1959,haag1967,wald1994} condition:
	\begin{equation}
		\frac{\widetilde{W}_{AA}(-\Omega)}{\widetilde{W}_{AA}(\Omega)} = \frac{\widetilde{W}_{BB}(-\Omega)}{\widetilde{W}_{BB}(\Omega)} = e^{-\beta_U \Omega}.
	\end{equation}
	
	Since the two detectors undergo identical motion in the $t$-$x$ plane, the squared spatial separation simply acquires an extra transverse term $|\Delta\mathbf{x}|^2 = \Delta x^2 + d^2$. The Wightman function between the two distinct detectors is then given by:
	\begin{equation}
		\begin{split}
			&W_{AB}(\Delta\tau) = W_{BA}(\Delta\tau) \\
			&= -\frac{a^2}{16\pi^2} \frac{1}{\sinh^2\left[ \frac{a}{2}\left( \Delta\tau - \mathrm{i}0^+ \right) \right] - \left( \frac{ad}{2} \right)^2}.
			\label{wab}
		\end{split}
	\end{equation}
	
	Taking the Fourier transform of Eq.~\eqref{wab}, we obtain (For details, see Appendix~\ref{C}):
	\begin{equation}
		\widetilde{W}_{AB}(\Omega) =  \widetilde{W}_{BA}(\Omega) = \frac{1}{2\pi d \sqrt{1 + \left( \frac{ad}{2} \right)^2}} \frac{\sin\left[ \frac{2\Omega}{a} \operatorname{arsinh} \frac{ad}{2} \right]}{1 - e^{-2\pi\Omega/a}}.
	\end{equation}
	
	We define the geometric correlation factor $\gamma_E$ as:
	\begin{equation}
		\gamma_E(\Omega, a, d) \equiv \frac{\widetilde{W}_{AB}(\Omega)}{\widetilde{W}_{AA}(\Omega)} =  \frac{\sin\left[ \frac{2\Omega}{a} \operatorname{arsinh} \frac{ad}{2} \right]}{\Omega d \sqrt{1 + \left( \frac{ad}{2} \right)^2}}.
	\end{equation}
	
	The geometric correlation factor $\gamma_E$ satisfies the following three fundamental properties (The proof is given in Appendix~\ref{D}):
	
	\textbf{Property 1 (Coincident-point limit).}  
	In the limit of vanishing detector separation $d \to 0$, the correlation factor approaches unity:
	\begin{equation}
		\gamma_E \to 1 \quad \text{as } d \to 0.
	\end{equation}
	
	\textbf{Property 2 (Inertial limit).}  
	In the limit $a\to0$, the result reduces to
	\begin{equation}
		\gamma_E \to \frac{\sin\Omega d}{\Omega d} \quad \text{as } a \to 0.
	\end{equation} 
	
	\textbf{Property 3 (Boundedness).}  
	The magnitude of the correlation factor is uniformly bounded by unity:
	\begin{equation}
		\left| \gamma_E \right| \le 1.
	\end{equation}
	
	The field we consider here is a Gaussian scalar field, for which all field expectation values are fully determined by the Wightman function defined in Eq.~\eqref{wXX}. We arrange the same-detector and cross-detector Wightman functions into the following matrix form:
	\begin{equation}
		\begin{aligned}
			\widetilde{W}_U(\Omega)
			&=
			\begin{pmatrix}
				\widetilde{W}_{AA}(\Omega) &
				\widetilde{W}_{AB}(\Omega) \\
				\widetilde{W}_{BA}(\Omega) &
				\widetilde{W}_{BB}(\Omega)
			\end{pmatrix}
			\\[6pt]
			&=
			\underbrace{
				\frac{\Omega}{2\pi}[n_U(\Omega)+1]
			}_{\substack{
					\text{Common local}\\
					\text{spectral factor}
			}}
			\underbrace{
				\begin{pmatrix}
					1 & \gamma_E(\Omega)\\
					\gamma_E(\Omega) & 1
				\end{pmatrix}
			}_{\substack{
					\text{Normalized spatial}\\
					\text{correlation matrix}
			}}.
		\end{aligned} \label{w31}
	\end{equation}

	\begin{equation}
		\begin{split}
			\widetilde{{W}}_U(-\Omega) &= 
			\begin{pmatrix}
				\widetilde{W}_{AA}(-\Omega) & \widetilde{W}_{AB}(-\Omega) \\
				\widetilde{W}_{BA}(-\Omega) & \widetilde{W}_{BB}(-\Omega)
			\end{pmatrix}\\ 
			&= \frac{\Omega}{2\pi}\, n_U(\Omega)
			\begin{pmatrix}
				1 & \gamma_E(\Omega) \\
				\gamma_E(\Omega) & 1
			\end{pmatrix}.
		\end{split}
	\end{equation}
	This also satisfies the detailed balance (KMS) condition for thermal equilibrium:
	\begin{equation}
		\widetilde{W}_U(-\Omega)=e^{-\beta_U\Omega}\widetilde{W}_U^{\,T}(\Omega).
	\end{equation}
	The transpose exchanges the two detector indices; the spectral matrix is symmetric for the present trajectories. We can also obtain its vacuum correlation matrix:
	\begin{equation}
		\widetilde{{W}}^{\text{vac}}_U(\Omega) = \frac{\Omega}{2\pi}
		\begin{pmatrix}
			1 & \gamma_U(\Omega) \\
			\gamma_U(\Omega) & 1
		\end{pmatrix}.
		\label{vac}
	\end{equation}
	
	We next construct a matched inertial thermal bath by considering
	two inertial detectors at rest in the same inertial frame,
	separated by a fixed distance $d$, with the massless scalar field
	prepared in a thermal state at temperature $T_U=a/(2\pi)$.
	The corresponding thermal occupation number is given by
	\begin{equation}
		n_T(\Omega)=n_U(\Omega).
	\end{equation}
	The geometric correlation factor is given by
	\begin{equation}
		\gamma_T(\Omega,d)
		=\lim_{a\to 0}\gamma_E(\Omega,a,d)
		=\frac{\sin(\Omega d)}{\Omega d}.
	\end{equation}
	The corresponding spectral matrix reads
	\begin{equation}
		\begin{aligned}
			\widetilde{W}_T(\Omega)
			&=
			\frac{\Omega}{2\pi}
			[n_U(\Omega)+1]
			\begin{pmatrix}
				1 & \gamma_T(\Omega,d)\\
				\gamma_T(\Omega,d) & 1
			\end{pmatrix}.
		\end{aligned}
	\end{equation}
	The local spectral contribution coincides exactly with
	the corresponding term in Eq.~\eqref{w31}.
	Consequently, the two environments have the same temperature
	and identical local spectra, whereas their cross spectra
	are generally different.
	
	Finally, we introduce an uncorrelated reference thermal bath, consisting of two independent local thermal reservoirs, each coupled to one detector. Both reservoirs are maintained at the same temperature \(T_U\), so that their thermal occupation numbers coincide with those of the previous two environments. Since the two reservoirs are statistically independent, the cross-correlation between the corresponding bath forces vanishes. Consequently, the geometric correlation factor is given by
	\begin{equation}
		\gamma_C=0.
	\end{equation}
	The corresponding spectral matrix is therefore given by
	\begin{equation}
		\widetilde{W}_C(\Omega)
		=
		\frac{\Omega}{2\pi}
		[n_U(\Omega)+1]
		\begin{pmatrix}
			1&0\\
			0&1
		\end{pmatrix}.
	\end{equation}
	In what follows, we use $E\in\{U,T,C\}$ to label the three environments.
	
	We now model each detector as a harmonic oscillator with natural frequency $\omega$. When the two detectors are directly coupled, the total system Hamiltonian is given by (Appendix~\ref{A})
	\begin{equation}
		\mathcal{H}_{S} = \frac{\omega}{2}\left(q_A^2 + p_A^2\right) + \frac{\omega}{2}\left(q_B^2 + p_B^2\right) + \lambda\left(q_A q_B - p_A p_B\right),\label{hs}
	\end{equation}
	where $\lambda$ parametrizes the strength of the direct bilinear coupling between the two detectors. \(q_i\) and \(p_i\) are the canonical coordinate and canonical momentum of the system, respectively, with \(i \in \{A,B\}\).

	From Eq.~\eqref{changfuction}, the Hamiltonian of the free field is given by
	\begin{equation}
		\mathcal{H}_{\text{field}} = \int \frac{d^3\mathbf{k}}{(2\pi)^3} \, \omega_{\mathbf{k}} \, a_{\mathbf{k}}^\dag a_{\mathbf{k}}.
	\end{equation}
	
	The interaction Hamiltonian between the detectors and the field is given by \cite{dewitt1979}
	\begin{equation}
		\mathcal{H}_{\text{int}}(\tau) = \kappa \chi(\tau) \left[ q_A(\tau) \hat{\phi}\left(x_A(\tau)\right) + q_B(\tau) \hat{\phi}\left(x_B(\tau)\right) \right],
	\end{equation}
	where $\kappa$ is the detector-field coupling constant and $\chi(\tau)$ is the proper-time switching modulation function (Appendix~\ref{B}).

	According to the Feynman--Vernon influence-functional theory
	\cite{feynman1963}, for a Gaussian field linearly coupled to Gaussian
	detectors, Wick's theorem \cite{wick1950} and the Gaussian
	influence-functional construction
	\cite{feynman1963,caldeira1983,grabert1988,hu1992}
	ensure that the reduced detector dynamics is entirely encoded in the
	two-point Wightman function \cite{wightman1956} along the detector
	worldlines; the resulting Gaussian detector dynamics can be expressed
	using the standard covariance-matrix formalism \cite{weedbrook2012}.
	A Gaussian environment with a prescribed two-point spectrum can
	equivalently be represented by a collection of harmonic oscillators
	through spectral-density matching
	\cite{caldeira1983,ford1988,grabert1988}; consequently, the explicit
	three-dimensional field may be replaced by an effective one-dimensional
	harmonic-oscillator bath without changing the detector dynamics.
	
	For each continuous frequency $\Omega \in [0, \infty)$, we introduce two bath oscillators at every frequency, with their canonical coordinates and momenta denoted by
	$
	\mathbf{Q}_\Omega =
	\begin{pmatrix}
		Q_{\Omega,+} ,
		Q_{\Omega,-}
	\end{pmatrix}^{\mathrm{T}},
	\mathbf{P}_\Omega =
	\begin{pmatrix}
		P_{\Omega,+} ,
		P_{\Omega,-}
	\end{pmatrix}^{\mathrm{T}}.
	$
	We denote the coordinate vector of the two-detector system as
	$
	\mathbf{q} = 
	\begin{pmatrix}
		q_{A} ,
		q_{B}
	\end{pmatrix}^{\mathrm{T}}.
	$
	The canonical commutation relations read:
	\[
	\left[\mathbf{Q}_\Omega, \mathbf{P}_{\Omega'}^\mathrm{T}\right] = \mathrm{i} I_2 \, \delta(\Omega - \Omega'),
	\left[\mathbf{Q}_\Omega, \mathbf{Q}_{\Omega'}^\mathrm{T}\right] = \left[\mathbf{P}_\Omega, \mathbf{P}_{\Omega'}^\mathrm{T}\right] = \mathbf{0},
	\]
	where $I_2$ denotes the unity matrix in $2\times2$ space.
	
	The bath Hamiltonian and the system-bath coupling Hamiltonian can then be written as:
	\begin{align}
		\mathcal{H_{\text{bath}}} &= \int_0^\infty d\Omega \, \frac{\Omega}{2} \sum_{\zeta=\pm} \left( Q_{\Omega,\zeta}^2 + P_{\Omega,\zeta}^2 \right), \\
		\mathcal{H}_{SB}(\tau) &= \chi(\tau) \int_0^\infty d\Omega \, \mathbf{q}^\mathrm{T} \mathbf{h}(\Omega) \mathbf{Q}_\Omega,
	\end{align}
	where $\mathbf{h}(\Omega)$ is a $2\times 2$ continuous coupling matrix: its rows correspond to detectors $A$ and $B$, and its columns correspond to bath channels $+$ and $-$.
	
	We define the spectral density matrix as:
	\begin{equation}
		\mathbf{J}_E(\Omega) = \frac{\pi}{2} \mathbf{h}(\Omega) \mathbf{h}(\Omega)^\mathrm{T}.
		\label{jf}
	\end{equation}
	
	The two-time correlation function of the collective bath coordinate
	\begin{equation}
		\mathbf{B}_E(\tau) = \int d\Omega \, \mathbf{h}(\Omega) \mathbf{Q}_\Omega(\tau)
	\end{equation}
	is computed from Eq.~\eqref{wfoction} as
	\begin{equation}
		\begin{split}
			&\mathcal{C}^{E}_{\text{bath}}(\Delta\tau) =	\\	 
			&\int_0^\infty \frac{d\Omega}{\pi} \mathbf{J}(\Omega) \left[ \left(n_U(\Omega)+1\right)e^{-\mathrm{i}\Omega\Delta\tau} + n_U(\Omega) e^{\mathrm{i}\Omega\Delta\tau} \right].
		\end{split}
		\label{cb}
	\end{equation}
	
	Since $\widetilde{W}_{AA}$ grows linearly at high frequencies (the intrinsic UV divergence of point-like detectors), we introduce a phenomenological regularization. The regularized field correlation function is then given by
	\begin{equation}
		\begin{split}
			\mathcal{C}^{E}_{\text{field}}(\Delta\tau) &= 
			\kappa^2 \int_0^\infty \frac{d\Omega}{2\pi} \, \widetilde{W}^{\text{vac}}(\Omega) \, e^{-\Omega/\Omega_c}\\
			& \left[ \left(n_U(\Omega)+1\right)e^{-\mathrm{i}\Omega\Delta\tau} + n_U(\Omega) e^{\mathrm{i}\Omega\Delta\tau} \right],
		\end{split}
		\label{cf}
	\end{equation}
	where $\Omega_c$ is the UV cutoff frequency.

	To ensure that our effective thermal bath produces the same physical effect as the quantum field, we set
	\begin{equation}
		\mathcal{C}^{E}_{\text{bath}} = \mathcal{C}^{E}_{\text{field}},
	\end{equation}
	which yields:
	\begin{equation}
		\mathbf{J}_E(\Omega) = \frac{\kappa^2}{2} \widetilde{W}^{\text{vac}}(\Omega) \, e^{-\Omega/\Omega_c}.
	\end{equation}
	Further substituting into Eq.~\eqref{vac} yields:
	\begin{equation}
		\mathbf{J}_E(\Omega) = J_0(\Omega)
		\begin{pmatrix}
			1 & \gamma_E(\Omega) \\
			\gamma_E(\Omega) & 1
		\end{pmatrix},\label{jmix}
	\end{equation}
	where $J_0(\Omega) = \frac{\kappa^2}{4\pi} \Omega \, e^{-\Omega/\Omega_c}$. We have $\det \mathbf{J}_E(\Omega) = J_0^2(\Omega)\left(1-\gamma_E^2(\Omega)\right) \geq 0$, which is guaranteed by $|\gamma_E(\Omega)| \leq 1$. Together with the non-negativity of its diagonal entries, we conclude that
	\begin{equation*}
		\mathbf{J}_E(\Omega) \succeq 0.
	\end{equation*}

	From Eq.~\eqref{jf}, we readily obtain:
	\begin{equation}
		\mathbf{h}(\Omega) = \frac{1}{\sqrt{2}}
		\begin{pmatrix}
			h_1(\Omega) & h_2(\Omega) \\
			h_1(\Omega) & -h_2(\Omega)
		\end{pmatrix},
	\end{equation}
	where\[
	h_1(\Omega)^2 = \frac{2}{\pi} J_0(\Omega)\left(1+\gamma_E(\Omega)\right),\]
	\[	h_2(\Omega)^2 = \frac{2}{\pi} J_0(\Omega)\left(1-\gamma_E(\Omega)\right).\]
	
	To discretize the bath, we introduce a set of positive-frequency
	nodes $\{\Omega_j\}_{j=1}^{N}$ together with the corresponding
	positive quadrature weights $\{w_j\}_{j=1}^{N}$, such that
	\begin{equation}
		\int_0^\infty f(\Omega)\,d\Omega
		\simeq
		\sum_{j=1}^{N}w_j f(\Omega_j).
	\end{equation}
	The dimensionless discrete canonical variables are then
	defined by
	\begin{equation}
		Q_{j,\zeta}=\sqrt{w_j}\,Q_{\Omega_j,\zeta},
		\qquad
		P_{j,\zeta}=\sqrt{w_j}\,P_{\Omega_j,\zeta},
		\qquad
		\zeta\in\{+,-\}.
	\end{equation}
	At each frequency node, we thus obtain two bath oscillators
	corresponding to the symmetric and antisymmetric channels.
	They satisfy the canonical commutation relations
	\begin{equation}
		[Q_{i,\zeta},P_{j,\zeta'}]
		=i\delta_{ij}\delta_{\zeta\zeta'},
		\qquad
		\zeta,\zeta'\in\{+,-\}.
	\end{equation}
	
	The discrete coupling matrix is defined as
	\begin{equation}
		\mathbf{g}_j = \sqrt{w_j}\, \mathbf{h}(\Omega_j).
	\end{equation}
	
	The discretized thermal bath Hamiltonian is then given by
	\begin{equation}
		\mathcal{H}_B = \sum_{j=1}^{N} \frac{\Omega_j}{2} \left( Q_{j,+}^2 + P_{j,+}^2 + Q_{j,-}^2 + P_{j,-}^2 \right).
	\end{equation}
	
	The system-bath coupling Hamiltonian reads
	\begin{equation}
		\begin{split}
			&\mathcal H_{SB}(\tau)=\\
			&\chi(\tau)\sum_{j=1}^{N}\biggl[
			\frac{g_{j,+}}{\sqrt2}\,(q_A+q_B)\,Q_{j,+}
			+\frac{g_{j,-}}{\sqrt2}\,(q_A-q_B)\,Q_{j,-}\biggr].
		\end{split}
	\end{equation}
	Here the discrete coupling strengths are fixed by spectral matching,
	\begin{equation}
		g_{j,+}^2 = \frac{2w_j}{\pi} J_0(\Omega_j)\left(1+\gamma_j\right),
		g_{j,-}^2 = \frac{2w_j}{\pi} J_0(\Omega_j)\left(1-\gamma_j\right),\label{gii}
	\end{equation}
	where $\gamma_j \equiv \gamma_E(\Omega_j)$.
	
	We arrange the discrete coupling strengths into a $2\times 2$ matrix
	(rows correspond to detectors $A,B$; columns correspond to bath channels 1, 2):
	\begin{equation}
		\mathbf{g}_j = \frac{1}{\sqrt{2}}
		\begin{pmatrix}
			g_{j,+} & g_{j,-} \\
			g_{j,+} & -g_{j,-}
		\end{pmatrix},
	\end{equation}
	which gives the matrix spectral matching condition
	\begin{equation}
		\frac{1}{2} \mathbf{g}_j \mathbf{g}_j^T = \frac{w_j}{\pi} \mathbf{J}_E(\Omega_j).
	\end{equation}
	
	Completing the bath squares produces a negative static detector interaction. We choose the following counterterm to compensate that interaction and keep the calibrated static detector Hamiltonian equal to $\mathcal H_S$:
	\begin{equation}
		\mathcal{H}_{CT}(\tau) 
		= \frac{\chi^2(\tau)}{2} \left[ \delta\omega_A \left(q_A^2 + q_B^2\right) + 2\delta\omega_{AB}\, q_A q_B \right], \label{ct}
	\end{equation}
	with the static counterterm coefficients given directly by the
	quadrature,
	\begin{equation}
		\delta\omega_A
		=\sum_{j=1}^{N}\frac{2w_j}{\pi}\frac{J_0(\Omega_j)}{\Omega_j}
		=\frac{\kappa^2}{2\pi^2}\sum_{j=1}^{N}w_j\,e^{-\Omega_j/\Omega_c},
	\end{equation}
	\begin{equation}
		\delta\omega_{AB}
		=\sum_{j=1}^{N}\frac{2w_j}{\pi}
		\frac{J_0(\Omega_j)\gamma_j}{\Omega_j}
		=\frac{\kappa^2}{2\pi^2}\sum_{j=1}^{N}w_j\,
		e^{-\Omega_j/\Omega_c}\gamma_j .
	\end{equation}
	
	The total Hamiltonian is then given by
	\begin{equation}
		\mathcal{H}_{\text{tot}}(\tau) = \mathcal{H}_S + \mathcal{H}_B + \mathcal{H}_{SB}(\tau) + \mathcal{H}_{CT}(\tau),
	\end{equation}
	where the total number of oscillators is given by
	\begin{equation}
		n = 2 + 2N.
	\end{equation}

\begin{figure}[!t]
	\centering
	\includegraphics[width=\columnwidth]{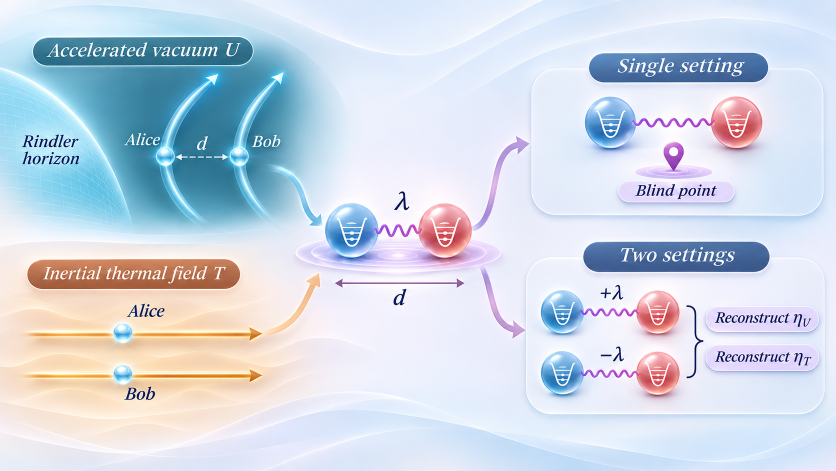}
\caption{Schematic of the two-detector setup in the
accelerated-vacuum ($U$) and matched inertial-thermal
($T$) environments. A single setting may exhibit a
blind point, whereas two opposite settings
($\pm\lambda$) enable the reconstruction of
spatial correlations.
	}
	\label{fig:schematic}
\end{figure}
	\section{the time evolution of the covariance matrix\label{xfc}}
	In this section, we evolve the state using the total Hamiltonian introduced in Sec.~\ref{hmd}, which is equivalent to propagating the corresponding covariance matrix.
	
	We first define the total coordinate vector
	\begin{equation}
		\begin{split}
			\mathbf{R} = \bigl(
			& q_A, q_B, Q_{1,+}, Q_{1,-}, \dots, Q_{N,+}, Q_{N,-}, \\
			& p_A, p_B, P_{1,+}, P_{1,-}, \dots, P_{N,+}, P_{N,-}
			\bigr)^{\mathrm{T}}.
		\end{split}
	\end{equation}
	
	Since all four Hamiltonians are quadratic forms of the canonical variables, the total Hamiltonian can be written compactly as (Appendix~\ref{A})
	\begin{equation}
		\mathcal{H}_{\mathrm{tot}}(\tau) = \frac{1}{2} \mathbf{R}^T H(\tau) \mathbf{R}.
	\end{equation}
	Here, $H(\tau)$ is a $2n \times 2n$ real symmetric matrix, referred to as the Hamiltonian matrix (quadratic form matrix). It takes the block-diagonal form
	\begin{equation}
		H(\tau) = H_{qq}(\tau) \oplus H_{pp},
	\end{equation}
	with
	\begin{equation}
		\begin{split}
			&H_{qq}(\tau) =
			\begin{pmatrix}
				H^{S}_q + \chi(\tau)^2 H_{CT} & \chi(\tau) C \\
				\chi(\tau) C^T & D
			\end{pmatrix},
			\\
			&H_{pp} =
			\begin{pmatrix}
				H^{S}_p & 0 \\
				0 & D
			\end{pmatrix}.
		\end{split}
	\end{equation}
	
	The individual submatrices are defined as follows.
	
	\textit{(1) Detector coordinate and momentum blocks (each of size $2\times 2$):}
	\begin{equation}
		H^{S}_q =
		\begin{pmatrix}
			\omega & \lambda \\
			\lambda & \omega
		\end{pmatrix},
		\qquad
		H^{S}_p =
		\begin{pmatrix}
			\omega & -\lambda \\
			-\lambda & \omega
		\end{pmatrix}.
	\end{equation}
	
	\textit{(2) Counterterm block ($2\times 2$, appearing only in the coordinate block):}
	\begin{equation}
		H_{CT} =
		\begin{pmatrix}
			\delta\omega_A & \delta\omega_{AB} \\
			\delta\omega_{AB} & \delta\omega_A
		\end{pmatrix}.
	\end{equation}
	
	\textit{(3) Bath frequency block $D$ ($2N \times 2N$ diagonal matrix): each frequency node hosts two channels, thus}
	\begin{equation}
		D = \operatorname{diag}(\Omega_1, \Omega_1, \Omega_2, \Omega_2, \dots, \Omega_N, \Omega_N).
	\end{equation}
	
	\textit{(4) System-bath coupling block $C$ ($2 \times 2N$): formed by horizontally concatenating $2\times 2$ column blocks for each frequency node,}
	\begin{equation}
		C = \bigl(\, \mathbf{g}_1\ \mathbf{g}_2\ \cdots\ \mathbf{g}_N \,\bigr).
	\end{equation}
	
	When the switching function satisfies $\chi(0) = 0$, the system and bath are fully decoupled at initial time, and the initial state is a product state, whose covariance matrix reads
	\begin{equation}
		\sigma(0) =
		\begin{pmatrix}
			\sigma^{S}_q & 0 & 0 & 0 \\
			0 & d_B & 0 & 0 \\
			0 & 0 & \sigma^{S}_p & 0 \\
			0 & 0 & 0 & d_B
		\end{pmatrix},
	\end{equation}
	where $\sigma^{S}_q$ and $\sigma^{S}_p$ denote the coordinate and momentum blocks, respectively, of the detector ground-state covariance matrix $\sigma_S(0)$.

	The initial state of the detectors is taken as the ground state of $\mathcal{H}_S$, whose covariance matrix is \cite{weedbrook2012}
	\begin{equation}
		\sigma_S(0) = \frac{1}{2\nu_S}
		\begin{pmatrix}
			\omega & -\lambda & 0 & 0 \\
			-\lambda & \omega & 0 & 0 \\
			0 & 0 & \omega & \lambda \\
			0 & 0 & \lambda & \omega
		\end{pmatrix},
	\end{equation}
	where
	\begin{equation*}
		\nu_S = \sqrt{\omega^2 - \lambda^2} \quad (|\lambda| < \omega).
	\end{equation*}
	
	The bath is initially prepared in a thermal state at the Unruh temperature
	\begin{equation}
		T_U = \frac{a}{2\pi},
	\end{equation}
	and its covariance matrix takes the block-diagonal form
	\begin{equation}
		\begin{split}
			&\sigma_B(0) =
			\begin{pmatrix}
				d_B & 0 \\
				0 & d_B
			\end{pmatrix},\\
			&d_B = \operatorname{diag}(v_1, v_1, v_2, v_2, \dots, v_N, v_N),
		\end{split}
	\end{equation}
	where
	\begin{equation*}
		v_j = \frac{1}{2}\coth\left(\frac{\pi\Omega_j}{a}\right).
	\end{equation*}
	
	For a quadratic Hamiltonian, linearizing the Heisenberg equations of motion yields
	\begin{equation}
		\dot{\mathbf{R}}(\tau) = A(\tau) \mathbf{R}(\tau),
	\end{equation}
	where $A(\tau) = \Omega_{\text{can}} H(\tau)$.
	
	We define the propagator $M(\tau)$ via
	\begin{equation}
		\dot{M}(\tau) = A(\tau) M(\tau), \qquad M(0) = I_{2n}.
	\end{equation}
	It is straightforward to verify that the propagator is a symplectic matrix, satisfying
	\begin{equation}
		M \Omega_{\text{can}} M^T = \Omega_{\text{can}}.
	\end{equation}
	
	It follows that the phase-space vector evolves according to
	\begin{equation}
		\mathbf{R}(\tau) = M(\tau) \mathbf{R}(0).
	\end{equation}
	
	The evolution equation for the total covariance matrix then reads
	\begin{equation}
		\dot{\sigma}(\tau) = A(\tau)\sigma(\tau) + \sigma(\tau) A(\tau)^T,
	\end{equation}
	or equivalently
	\begin{equation}
		\sigma(\tau) = M(\tau)\sigma(0)M(\tau)^T.
	\end{equation}
	
	Since the detector coordinates correspond to the first two entries and the detector momenta correspond to entries $n+1$ and $n+2$ of the full phase-space vector, we define a $4 \times 2n$ projection matrix $P_S$ to extract the reduced covariance matrix of the system at arbitrary time $\tau$:
	\begin{equation}
		\sigma_S(\tau) = P_S \sigma(\tau) P_S^T.
	\end{equation}
	\section{Gaussian ergotropy at infinite proper time
		\label{tau}}
	
	In this section, we derive the analytical expression for the Gaussian ergotropy in the limit of infinite proper time. For convenience, we let \(\zeta = \pm\).
	
	Gaussian ergotropy is the maximal amount of work that cyclic Gaussian Hamiltonian processes can extract from a Gaussian state. It is computed as follows \cite{Hovhannisyan2020}:
	\begin{equation}
		\mathcal{G}=\frac{1}{2}\operatorname{Tr}\left[\sigma H\right] - \sum_{k=1}^{n_S} s_k^\uparrow l_k^\downarrow,\label{gsn}
	\end{equation}
	where $n_S=2$ is the number of detector modes, and \(s_k\) and \(l_k\) denote the symplectic eigenvalues of the covariance matrix and the quadratic Hamiltonian matrix, respectively, with the upward and downward superscript arrows indicating ascending and descending order.
	
	Next, we investigate the Gaussian ergotropy with respect to the Hamiltonian $\mathcal{H}_{S}$. The minimization allows joint Gaussian unitaries on both detector modes, including collective-mode interchange; it is restricted to Gaussian operations.
	
	First, we introduce the orthogonal transformation
	\begin{equation}
		q_{\zeta} = \frac{q_A \pm q_B}{\sqrt{2}}, \qquad
		p_{\zeta} = \frac{p_A \pm p_B}{\sqrt{2}}. \label{zjbh}
	\end{equation}
	The canonical commutation relations are preserved:
	\begin{equation}
		[q_{\zeta}, p_{\zeta}] = \mathrm{i}, \qquad [q_+, p_-] = 0.
	\end{equation}
	
	Substituting Eq.~\eqref{zjbh} into Eq.~\eqref{hs}($\mathcal{H}_s$) yields
	\begin{equation}
		\begin{split}
			&\mathcal{H}_S = \\
			&\frac{\omega}{2}\left(q_+^2 + q_-^2 + p_+^2 + p_-^2\right) + \frac{\lambda}{2}\left[\left(q_+^2 - q_-^2\right) - \left(p_+^2 - p_-^2\right)\right] \\
			&=
			\frac{\omega + \lambda}{2} q_+^2 + \frac{\omega - \lambda}{2} p_+^2 + \frac{\omega - \lambda}{2} q_-^2 + \frac{\omega + \lambda}{2} p_-^2 \\
			&= \mathcal{H}_+ + \mathcal{H}_-,\label{hsfj}
		\end{split}
	\end{equation}
	where 
	\begin{equation}
		\mathcal{H}_\zeta = \frac{\alpha_\zeta}{2} q_\zeta^2 + \frac{\beta_\zeta}{2} p_\zeta^2, \qquad
		\alpha_\zeta = \omega \pm \lambda, \qquad
		\beta_\zeta = \omega \mp \lambda.
	\end{equation}
	
	We thus find that the two modes are exactly decoupled. From the Heisenberg equations of motion
	\begin{equation}
		\dot{q}_\zeta = \mathrm{i}[\mathcal{H}_\zeta, q_\zeta] = \beta_\zeta p_\zeta, \qquad
		\dot{p}_\zeta = \mathrm{i}[\mathcal{H}_\zeta, p_\zeta],
	\end{equation}
	we find that the two modes share the same normal-mode frequency:
	\begin{equation}
		\nu_S = \sqrt{\omega^2 - \lambda^2}.
	\end{equation}
	
	Diagonalizing the spectral density matrix \eqref{jmix} yields:
	\begin{equation}
		J_{E\zeta}(\Omega) = J_0(\Omega)\left[1 \pm \gamma_E(\Omega)\right].
	\end{equation}
	
	Under the new transformation, Eq.~\eqref{ct} becomes:
	\begin{equation}
		\begin{split}
			\mathcal{H}_{\text{CT}}
			&= \frac{1}{2} \left[ (\delta\omega_A + \delta\omega_{AB}) q_+^2 +  (\delta\omega_A - \delta\omega_{AB}) q_-^2\right] \\
			&= \frac{1}{2} \left( \delta\omega_+  q_+^2 +  \delta\omega_- q_-^2 \right),
		\end{split}
	\end{equation}
	where
	\begin{equation}
		\delta\omega_\zeta = \delta\omega_A \pm \delta\omega_{AB}.
	\end{equation}
	
	Now taking the limits \(N\to\infty\), we arrive at the closed-form expression for \(\delta\omega_\zeta\) in the continuous limit.
	\begin{equation}
		\begin{split}
			&\delta\omega_\zeta = \frac{\kappa^2}{2\pi^2} \int_0^\infty d\Omega \, e^{-\Omega/\Omega_c} \left[1 \pm \gamma_U(\Omega)\right] \\
			&= \frac{\kappa^2}{2\pi^2} \left[ \Omega_c \pm \frac{1}{d\sqrt{1+(ad/2)^2}} \arctan\left( \frac{2\Omega_c}{a} \operatorname{arsinh}\frac{ad}{2} \right) \right].
		\end{split}
	\end{equation}
	
	The total Hamiltonian can then be decomposed into two sectors, each of which corresponds to a fully quadratic model consisting of one harmonic oscillator coupled to an independent oscillator bath.
	\begin{equation}
		\begin{split}
			&\mathcal{H}_{\text{tot}} = \sum_{\zeta=\pm}\mathcal{H}_\zeta^{\text{tot}}, \\
			&\mathcal{H}_\zeta^{\text{tot}} = \\
			&\frac{\alpha_\zeta + \delta\omega_\zeta}{2} q_\zeta^2 + \frac{\beta_\zeta}{2} p_\zeta^2 + \sum_j \frac{\Omega_j}{2}\left(Q_{j,\zeta}^2 + P_{j,\zeta}^2\right) + \sum_j g_{j,\zeta} q_\zeta Q_{j,\zeta}.
		\end{split}
	\end{equation}
	
	Next, we carry out a scaling transformation for the system Hamiltonian \(\mathcal{H}_S\)\eqref{hsfj}:
	\begin{equation}
		q'_\zeta = \left( \frac{\alpha_\zeta}{\beta_\zeta} \right)^{1/4} q_\zeta, \qquad
		p'_\zeta = \left( \frac{\beta_\zeta}{\alpha_\zeta} \right)^{1/4} p_\zeta.
	\end{equation}
	so
	\begin{equation}
		\mathcal{H}_S=\frac{\nu_S}{2}\sum_{\zeta=\pm1} \left( q_\zeta'^2 + p_\zeta'^2 \right).
	\end{equation}
	It is then straightforward to derive:
	\begin{equation}
		\begin{split}
			&\langle q_\zeta^2 \rangle_0 = \frac{\beta_\zeta}{2\nu_S}, \qquad
			\langle p_\zeta^2 \rangle_0 = \frac{\alpha_\zeta}{2\nu_S}, \\
			&\langle q_+ q_- \rangle_0 = \langle p_+ p_- \rangle_0 = 0, \\
&\frac12\langle q_\zeta p_\zeta+p_\zeta q_\zeta\rangle_0=0.
		\end{split}
	\end{equation}
	
	The canonical commutator gives $\langle q_\zeta p_\zeta\rangle_0=i/2$; the covariance uses the symmetrized moment displayed above. The covariance matrix for each sector is given by:
	\begin{equation}
		\sigma_\zeta = \begin{pmatrix}
			\langle q_\zeta^2 \rangle_0 & 0 \\
			0 & \langle p_\zeta^2 \rangle_0
		\end{pmatrix}.
	\end{equation}
	
	We obtain the stationary work from the reduced equilibrium state of the interacting detector--bath Hamiltonian. Local relaxation requires the continuum-reservoir limit before the long-time limit: a finite closed oscillator bath instead has quasiperiodic dynamics. 

	More precisely, the reduced equilibrium state is formally
	defined as
	\begin{equation}
		\rho_S^{\mathrm{eq}}
		=
		\frac{
			\operatorname{Tr}_B
			\left[
			e^{-\mathcal{H}_{\mathrm{tot}}/T_U}
			\right]
		}{Z_{\mathrm{tot}}},
		\label{rhoeq}
	\end{equation}
	where the total partition function is
	\begin{equation}
		Z_{\mathrm{tot}}
		=
		\operatorname{Tr}_{S\otimes B}
		\left[
		e^{-\mathcal{H}_{\mathrm{tot}}/T_U}
		\right].
		\label{ztot}
	\end{equation}
	Here, $\operatorname{Tr}_B$ denotes the partial trace over
	all bath modes, whereas
	$\operatorname{Tr}_{S\otimes B}$ denotes the full trace over
	the Hilbert space of the combined detector system and bath.
	The above continuum expressions are understood as the
	limits of the corresponding finite-bath equilibrium
	quantities.
	
	We now calculate the equilibrium covariance matrix by
	employing the Euclidean action. Setting the interaction
	switching function to its final value, $\chi=1$, the
	Euclidean action for a single collective sector is given by
	\cite{grabert1988}
	\begin{equation}
		\begin{aligned}
			S_E^{(\zeta)}
			=\int_0^{1/T_U}\mathrm{d}u
			\Bigg[
			&\frac{\dot q_\zeta^2}{2\beta_\zeta}
			+\frac{\alpha_\zeta}{2}q_\zeta^2
			+\frac{\delta\omega_\zeta}{2}q_\zeta^2
			\\
			&+\sum_j
			\left(
			\frac{\dot Q_{j,\zeta}^2}{2\Omega_j}
			+\frac{\Omega_j}{2}Q_{j,\zeta}^2
			+g_{j,\zeta}q_\zeta Q_{j,\zeta}
			\right)
			\Bigg],
			\label{se}
		\end{aligned}
	\end{equation}
	where $u$ denotes imaginary time, and the dot represents
	differentiation with respect to $u$. The static counterterm
	is determined by
	\begin{equation}
		\delta\omega_\zeta
		=
		\sum_j\frac{g_{j,\zeta}^2}{\Omega_j}.
	\end{equation}
	
	The Gaussian integration over the bath coordinates
	$Q_{j,\zeta}$ is determined by the Euclidean equation
	\begin{equation}
		\left(
		-\partial_u^2+\Omega_j^2
		\right)Q_{j,\zeta}
		=
		-g_{j,\zeta}\Omega_j q_\zeta,
	\end{equation}
	whose solution in the Matsubara frequency domain reads
	\begin{equation}
		Q_{j,\zeta}(\mathrm{i}\nu_n)
		=
		-\frac{g_{j,\zeta}\Omega_j}
		{\nu_n^2+\Omega_j^2}
		q_\zeta(\mathrm{i}\nu_n),
		\label{qjz}
	\end{equation}
	where the bosonic Matsubara frequencies are defined as
	$\nu_n=2\pi nT_U$, with $n\in\mathbb{Z}$.
	
	Substituting this result back into Eq.~\eqref{se} yields
	the exact inverse Green function:
	\begin{equation}
		G_\zeta^{-1}(\mathrm{i}\nu_n)
		=
		D_{0,\zeta}(\nu_n)
		+\Sigma_\zeta(\nu_n),
		\label{g-1}
	\end{equation}
	where
	\begin{equation}
		\begin{aligned}
			D_{0,\zeta}(\nu_n)
			&=
			\alpha_\zeta+\frac{\nu_n^2}{\beta_\zeta},
			\\
			\Sigma_\zeta(\nu_n)
			&=
			\nu_n^2\Lambda_\zeta(\nu_n),
			\\
			\Lambda_\zeta(\nu_n)
			&\equiv
			\sum_j
			\frac{g_{j,\zeta}^2}
			{\Omega_j(\nu_n^2+\Omega_j^2)}.
			\label{deta}
		\end{aligned}
	\end{equation}
	The static counterterm compensates the zero-frequency
	bath-induced displacement. Consequently, the self-energy
	vanishes at zero frequency, while its frequency-dependent
	part is retained in $\Sigma_\zeta$.
	
	Starting from Eq.~\eqref{g-1}, we carry out a geometric
	series expansion. Since $\Sigma_\zeta=O(\kappa^2)$, we
	obtain
	\begin{equation}
		G_\zeta(\mathrm{i}\nu_n)
		=
		G_{0,\zeta}(\mathrm{i}\nu_n)
		-
		\nu_n^2\Lambda_\zeta(\nu_n)
		G_{0,\zeta}^2(\mathrm{i}\nu_n)
		+O(\kappa^4),
		\label{gz}
	\end{equation}
	where $G_{0,\zeta}\equiv D_{0,\zeta}^{-1}$.
	
	The steady-state coordinate variance can be directly
	obtained from the Green function:
	\begin{equation}
		Q_\zeta^{\mathrm{eq}}
		=
		T_U\sum_{n\in\mathbb{Z}}
		G_\zeta(\mathrm{i}\nu_n).
		\label{qeq}
	\end{equation}
	
	To obtain the momentum covariance, we employ the
	stationary equation for the symmetrized coordinate-momentum
	correlator. In the steady state, its time derivative
	vanishes in the long-time limit:
	\begin{equation}
		\frac{\mathrm{d}}{\mathrm{d}\tau}
		\left\langle
		\frac{\{q_\zeta,p_\zeta\}}{2}
		\right\rangle^{\mathrm{eq}}
		=0.
	\end{equation}

	Substituting this relation into the Heisenberg equations
	of motion,
	\begin{equation}
		\begin{aligned}
			\frac{\mathrm{d}q_\zeta}{\mathrm{d}\tau}
			&=
			\mathrm{i}
			[\mathcal{H}_{\mathrm{tot}},q_\zeta]
			=
			\beta_\zeta p_\zeta,
			\\
			\frac{\mathrm{d}p_\zeta}{\mathrm{d}\tau}
			&=
			\mathrm{i}
			[\mathcal{H}_{\mathrm{tot}},p_\zeta]
			\\
			&=
			-(\alpha_\zeta+\delta\omega_\zeta)q_\zeta
			-\sum_jg_{j,\zeta}Q_{j,\zeta},
		\end{aligned}
	\end{equation}
	we immediately obtain the virial identity for the
	$\zeta$ sector:
	\begin{equation}
		\begin{aligned}
			\beta_\zeta P_\zeta^{\mathrm{eq}}
			=
			(\alpha_\zeta+\delta\omega_\zeta)
			Q_\zeta^{\mathrm{eq}}	
			+
			\sum_jg_{j,\zeta}
			\langle q_\zeta Q_{j,\zeta}\rangle^{\mathrm{eq}}.
		\end{aligned}
		\label{bp}
	\end{equation}

	The system-bath cross-correlation can be evaluated
	using Eq.~\eqref{qjz}:
	\begin{equation}
		\langle q_\zeta Q_{j,\zeta}\rangle^{\mathrm{eq}}
		=
		-T_U\sum_n
		\frac{g_{j,\zeta}\Omega_j}
		{\nu_n^2+\Omega_j^2}
		G_\zeta(\mathrm{i}\nu_n).
	\end{equation}
	
	Using the counterterm relation and the definition of
	$\Lambda_\zeta$, we find
	\begin{equation}
		\begin{aligned}
			&\sum_jg_{j,\zeta}
			\langle q_\zeta Q_{j,\zeta}\rangle^{\mathrm{eq}}
			\\
			&=
			-\delta\omega_\zeta Q_\zeta^{\mathrm{eq}}
			+
			T_U\sum_n
			\nu_n^2\Lambda_\zeta(\nu_n)
			G_\zeta(\mathrm{i}\nu_n).
		\end{aligned}
	\end{equation}
	
	Combining this result with the virial identity
	Eq.~\eqref{bp}, we obtain
	\begin{equation}
		P_\zeta^{\mathrm{eq}}
		=
		\frac{T_U}{\beta_\zeta}
		\sum_n
		\left[
		\alpha_\zeta+\Sigma_\zeta(\nu_n)
		\right]
		G_\zeta(\mathrm{i}\nu_n).
		\label{peq}
	\end{equation}
	
	Moreover, the equilibrium state is invariant under
	time reversal, so that the symmetrized coordinate-momentum
	correlator vanishes:
	\begin{equation}
		\frac{1}{2}
		\langle\{q_\zeta,p_\zeta\}\rangle^{\mathrm{eq}}
		=0.
		\label{ceq}
	\end{equation}
	
	We next introduce the uncoupled thermal reference
	covariances. Using
	\[
	\nu_S=\sqrt{\alpha_\zeta\beta_\zeta}
	=\sqrt{\omega^2-\lambda^2},
	\]
	and the Matsubara summation identity
	\cite{FetterWalecka1971},
	\begin{equation}
		T_U\sum_n
		G_{0,\zeta}(\mathrm{i}\nu_n)
		=
		\frac{\beta_\zeta}{2\nu_S}
		\coth\left(\frac{\nu_S}{2T_U}\right),
		\label{ms}
	\end{equation}
	we define
	\begin{equation}
		Q_{0,\zeta}
		=
		\frac{\beta_\zeta c_U}{2\nu_S},
		\qquad
		P_{0,\zeta}
		=
		\frac{\alpha_\zeta c_U}{2\nu_S},
		\qquad
		c_U=
		\coth\left(\frac{\nu_S}{2T_U}\right).
		\label{q0p0}
	\end{equation}
	These are the uncoupled thermal reference covariances
	used in the equilibrium expansion, rather than the
	initial ground-state covariances of the dynamical
	evolution.
	
	Writing
	\[
	\delta Q_\zeta
	\equiv
	Q_\zeta^{\mathrm{eq}}-Q_{0,\zeta},
	\qquad
	\delta P_\zeta
	\equiv
	P_\zeta^{\mathrm{eq}}-P_{0,\zeta},
	\]
	and combining Eqs.~\eqref{gz} and \eqref{ms}, we obtain
	the coordinate variance correction:
	\begin{equation}
		\delta Q_\zeta
		=
		-T_U\sum_n
		\frac{
			\nu_n^2\Lambda_\zeta(\nu_n)
		}{
			D_{0,\zeta}(\nu_n)^2
		}
		+O(\kappa^4).
		\label{qwuqun}
	\end{equation}
	
	Similarly, expanding Eq.~\eqref{peq} to the same order
	gives the momentum variance correction:
	\begin{equation}
		\delta P_\zeta
		=
		\frac{\alpha_\zeta}{\beta_\zeta}
		\delta Q_\zeta
		+
		\frac{T_U}{\beta_\zeta}
		\sum_n
		\frac{
			\nu_n^2\Lambda_\zeta(\nu_n)
		}{
			D_{0,\zeta}(\nu_n)
		}
		+O(\kappa^4).
		\label{pwupun}
	\end{equation}
	
	We now turn our attention to the Matsubara sums
	\begin{equation}
		T_U\sum_n
		\frac{
			\nu_n^2\Lambda_\zeta(\nu_n)
		}{
			D_{0,\zeta}(\nu_n)^k
		},
		\qquad k=1,2.
	\end{equation}
	Interchanging the order of the bath summation and
	the Matsubara summation, we obtain
	\begin{equation}
		\begin{aligned}
			T_U\sum_n
			\frac{
				\nu_n^2\Lambda_\zeta(\nu_n)
			}{
				D_{0,\zeta}(\nu_n)^k
			}
			=
			\sum_j
			\frac{g_{j,\zeta}^2}{\Omega_j}
			\beta_\zeta^{k}B_k(\Omega_j),
		\end{aligned}
	\end{equation}
	where
	\begin{equation}
		B_k(\Omega)
		\equiv
		T_U\sum_{n\in\mathbb{Z}}
		\frac{\nu_n^2}
		{(\nu_n^2+\Omega^2)
			(\nu_n^2+\nu_S^2)^k},
		\qquad k=1,2.
		\label{bk}
	\end{equation}
	
	Evaluating the Matsubara summation explicitly, we obtain
	\begin{equation}
		B_1(\Omega)
		=
		\frac{
			\Omega\coth\left(\frac{\Omega}{2T_U}\right)
			-\nu_S c_U
		}{
			2(\Omega^2-\nu_S^2)
		}.
	\end{equation}
	The second function can be expressed more compactly
	as
	\begin{equation}
		B_2(\Omega)
		=
		-\frac{1}{2\nu_S}
		\frac{\partial B_1(\Omega)}{\partial\nu_S},
		\label{b2}
	\end{equation}
	where the derivative is taken at fixed $T_U$ and
	$\Omega$. The apparent singularity at $\Omega=\nu_S$
	is removable. In particular,
	\begin{equation}
		B_1(\nu_S)
		=
		\frac{
			c_U-\dfrac{\nu_S}{2T_U}
			\operatorname{csch}^2
			\left(\dfrac{\nu_S}{2T_U}\right)
		}{4\nu_S}.
	\end{equation}
	At zero temperature, these functions reduce to
	\begin{equation}
		B_1(\Omega)
		=
		\frac{1}{2(\Omega+\nu_S)},
		\qquad
		B_2(\Omega)
		=
		\frac{1}{4\nu_S(\Omega+\nu_S)^2}.
	\end{equation}
	
	We now take the continuum-reservoir limit. For the
	three environments $E\in\{U,T,C\}$, the spectral-matching
	relation gives
	\begin{equation}
		J_{E,\zeta}(\Omega)
		=
		\frac{\kappa^2\Omega e^{-\Omega/\Omega_c}}{4\pi}
		[1 \pm \gamma_E(\Omega)],
	\end{equation}
	where $\zeta=+1$ and $\zeta=-1$ denote the symmetric
	and antisymmetric channels, respectively. The discrete
	couplings satisfy
	\begin{equation}
		g_{j,\zeta}^2
		=
		\frac{2w_j}{\pi}
		J_{E,\zeta}(\Omega_j).
	\end{equation}
	Consequently, for a sufficiently dense frequency grid,
	\begin{equation}
		\begin{aligned}
			&\sum_jg_{j,\zeta}^2f(\Omega_j)
			\\
			&\longrightarrow
			\frac{\kappa^2}{2\pi^2}
			\int_0^\infty
			\mathrm{d}\Omega\,
			\Omega e^{-\Omega/\Omega_c}
			[1 \pm \gamma_E(\Omega)]
			f(\Omega).
		\end{aligned}
		\label{continuum}
	\end{equation}
	
	Substituting Eq.~\eqref{continuum} into
	Eqs.~\eqref{qwuqun} and \eqref{pwupun}, we obtain
	\begin{align}
		\delta Q_\zeta
		&=
		-\frac{\kappa^2\beta_\zeta^2}{2\pi^2}
		\int_0^\infty
		\mathrm{d}\Omega\,
		e^{-\Omega/\Omega_c}
		[1 \pm \gamma_E(\Omega)]
		B_2(\Omega)
		+O(\kappa^4),
		\label{lxdp1}
		\\
		\delta P_\zeta
		&=
		\frac{\alpha_\zeta}{\beta_\zeta}
		\delta Q_\zeta
		\nonumber\\
		&\quad+
		\frac{\kappa^2}{2\pi^2}
		\int_0^\infty
		\mathrm{d}\Omega\,
		e^{-\Omega/\Omega_c}
		[1 \pm \gamma_E(\Omega)]
		B_1(\Omega)
		+O(\kappa^4).
		\label{lxdp2}
	\end{align}
	
	Since each collective sector is in a single-mode
	Gaussian equilibrium state, its Gaussian ergotropy
	relative to the corresponding sector Hamiltonian
	$\mathcal{H}_\zeta$ is
	\begin{equation}
		\mathcal{G}_{E,\zeta}^{\mathrm{eq}}
		=
		\frac{
			\alpha_\zeta Q_\zeta^{\mathrm{eq}}
			+\beta_\zeta P_\zeta^{\mathrm{eq}}
		}{2}
		-
		\nu_S
		\sqrt{
			Q_\zeta^{\mathrm{eq}}P_\zeta^{\mathrm{eq}}
			-(\frac{1}{2}
			\langle\{q_\zeta,p_\zeta\}\rangle^{\mathrm{eq}})^2
		}.
	\end{equation}
	The total Gaussian ergotropy is obtained by summing
	over the two collective sectors:
	\begin{equation}
		\mathcal{G}_{E}^{\mathrm{eq}}
		=
		\sum_{\zeta=\pm1}
		\mathcal{G}_{E,\zeta}^{\mathrm{eq}}.
	\end{equation}
	Here the extraction is defined with respect to the
	full detector Hamiltonian $\mathcal{H}_S$ and allows collective
	Gaussian unitaries on the two detectors.
	
	To obtain the leading weak-coupling behavior, we
	introduce the normalized covariance corrections
	\begin{equation}
		x_\zeta
		\equiv
		\frac{\delta Q_\zeta}{Q_{0,\zeta}},
		\qquad
		y_\zeta
		\equiv
		\frac{\delta P_\zeta}{P_{0,\zeta}}.
	\end{equation}
	Since $\frac{1}{2}
	\langle\{q_\zeta,p_\zeta\}\rangle^{\mathrm{eq}}=0$, the ergotropy of
	each sector can be rewritten as
	\begin{equation}
		\mathcal{G}_{E,\zeta}^{\mathrm{eq}}
		=
		\frac{\nu_S c_U}{4}
		\left[
		2+x_\zeta+y_\zeta
		-2\sqrt{(1+x_\zeta)(1+y_\zeta)}
		\right].
	\end{equation}
	
	Expanding the square root, the terms linear in
	$x_\zeta$ and $y_\zeta$ cancel. Therefore,
	\begin{equation}
		\mathcal{G}_{E}^{\mathrm{eq}}
		=
		\frac{\nu_S c_U}{16}
		\sum_{\zeta=\pm1}
		(x_\zeta-y_\zeta)^2
		+O(\kappa^6).
		\label{gen}
	\end{equation}
	
	Moreover, since
	\[
	\frac{P_{0,\zeta}}{Q_{0,\zeta}}
	=
	\frac{\alpha_\zeta}{\beta_\zeta},
	\]
	the contribution proportional to $\delta Q_\zeta$
	in Eqs.~\eqref{lxdp1} and \eqref{lxdp2} cancels
	in the difference $x_\zeta-y_\zeta$. We consequently
	obtain
	\begin{equation}
		x_\zeta-y_\zeta
		=
		-\frac{\kappa^2\nu_S}
		{\pi^2\alpha_\zeta c_U}
		I_{E,\zeta}
		+O(\kappa^4),
		\label{xy}
	\end{equation}
	where
	\begin{equation}
		I_{E,\zeta}
		\equiv
		\int_0^\infty
		\mathrm{d}\Omega\,
		e^{-\Omega/\Omega_c}
		[1 \pm \gamma_E(\Omega)]
		B_1(\Omega).
		\label{IE}
	\end{equation}
	
	Substituting Eq.~\eqref{xy} into Eq.~\eqref{gen},
	we finally obtain
	\begin{equation}
		\mathcal{G}_{E}^{\mathrm{eq}}
		=
		\kappa^4 \aleph_E+O(\kappa^6),
		\label{GE}
	\end{equation}
	where
	\begin{equation}
		\aleph_E
		=
		\frac{\nu_S^3}{16\pi^4c_U}
		\sum_{\zeta=\pm1}
		\frac{I_{E,\zeta}^2}{\alpha_\zeta^2}.
		\label{NE}
	\end{equation}
	Equivalently,
	\begin{equation}
		\aleph_E
		=
		\lim_{\kappa\to0}
		\frac{\mathcal{G}_{E}^{\mathrm{eq}}}{\kappa^4}.
	\end{equation}
	
	Thus, in the weak-coupling regime, the equilibrium
	Gaussian ergotropy exhibits a quartic dependence on
	the detector--bath coupling:
	\begin{equation}
			\mathcal{G}_{E}^{\mathrm{eq}}\propto\kappa^4.
	\end{equation}
	\section{Gaussian Work as a Probe of Spatial Correlations}
	\label{cr:main}

	The accelerated vacuum and the matched inertial thermal field share
	the same local spectrum. Their distinction is encoded in the cross
	spectrum sampled by the two detectors. We now determine how this
	difference enters the leading-order work coefficient
	$\aleph_E$ derived in Sec.~\ref{tau}. Throughout this analysis the static interaction is fully compensated and the extraction reference is the common Hamiltonian $\mathcal H_S$.

	For convenience, we write the geometric factors obtained in Sec.~\ref{hmd} as
	\begin{align}
		L(a,d)&=\frac{2}{a}\operatorname{arsinh}\frac{ad}{2},\notag\\
		D(a,d)&=d\sqrt{1+(ad/2)^2},\notag\\
		\gamma_U(\Omega,a,d)&=\frac{\sin[\Omega L(a,d)]}{\Omega D(a,d)},\qquad
		\gamma_T(\Omega,d)=\frac{\sin(\Omega d)}{\Omega d}.
		\label{add:geometry}
	\end{align}
	Defining $b(a,d)\equiv L(a,d)/D(a,d)$,
	we can write
	\begin{equation}
		\gamma_U(\Omega,a,d)
		=b(a,d)\,\mathrm{sinc}\!\left[\Omega L(a,d)\right],
	\end{equation}
	where $\mathrm{sinc}(x)\equiv\sin x/x$, with
	$\mathrm{sinc}(0)=1$.
	
	We choose the exponential regulator as
	\begin{equation}
	r_{\mathrm{exp}}(\Omega)
	= e^{-\Omega/\Omega_c}.
	\end{equation} To examine the dependence on its form, we also introduce a Gaussian regulator:
	\begin{equation}
		r_G(\Omega)=e^{-\pi\Omega^2/(4\Omega_c^2)}.
		\label{add:regulators}
	\end{equation}
	Both satisfy $\int_0^\infty r(\Omega)d\Omega=\Omega_c$. In the following formulas, $r$ denotes either specified frequency regulator; setting $r=r_{\exp}$ recovers the preceding model.
	\subsection{The low-frequency collective response}
	The symmetric and antisymmetric channels sample the combinations $1\pm\gamma_E$. Their low-frequency behavior therefore exposes how the same local spectrum can produce different collective fluctuations. Put $b=L/D$, so that $0<b<1$ for $a,d>0$. Expanding the sine factors at fixed geometry gives
	\begin{align}
		J_{U-}(\Omega)&=\frac{\kappa^2r(\Omega)}{4\pi}
		\left[(1-b)\Omega+\frac{bL^2}{6}\Omega^3+O(\Omega^5)\right],\notag\\
		J_{T-}(\Omega)&=\frac{\kappa^2r(\Omega)}{4\pi}
		\left[\frac{d^2}{6}\Omega^3+O(\Omega^5)\right].
		\label{cr:infrared}
	\end{align}
	For a common regulator continuous at the origin with $r(0)=1$, the antisymmetric accelerated channel is therefore Ohmic, while the matched inertial thermal channel is super-Ohmic with exponent three. The symmetric channels are Ohmic in both environments.
	
	The symmetrized force spectrum is $S_{E-}(\Omega)=J_{E-}(\Omega)\coth[\Omega/(2T_U)]$. At fixed $T_U>0$, Eq.~\eqref{cr:infrared} gives
	\begin{align}
		S_{U-}(\Omega)&\longrightarrow\frac{\kappa^2T_U}{2\pi}(1-b),\notag\\
		S_{T-}(\Omega)&\sim\frac{\kappa^2T_Ud^2}{12\pi}\Omega^2.
		\label{cr:infrared_noise}
	\end{align}
	Thus the antisymmetric coordinate samples a finite low-frequency noise floor in U, whereas its noise in T is suppressed quadratically. Gaussian work, however, samples the full frequency-dependent detector response rather than the infrared limit alone. 
	
	For small separation at fixed $a$ and $\Omega$, the same expansion takes the form
	\begin{align}
		J_{U-}(\Omega)&=\frac{\kappa^2r(\Omega)d^2}{24\pi}
		(a^2\Omega+\Omega^3)+O(d^4),\notag\\
		J_{T-}(\Omega)&=\frac{\kappa^2r(\Omega)d^2}{24\pi}\Omega^3+O(d^4).
		\label{cr:near_dark}
	\end{align}
	 At exact coincidence the antisymmetric spectrum vanishes in both environments, and at $a=0$ the two spectra agree. The Ohmic and cubic responses refer to fixed $a,d>0$. 
	\subsection{From the weighted correlation to work}
	To isolate the spatial input from the common local response, we separate the local and cross contributions to the integral in Eq.~\eqref{IE}:
	\begin{align}
		I_0&=\int_0^\infty d\Omega\,r(\Omega)B_1(\Omega),\notag\\
		I_{c,E}&=\int_0^\infty d\Omega\,r(\Omega)\gamma_E(\Omega)B_1(\Omega),\notag\\
		\eta_E&=\frac{I_{c,E}}{I_0},\qquad E\in\{U,T,C\}.
		\label{cr:integrals}
	\end{align}
	The weight $r(\Omega)B_1(\Omega)$ is fixed by the detector frequency, the magnitude of the direct coupling, the temperature, and the regulator. We hold $(\omega,|\lambda|,T_U,r)$ fixed when comparing the spatial correlations; changing the sign of $\lambda$ leaves $\nu_S$ and this weight unchanged. Thus $\eta_E$ is one response-weighted spatial correlation. 

	 Since $B_1$ is positive and $|\gamma_E|\leq1$, the weighted correlation satisfies $|\eta_E|\leq1$. We have
	\begin{equation}
		I_{E,\zeta}=I_0(1\pm \eta_E),
	\end{equation}
	and $\eta_C=0.$

	Using the geometric factor, we obtain the following relation between the weighted spatial correlations:
	\begin{equation}
	\eta_U(a,d)=b(a,d)\eta_T(T_U,L(a,d)).
	\label{cr:distance_identity}
	\end{equation}
	Both sides are evaluated at the same temperature,
	$T_U=a/(2\pi)$, showing that acceleration modifies the
	structure of the weighted correlation. This identity
	separates the geometric effects from the common local
	thermal response.

	Substituting these expressions into Eq.~\eqref{NE}, we obtain
	\begin{align}
		\aleph_C&=\frac{I_0^2(\omega^2+\lambda^2)}{8\pi^4\nu_Sc_U},\notag\\
		R_E\equiv\frac{\aleph_E}{\aleph_C}
		&=1+\eta_E^2-s\eta_E,\qquad
		s=\frac{4\omega\lambda}{\omega^2+\lambda^2}.
		\label{cr:criterion}
	\end{align}
	Thus, the cross correlation enters the leading-order work coefficient through a quadratic dependence on the weighted correlation. In particular, for $\lambda=0$ we find $R_E=1+\eta_E^2$. For positive $\lambda$ and $0<\eta_E<1$, correlations enhance the coefficient relative to $C$ when $\eta_E>s$, and suppress it when $\eta_E<s$. The minimum occurs at
	\begin{equation}
		\eta_E=\frac{s}{2},\qquad
		R_{\min}=1-\frac{s^2}{4}
		=\frac{(\omega^2-\lambda^2)^2}{(\omega^2+\lambda^2)^2}.
		\label{cr:minimum}
	\end{equation}
	
	This quadratic dependence establishes how the weighted correlation
	is encoded in the work and determines what can be inferred from
	work measurements. We now apply this relation to investigate the
	distinguishability between the accelerated and inertial environments.
	\section{Equilibrium correlation readout and finite-coupling tests}
	\label{sec:readout}
	\subsection{The unique finite-distance blind point}
	Subtracting the two physical coefficients in Eq.~\eqref{cr:criterion} gives
	\begin{equation}
		\frac{\aleph_U-\aleph_T}{\aleph_C}
		=(\eta_U-\eta_T)(\eta_U+\eta_T-s).
		\label{cr:contrast}
	\end{equation}
	For the exponential and Gaussian regulators in Eq.~\eqref{add:regulators}, the weighted correlations satisfy
	\begin{equation}
		0<\eta_U(a,d)<\eta_T(T_U,d)<1,
		\qquad a,d>0.
		\label{cr:ordering}
	\end{equation}
	Both weighted correlations decrease strictly with separation at fixed $a>0$. Appendix~\ref{cr:proof_order} proves these properties by expressing the cross integrals through a positive, decreasing cosine transform. 
	
	Since $\eta_U+\eta_T$ decreases continuously and monotonically from $2$ to $0$ as the separation $d$ increases, and $0<s<2$ for $0<\lambda<\omega$, there exists a unique finite and nonzero separation $d_0>0$, which we refer to as the blind point, at which the two environments cannot be distinguished by a single-setting work measurement. It is determined by
	\begin{equation}
		\eta_U(a,d_0)+\eta_T(T_U,d_0)=s.
		\label{cr:blind}
	\end{equation}
	At this separation the leading work contrast vanishes although the field correlations remain different. For $0<d<d_0$ we have $\aleph_U<\aleph_T$, whereas for $d>d_0$ we have $\aleph_U>\aleph_T$. Each coefficient has one distance minimum, and their positions obey
	\begin{equation}
		d_{\min,U}<d_0<d_{\min,T}.
		\label{cr:min_order}
	\end{equation}
	For $\lambda\leq0$, the work ordering is $\aleph_U<\aleph_T$ at every finite $a,d>0$. At $d=0$ the two environments share $\eta_U=\eta_T=1$, and their work contrast vanishes. The contrast also tends to zero as $d\to\infty$ and is identically zero at $a=0$. The point $d_0$ is therefore the unique interior zero for positive direct coupling: two different weighted correlations give the same leading work.
	\subsection{Two-setting reconstruction of the weighted correlation}
	In this subsection, we show how to overcome this blind point. We take $\lambda=+g$ in the first equilibrium preparation and $\lambda=-g$ in the second, with $0<g<\omega$ and all other parameters fixed. The normalized work difference between the two settings is
	\begin{align}
		\mathcal A_E&=
		\frac{\aleph_E(-g)-\aleph_E(+g)}{\aleph_E(-g)+\aleph_E(+g)}\notag
		=\frac{s_g\eta_E}{1+\eta_E^2},\\
		s_g&=\frac{4\omega g}{\omega^2+g^2}.
		\label{cr:asymmetry}
	\end{align}
	The common factor containing $I_0^2$ cancels. Differentiating the last expression gives
	\begin{equation}
		\frac{d\mathcal A}{d\eta}
		=\frac{s_g(1-\eta^2)}{(1+\eta^2)^2}>0,
		\qquad 0<\eta<1.
		\label{cr:asymmetry_derivative}
	\end{equation}
	Combining Eqs.~\eqref{cr:ordering} and \eqref{cr:asymmetry_derivative}, we find
	\begin{equation}
		0<\mathcal A_U<\mathcal A_T<\frac{s_g}{2}.
	\end{equation}
	Thus the two settings remove the finite-distance blind point of the leading coefficient. 
	
	Solving Eq.~\eqref{cr:asymmetry} for the root in $0<\eta<1$, we obtain
	\begin{equation}
		\eta_E=\frac{2\mathcal A_E}
		{s_g+\sqrt{s_g^2-4\mathcal A_E^2}}.
		\label{cr:inverse}
	\end{equation}
	The common factor $I_0^2$ cancels in the normalized work
	asymmetry, allowing the weighted spatial correlation to be
	determined without separately evaluating the local spectral
	response integral.
\subsection{Matched calculations and full Gibbs tests}
In practice, the coupling strength $\kappa$ cannot be taken arbitrarily close to zero and must remain finite. This raises the question of whether the preceding theoretical relations can still distinguish the two environments at finite coupling. To address this question, we retain the full interacting Green function without performing a weak-coupling expansion, obtain the covariance matrix of the full interacting Gibbs state, and calculate the Gaussian ergotropy directly from this covariance matrix. We first examine both the single-setting and two-setting readout schemes at the blind-point separation $d_0$ predicted by the weak-coupling theory. At this separation, we examine whether the full Gibbs calculation can still distinguish the two environments and whether the two-setting scheme improves the distinction. We further substitute the finite-coupling work values into the inversion formula derived in the weak-coupling regime and compare the reconstructed weighted correlations $\widehat{\eta}_E$ with the theoretical values $\eta_E$ obtained directly from their defining integrals. These comparisons allow us to assess the applicability of the preceding weak-coupling relations at finite coupling.

In the following calculations, we adopt the exponential regulator and set $\omega=1$, $a=2\pi$, $T_U=1$, $\Omega_c=5$, and $g=0.1$. All Gaussian ergotropies are evaluated with respect to the same detector Hamiltonian $\mathcal{H}_S$.

\begin{figure*}[!t]
\centering
\includegraphics[width=.98\textwidth]{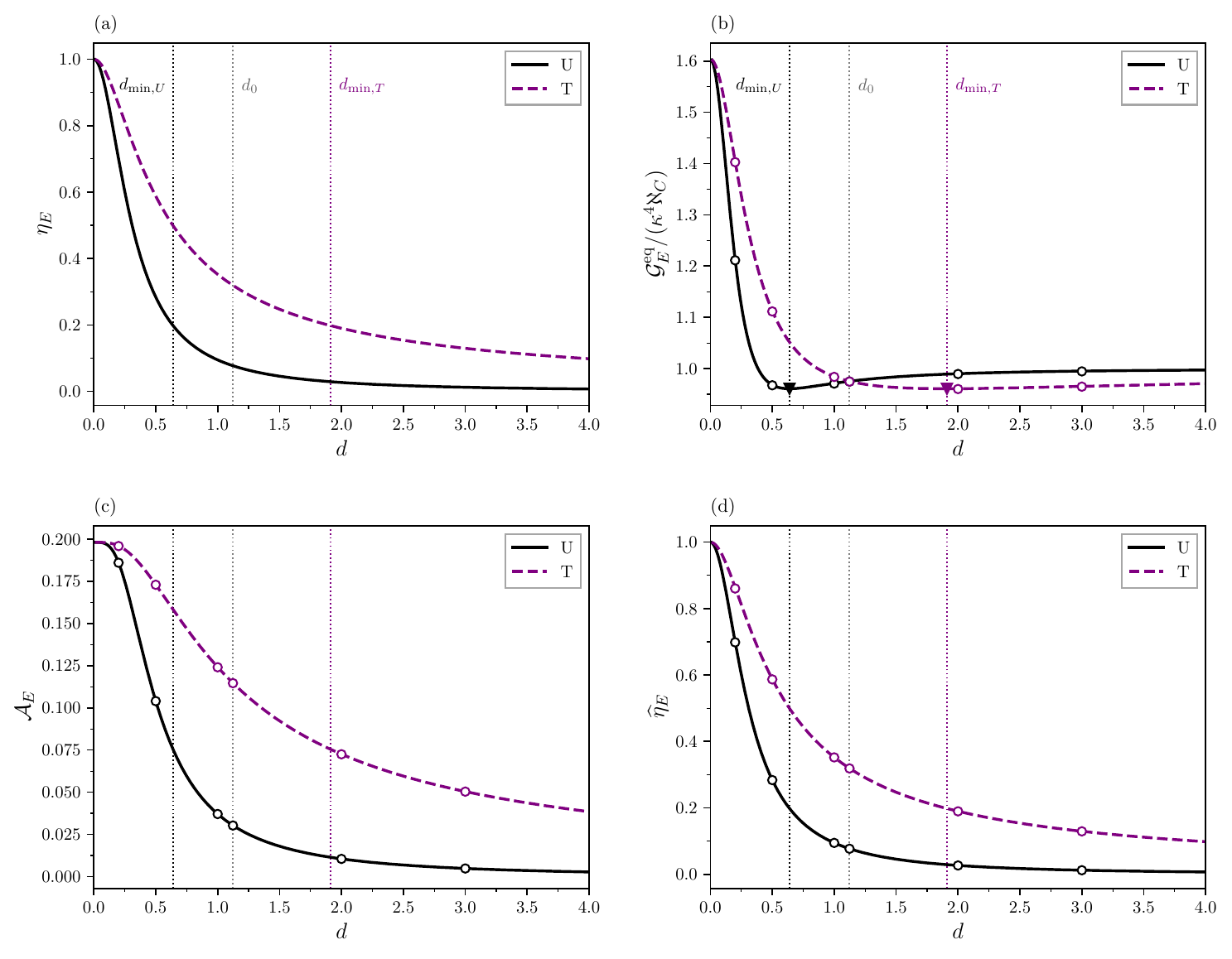}
\caption{Dependence of the weighted correlation and Gaussian ergotropy on the detector separation $d$ in the accelerated vacuum (U) and the inertial thermal field (T), with fixed parameters $\omega=1$, $a=2\pi$, and $\Omega_c=5$. (a) Theoretical weighted correlation $\eta_E$. (b) Normalized Gaussian ergotropy for $\lambda=0.1$. (c) $\mathcal A_E$ for $\lambda=\pm0.1$. (d) Weighted correlation $\widehat{\eta}_E$ reconstructed from the work asymmetry. The solid and dashed curves represent the weak-coupling results, while the open circles show the finite-coupling results at $\kappa^2=0.05$, obtained using the full interacting Green function without a weak-coupling expansion. The vertical dotted lines mark $d_{\min,U}$, $d_0$, and $d_{\min,T}$.}
\label{fig:matched_readout}
\end{figure*}

Figure~\ref{fig:matched_readout} displays the complete chain from the weighted correlation to work and back.
For a single interaction setting, the two work coefficients
intersect at the blind-point separation $d_0$, while their
respective minima occur at $d_{\min,U}$ and $d_{\min,T}$.
Numerically, we obtain
\begin{equation}
	\begin{aligned}
		d_0 &= 1.1237119953,\\
		d_{\min,U} &= 0.6396353494,\\
		d_{\min,T} &= 1.9104614104.
	\end{aligned}
	\label{num:blindlocations}
\end{equation}
These values satisfy the analytical ordering in Eq.~\eqref{cr:min_order}. At $d_0$, the leading correlations are $\eta_U=0.0770176168$ and $\eta_T=0.3190219871$: the equality of the single-setting work occurs between distinct spatial correlations.

The solid and dashed curves in Fig.~\ref{fig:matched_readout}
are obtained from the leading-order weak-coupling expressions.
In contrast, the open circles are calculated at finite coupling
$\kappa^2=0.05$, retaining the full interacting Green function
without performing a weak-coupling expansion. At the blind-point
separation $d_0$ predicted by the weak-coupling analytical
expressions, the Gaussian ergotropies obtained from the full
interacting Gibbs covariance are
\begin{align}
\mathcal G_U(+g)&=1.47351849\times10^{-7},\notag\\
\mathcal G_T(+g)&=1.47344610\times10^{-7},\notag\\
\mathcal G_U(-g)&=1.56559763\times10^{-7},\notag\\
\mathcal G_T(-g)&=1.85483574\times10^{-7}.
\label{num:blindwork}
\end{align}
At the blind point predicted by the weak-coupling theory,
we find that $\mathcal G_U(+g)$ and $\mathcal G_T(+g)$
are nearly identical. Using both interaction settings,
we obtain $\mathcal A_{U,W}=0.03029800$ and
$\mathcal A_{T,W}=0.11459055$. This demonstrates that
the two-setting measurement retains a finite ability to
distinguish the two environments precisely where the
single-setting contrast vanishes.

Substituting the work values in Eq.~\eqref{num:blindwork} into Eq.~\eqref{cr:inverse}, we reconstruct
$\widehat{\eta}_U=0.07695551$ and
$\widehat{\eta}_T=0.31873607$.
These values are in close agreement with the theoretical weighted correlations, $\eta_U=0.07701762$ and $\eta_T=0.31902199$, indicating that the two-setting reconstruction scheme retains its ability to distinguish the two environments at finite coupling.

\begin{figure*}[!t]
\centering
\includegraphics[width=.98\textwidth]{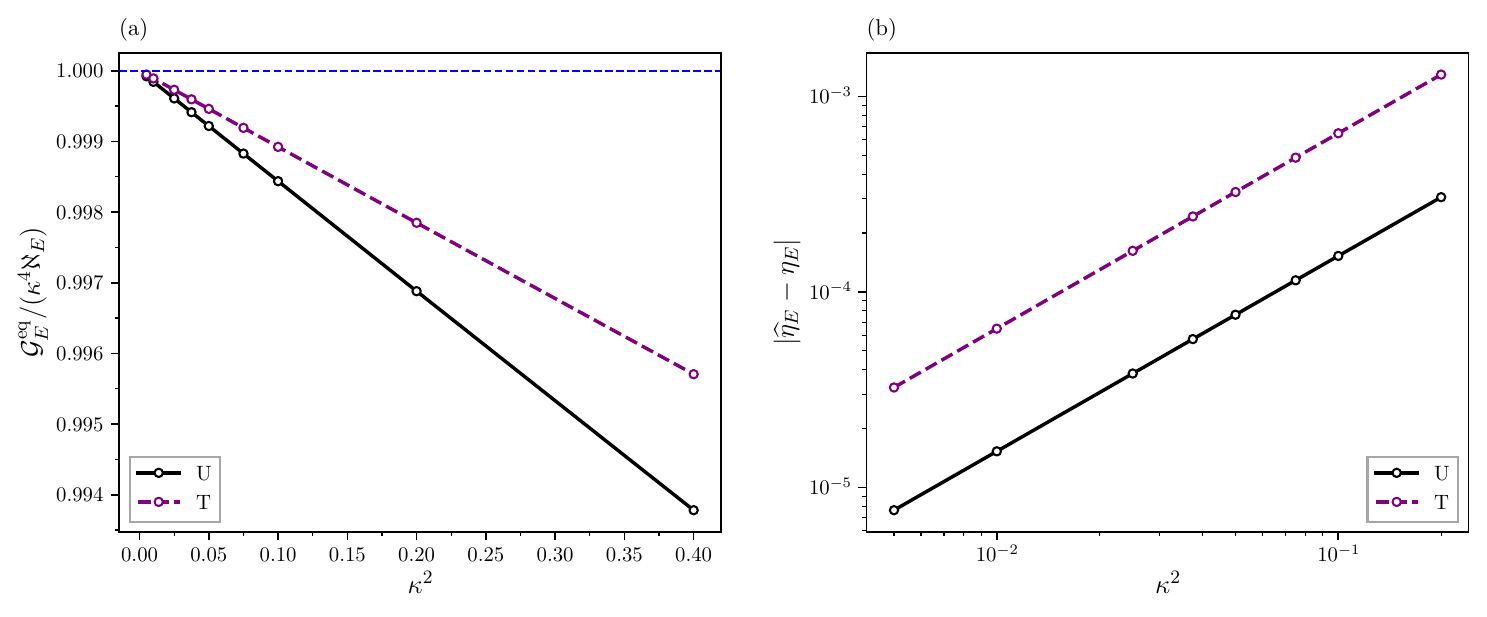}
\caption{Dependence of the equilibrium Gaussian ergotropy and the correlation reconstruction error on the detector--field coupling strength $\kappa^2$ in the accelerated vacuum (U) and the inertial thermal field (T), with fixed parameters $\omega=1$, $d=1$, $a=2\pi$, and $\Omega_c=5$. (a) Equilibrium Gaussian ergotropy normalized by its leading-order contribution $\kappa^4\mathcal N_E$, for $\lambda=0.5$; the blue dashed line indicates the weak-coupling limit. (b) Absolute reconstruction error $|\widehat{\eta}_E-\eta_E|$ obtained from the two settings $\lambda=\pm0.1$. The solid curves are obtained using the full interacting Green function without a weak-coupling expansion.}
\label{fig:matched_finite}
\end{figure*}

Figure~\ref{fig:matched_finite}(a) shows the equilibrium
Gaussian ergotropy normalized by its leading-order
weak-coupling contribution $\kappa^4\aleph_E$.
For both environments, the normalized ergotropy
decreases monotonically with increasing $\kappa^2$
and remains below unity over the displayed range.
This indicates that the full Gibbs ergotropy is lower
than the leading-order prediction at finite coupling.
The decrease is more pronounced for the accelerated
vacuum than for the inertial thermal field, revealing
a stronger finite-coupling correction in the former.
As $\kappa^2$ decreases, both curves approach unity,
and the deviation from the leading-order prediction
gradually disappears. Figure~\ref{fig:matched_finite}(b) shows the deviations of the reconstructed weighted correlations from their theoretical values. The reconstruction errors decrease as the detector--field coupling becomes weaker, exhibiting an approximately linear dependence on $\kappa^2$. The error for the inertial thermal field remains larger than that for the accelerated vacuum, indicating different finite-coupling corrections in the two environments. Both errors vanish in the weak-coupling limit, consistent with the $O(\kappa^2)$ correction predicted by the leading-order inversion formula. Taken together, these results demonstrate that the weak-coupling theory provides a reliable description of the full Gibbs calculation and that the two-setting scheme remains effective for reconstructing the weighted spatial correlations at finite coupling.

\section{Spatial correlations and the additional global work}
\label{st:global}
The equilibrium work readout established above relies on joint Gaussian operations on the two detectors. This raises a further question: how much of the extracted work can already be obtained through independent local operations on the two detectors, and how much constitutes an additional contribution accessible only through joint operations? To distinguish these two contributions, we set $\lambda=0$, so that the extraction Hamiltonian becomes additive,
\begin{equation}
	\mathcal{H}_S=\frac{\omega}{2}
	\left(q_A^2+p_A^2+q_B^2+p_B^2\right)
	=\frac{\omega}{2}
	\left(q_+^2+p_+^2+q_-^2+p_-^2\right).
	\label{eq:HS_additive}
\end{equation}

At equilibrium, we define the second moments of the collective coordinates and momenta as
\begin{equation}
	Q_\zeta^{\mathrm{eq}}
	=\langle q_\zeta^2\rangle_{\mathrm{eq}},
	\qquad
	P_\zeta^{\mathrm{eq}}
	=\langle p_\zeta^2\rangle_{\mathrm{eq}},
	\qquad \zeta=\pm.
	\label{eq:equilibrium_moments}
\end{equation}
The corresponding collective covariance matrices are
\begin{equation}
	\sigma_\zeta^{\mathrm{eq}}
	=
	\begin{pmatrix}
		Q_\zeta^{\mathrm{eq}}&0\\
		0&P_\zeta^{\mathrm{eq}}
	\end{pmatrix}.
	\label{eq:collective_covariance}
\end{equation}
Transforming back to the detector basis, the equilibrium covariance matrix takes the form
\begin{equation}
	\sigma_{AB}^{\mathrm{eq}}
	=
	\begin{pmatrix}
		\bar{\sigma}^{\mathrm{eq}}&\sigma_c^{\mathrm{eq}}\\
		\sigma_c^{\mathrm{eq}}&\bar{\sigma}^{\mathrm{eq}}
	\end{pmatrix},
	\label{eq:detector_covariance}
\end{equation}
where
\begin{equation}
	\bar{\sigma}^{\mathrm{eq}}
	=\frac{\sigma_+^{\mathrm{eq}}+\sigma_-^{\mathrm{eq}}}{2},
	\qquad
	\sigma_c^{\mathrm{eq}}
	=\frac{\sigma_+^{\mathrm{eq}}-\sigma_-^{\mathrm{eq}}}{2}.
	\label{eq:local_cross_covariance}
\end{equation}
Here, $\bar{\sigma}^{\mathrm{eq}}$ is the covariance matrix of either local detector state, while $\sigma_c^{\mathrm{eq}}$ describes the inter-detector covariance. The equality of the two local covariance matrices follows from the symmetry between the detectors.

We first consider local work extraction, for which the allowed Gaussian unitary operations are restricted to independent transformations on the two detectors. The symplectic eigenvalue of either local covariance matrix is
\begin{equation}
	\bar{\nu}^{\mathrm{eq}}
	=\sqrt{\det\bar{\sigma}^{\mathrm{eq}}}.
	\label{eq:local_symplectic_eigenvalue}
\end{equation}
For a single detector with the extraction Hamiltonian
$\mathcal{H}_A=\omega(q_A^2+p_A^2)/2$, the minimum energy attainable under Gaussian unitary operations is $\omega\bar{\nu}^{\mathrm{eq}}$. The same result holds for detector B. Consequently, the total local passive energy is
\begin{equation}
	E_{\mathrm{pas}}^{\mathrm{loc,eq}}
	=2\omega\bar{\nu}^{\mathrm{eq}}
	=2\omega\sqrt{\det\bar{\sigma}^{\mathrm{eq}}}.
	\label{eq:local_passive_energy}
\end{equation}
The initial energy of the equilibrium detector state, evaluated with respect to $\mathcal{H}_S$, is
\begin{equation}
	E_S^{\mathrm{eq}}
	=\frac{\omega}{2}
	\left(
	Q_+^{\mathrm{eq}}+P_+^{\mathrm{eq}}
	+Q_-^{\mathrm{eq}}+P_-^{\mathrm{eq}}
	\right).
	\label{eq:equilibrium_detector_energy}
\end{equation}
It follows that the local Gaussian ergotropy is
\begin{equation}
\mathcal G_{\mathrm{loc}}^{\mathrm{eq}}
		=E_S^{\mathrm{eq}}
		-2\omega\sqrt{\det\bar{\sigma}^{\mathrm{eq}}}.	
	\label{eq:local_ergotropy}
\end{equation}

We now turn to global work extraction, where joint Gaussian unitary operations are allowed on the entire two-detector system. In the collective-mode representation, the symplectic eigenvalues of the equilibrium covariance matrix are
\begin{equation}
	\nu_\pm^{\mathrm{eq}}
	=\sqrt{\det\sigma_\pm^{\mathrm{eq}}}
	=\sqrt{Q_\pm^{\mathrm{eq}}P_\pm^{\mathrm{eq}}}.
	\label{eq:global_symplectic_eigenvalues}
\end{equation}
Since both collective modes have the same frequency $\omega$, the global passive energy is given by the sum of their minimum energies,
\begin{equation}
	E_{\mathrm{pas}}^{\mathrm{glob,eq}}
	=\omega\left(\nu_+^{\mathrm{eq}}+\nu_-^{\mathrm{eq}}\right)
	=\omega\left(
	\sqrt{\det\sigma_+^{\mathrm{eq}}}
	+\sqrt{\det\sigma_-^{\mathrm{eq}}}
	\right).
	\label{eq:global_passive_energy}
\end{equation}
The corresponding global Gaussian ergotropy is therefore
\begin{equation}
		\mathcal G_{\mathrm{glob}}^{\mathrm{eq}}
		=E_S^{\mathrm{eq}}
		-\omega\left(
		\sqrt{\det\sigma_+^{\mathrm{eq}}}
		+\sqrt{\det\sigma_-^{\mathrm{eq}}}
		\right).
	\label{eq:global_ergotropy}
\end{equation}

Comparing the two extraction protocols, we define the additional work accessible through joint operations as
\begin{equation}
	\Delta\mathcal G_{\mathrm{corr}}^{\mathrm{eq}}
	=\mathcal G_{\mathrm{glob}}^{\mathrm{eq}}
	-\mathcal G_{\mathrm{loc}}^{\mathrm{eq}}.
	\label{eq:additional_work_definition}
\end{equation}
Substituting the above expressions, the common initial energy $E_S^{\mathrm{eq}}$ cancels, yielding
\begin{equation}
		\Delta\mathcal G_{\mathrm{corr}}^{\mathrm{eq}}
		=\omega\left[
		2\sqrt{\det\bar{\sigma}^{\mathrm{eq}}}
		-\sqrt{\det\sigma_+^{\mathrm{eq}}}
		-\sqrt{\det\sigma_-^{\mathrm{eq}}}
		\right].
		\label{eq:additional_global_work}
\end{equation}

The non-negativity of this quantity can be established explicitly. Define
\begin{equation}
	\begin{aligned}
		A_{\mathrm{eq}}
		&=\sqrt{
			(Q_+^{\mathrm{eq}}+Q_-^{\mathrm{eq}})
			(P_+^{\mathrm{eq}}+P_-^{\mathrm{eq}})
		},\\
		B_{\mathrm{eq}}
		&=\sqrt{Q_+^{\mathrm{eq}}P_+^{\mathrm{eq}}}
		+\sqrt{Q_-^{\mathrm{eq}}P_-^{\mathrm{eq}}}.
	\end{aligned}
	\label{eq:AB_definitions}
\end{equation}
Equation~\eqref{eq:additional_global_work} then gives
\begin{equation}
	\frac{\Delta\mathcal G_{\mathrm{corr}}^{\mathrm{eq}}}{\omega}
	=A_{\mathrm{eq}}-B_{\mathrm{eq}}.
\end{equation}
Using
\begin{equation}
	\begin{aligned}
		A_{\mathrm{eq}}^2-B_{\mathrm{eq}}^2
		={}&Q_+^{\mathrm{eq}}P_-^{\mathrm{eq}}
		+Q_-^{\mathrm{eq}}P_+^{\mathrm{eq}}\\
		&-2\sqrt{
			Q_+^{\mathrm{eq}}P_+^{\mathrm{eq}}
			Q_-^{\mathrm{eq}}P_-^{\mathrm{eq}}
		}\\
		={}&\left(
		\sqrt{Q_+^{\mathrm{eq}}P_-^{\mathrm{eq}}}
		-\sqrt{Q_-^{\mathrm{eq}}P_+^{\mathrm{eq}}}
		\right)^2,
	\end{aligned}
\end{equation}
we obtain
\begin{equation}
		\frac{\Delta\mathcal G_{\mathrm{corr}}^{\mathrm{eq}}}{\omega}
		=
		\frac{
			\left(
			\sqrt{Q_+^{\mathrm{eq}}P_-^{\mathrm{eq}}}
			-\sqrt{Q_-^{\mathrm{eq}}P_+^{\mathrm{eq}}}
			\right)^2
		}{
			A_{\mathrm{eq}}+B_{\mathrm{eq}}
		}
		\geq0.
	\label{eq:positive_square}
\end{equation}
Thus, joint Gaussian operations can never yield less extractable work than independent local operations. Equality holds precisely when
\begin{equation}
	Q_+^{\mathrm{eq}}P_-^{\mathrm{eq}}
	=Q_-^{\mathrm{eq}}P_+^{\mathrm{eq}},
\end{equation}
or, equivalently,
\begin{equation}
		\frac{Q_+^{\mathrm{eq}}}{P_+^{\mathrm{eq}}}
		=
		\frac{Q_-^{\mathrm{eq}}}{P_-^{\mathrm{eq}}}.
	\label{eq:equal_squeezing_condition}
\end{equation}
This condition means that the two collective covariance matrices are proportional. Hence, a difference in the overall fluctuation amplitudes of the two collective modes is not sufficient to produce additional global work; the two modes must also exhibit different ratios of position and momentum fluctuations.

To establish the connection between the additional global
work and the spatial correlations induced by the common
environment, we consider the weak-coupling expansion.
At zero detector--field coupling, the two collective modes
reduce to identical thermal states with
$Q_\pm^{\mathrm{eq}}=P_\pm^{\mathrm{eq}}=c_U/2$,
where $c_U=\coth(\omega/2T_U)$.
For weak but finite coupling, the equilibrium variances
therefore admit the expansions
\begin{align}
	Q_\zeta^{\mathrm{eq}}
	&=\frac{c_U}{2}
	+\kappa^2q_\zeta^{(1)}+O(\kappa^4),\notag\\
	P_\zeta^{\mathrm{eq}}
	&=\frac{c_U}{2}
	+\kappa^2p_\zeta^{(1)}+O(\kappa^4),
	\qquad \zeta=\pm,
	\label{secVII:collectiveseries}
\end{align}
where the first-order Taylor coefficients are decomposed
into their average and half-difference as
\begin{equation}
	\begin{aligned}
		q_0&=\frac{q_+^{(1)}+q_-^{(1)}}{2},
		&
		q_c&=\frac{q_+^{(1)}-q_-^{(1)}}{2},
		\\
		p_0&=\frac{p_+^{(1)}+p_-^{(1)}}{2},
		&
		p_c&=\frac{p_+^{(1)}-p_-^{(1)}}{2}.
	\end{aligned}
\end{equation}
Thus, $q_0$ and $p_0$ represent the average first-order
corrections of the two collective modes, while $q_c$
and $p_c$ quantify half the differences between their
respective corrections. Accordingly, the average
corrections affect both modes equally, whereas the
difference corrections enter with opposite signs for
the $+$ and $-$ modes.

For either collective mode, the corresponding single-mode
Gaussian ergotropy is
\begin{equation}
	\begin{aligned}
		\mathcal G_\zeta^{\mathrm{eq}}
		&=\omega\left[
		\frac{Q_\zeta^{\mathrm{eq}}+P_\zeta^{\mathrm{eq}}}{2}
		-\sqrt{Q_\zeta^{\mathrm{eq}}P_\zeta^{\mathrm{eq}}}
		\right]\\
		&=\frac{\omega\kappa^4}{4c_U}
		\left(p_\zeta^{(1)}-q_\zeta^{(1)}\right)^2
		+O(\kappa^6).
	\end{aligned}
	\label{secVII:single-mode-work}
\end{equation}
The term linear in $\kappa^2$ vanishes because the first
derivatives of the work function with respect to $Q$ and
$P$ vanish at the reference thermal covariance
$Q=P=c_U/2$. Consequently, higher-order corrections
to the covariance matrices do not affect the leading
$O(\kappa^4)$ work.

Summing the contributions from the two collective modes, the global Gaussian ergotropy becomes
\begin{equation}
	\begin{aligned}
		\mathcal G_{\mathrm{glob}}^{\mathrm{eq}}
		&=\sum_{\zeta=\pm}\mathcal G_\zeta^{\mathrm{eq}}\\
		&=\frac{\omega\kappa^4}{2c_U}
		\left[
		(p_0-q_0)^2+(p_c-q_c)^2
		\right]+O(\kappa^6).
	\end{aligned}
	\label{secVII:global-weak}
\end{equation}
For local extraction, the two detectors have identical marginal variances,
\begin{align}
	\bar Q^{\mathrm{eq}}
	&=\frac{Q_+^{\mathrm{eq}}+Q_-^{\mathrm{eq}}}{2}
	=\frac{c_U}{2}+\kappa^2q_0+O(\kappa^4),\notag\\
	\bar P^{\mathrm{eq}}
	&=\frac{P_+^{\mathrm{eq}}+P_-^{\mathrm{eq}}}{2}
	=\frac{c_U}{2}+\kappa^2p_0+O(\kappa^4).
	\label{secVII:local-marginals}
\end{align}
The mode-dependent corrections cancel in the local marginals. Since the two detectors contribute equally, their total local Gaussian ergotropy is
\begin{equation}
	\begin{aligned}
		\mathcal G_{\mathrm{loc}}^{\mathrm{eq}}
		&=2\omega\left[
		\frac{\bar Q^{\mathrm{eq}}+\bar P^{\mathrm{eq}}}{2}
		-\sqrt{\bar Q^{\mathrm{eq}}\bar P^{\mathrm{eq}}}
		\right]\\
		&=\frac{\omega\kappa^4}{2c_U}
		(p_0-q_0)^2+O(\kappa^6).
	\end{aligned}
	\label{secVII:local-weak}
\end{equation}
Subtracting Eq.~\eqref{secVII:local-weak} from Eq.~\eqref{secVII:global-weak}, we obtain
\begin{equation}
		\Delta\mathcal G_{\mathrm{corr}}^{\mathrm{eq}}
		=\frac{\omega\kappa^4}{2c_U}
		(p_c-q_c)^2+O(\kappa^6).
	\label{secVII:gap-weak}
\end{equation}
This result shows that the additional global work is determined entirely by the difference between the mode-dependent momentum and position corrections.

Using the previously defined weighted spatial correlation
\begin{equation}
	\eta_E=\frac{I_{c,E}}{I_0},
	\label{secVII:weighted-correlation}
\end{equation}
and introducing the leading-order local-work coefficient
for $\lambda=0$,
\begin{equation}
	\aleph_C=\frac{\omega I_0^2}{8\pi^4c_U},
\end{equation}
the local and global Gaussian ergotropies can be expressed as
\begin{align}
	\mathcal G_{\mathrm{loc}}^{\mathrm{eq}}
	&=\kappa^4\aleph_C+O(\kappa^6),\notag\\
	\mathcal G_{\mathrm{glob}}^{\mathrm{eq}}
	&=\kappa^4\aleph_C(1+\eta_E^2)
	+O(\kappa^6).
	\label{secVII:leading-works}
\end{align}
The difference between the global and local Gaussian
ergotropies is therefore given by
\begin{equation}
		\Delta\mathcal G_{\mathrm{corr}}^{\mathrm{eq}}
		=\kappa^4\aleph_C\eta_E^2+O(\kappa^6).
	\label{secVII:correlation-work}
\end{equation}
This relation establishes a direct connection between the spatial correlation induced by the common environment and the additional work accessible through joint Gaussian operations. The common response of the two collective modes determines the leading local contribution, whereas their different responses give rise to the additional global contribution.

We finally examine whether the additional global work requires entanglement between the two detectors. For a two-mode Gaussian state, this can be assessed using the positive-partial-transpose (PPT) criterion. Partial transposition with respect to detector $B$ reverses the sign of $p_B$. In the collective-mode representation, this operation interchanges $p_+$ and $p_-$ while leaving $q_+$ and $q_-$ unchanged. The partially transposed equilibrium covariance matrix therefore takes the form
\begin{equation}
	\sigma_{AB}^{\mathrm{eq,PT}}
	\sim
	\operatorname{diag}
	\left(
	Q_+^{\mathrm{eq}},
	P_-^{\mathrm{eq}},
	Q_-^{\mathrm{eq}},
	P_+^{\mathrm{eq}}
	\right).
	\label{secVII:pt-covariance}
\end{equation}
Its symplectic eigenvalues are
\begin{equation}
	\widetilde\nu_1^{\mathrm{eq}}
	=\sqrt{Q_+^{\mathrm{eq}}P_-^{\mathrm{eq}}},
	\qquad
	\widetilde\nu_2^{\mathrm{eq}}
	=\sqrt{Q_-^{\mathrm{eq}}P_+^{\mathrm{eq}}}.
	\label{secVII:pt-eigenvalues}
\end{equation}
In our covariance convention, the two-mode Gaussian state is separable if and only if
\begin{equation}
		\min\left(
		\widetilde\nu_1^{\mathrm{eq}},
		\widetilde\nu_2^{\mathrm{eq}}
		\right)\geq\frac12.
	\label{secVII:ppt}
\end{equation}

At zero detector--field coupling, the two collective modes reduce to identical thermal states, with
\begin{equation}
	Q_\pm^{\mathrm{eq}}(0)
	=P_\pm^{\mathrm{eq}}(0)
	=\frac{c_U}{2}.
\end{equation}
The corresponding PPT symplectic eigenvalues therefore satisfy
\begin{equation}
	\begin{aligned}
		\min\left(
		\widetilde\nu_1^{\mathrm{eq}},
		\widetilde\nu_2^{\mathrm{eq}}
		\right)-\frac12
		&=\frac12\coth\left(\frac{\omega}{2T_U}\right)-\frac12\\
		&=\frac{1}{e^{\omega/T_U}-1}>0.
	\end{aligned}
	\label{secVII:ppt-margin}
\end{equation}
At zero detector--field coupling, the smallest PPT
symplectic eigenvalue is strictly larger than the
separability threshold $1/2$ for $T_U>0$. Since the
regulated equilibrium covariance matrix and its
partially transposed symplectic eigenvalues depend
continuously on $\kappa^2$, the smallest PPT symplectic
eigenvalue remains above $1/2$ for sufficiently weak
but finite coupling. Therefore, there exists a
$\kappa_*^2>0$ such that the two-detector equilibrium
state remains separable for
$0\leq\kappa^2<\kappa_*^2$.

Since $I_{c,E}>0$, Eq.~\eqref{secVII:correlation-work}
implies that the additional global work is positive
for sufficiently weak but nonzero coupling. At the same
time, the PPT criterion shows that the two-detector
equilibrium state remains separable over a finite interval
of weak coupling for $T_U>0$. Hence, within this interval,
the detectors can remain separable, while joint Gaussian
operations still extract more work than independent local
operations. This demonstrates that entanglement between
the detectors is not necessary for the additional global
work. At $T_U=0$, the PPT symplectic eigenvalues equal $1/2$ at zero coupling, exactly at the separability boundary. 
		
\section{Finite-Time Dynamics of Gaussian Ergotropy}
\label{sec:dynamics}

Having established the equilibrium relation between spatial correlations and the additional global work, we now investigate the instantaneous dynamics of the Gaussian ergotropy in the accelerated-vacuum environment. Using the covariance-matrix propagation developed in Sec.~\ref{xfc}, we calculate the time-dependent Gaussian ergotropy $\mathcal{G}_U(\tau)$ of the detector pair with respect to the detector Hamiltonian $\mathcal{H}_S$, as defined in Eq.~\eqref{gsn}.

For the numerical propagation, we specify the bath discretization introduced in Sec.~\ref{hmd} by employing an $N$-point Gauss--Legendre quadrature on the finite frequency interval $[\Omega_{\min},\Omega_{\max}]$. The quadrature nodes $z_j$ are the zeros of the Legendre polynomial $P_N(z)$ on $[-1,1]$:
\begin{equation}
	P_N(z_j)=0,\qquad
	-1<z_1<z_2<\cdots<z_N<1.
\end{equation}
The corresponding reference weights are
\begin{equation}
	\widetilde{w}_j
	=
	\frac{2}
	{(1-z_j^2)[P_N'(z_j)]^2}.
\end{equation}
Mapping the reference interval onto the physical frequency interval gives
\begin{equation}
	\begin{aligned}
		\Omega_j
		&=
		\frac{\Omega_{\max}-\Omega_{\min}}{2}z_j
		+\frac{\Omega_{\max}+\Omega_{\min}}{2},
		\\
		w_j
		&=
		\frac{\Omega_{\max}-\Omega_{\min}}{2}
		\widetilde{w}_j .
	\end{aligned}
	\label{eq:dynamics_GL_nodes}
\end{equation}
The frequency integrals are then evaluated according to
\begin{equation}
	\int_{\Omega_{\min}}^{\Omega_{\max}}
	f(\Omega)\,d\Omega
	\simeq
	\sum_{j=1}^{N}w_j f(\Omega_j).
	\label{eq:dynamics_GL_rule}
\end{equation}
We take $\Omega_{\min}=10^{-5}$ and $\Omega_{\max}=50$. The numerical results use the exponential frequency regulator, with $\Omega_c=5$ unless otherwise specified.

\begin{figure}[!t]
	\centering
	\includegraphics[width=\columnwidth]
	{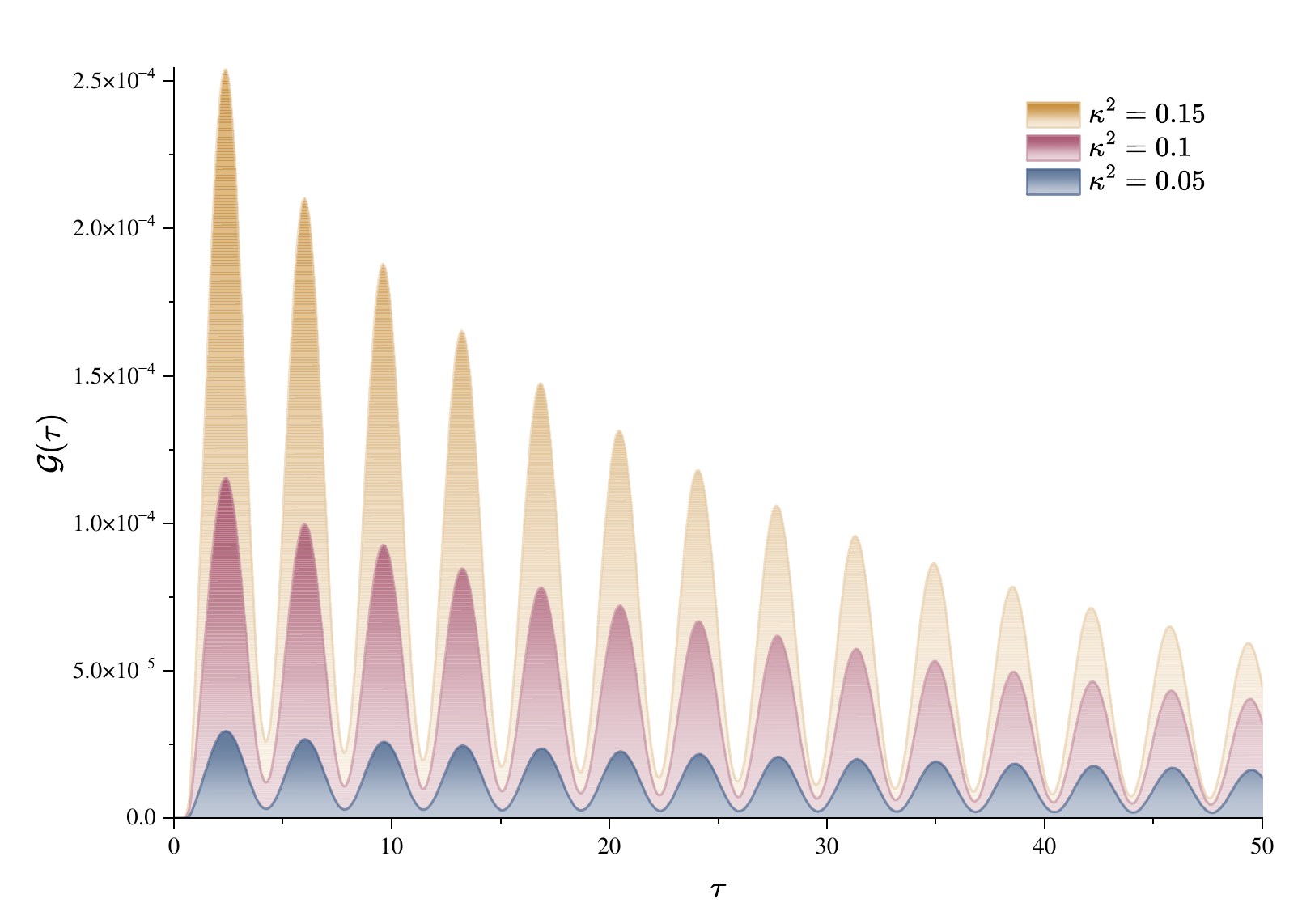}
	\caption{Time evolution of the Gaussian ergotropy for different values of the detector--field coupling strength $\kappa^2$, with fixed parameters $\lambda=0.5$, $\omega=1$, $d=1$, $a=2\pi$, and $N=20000$.}
	\label{fig:dyn_kappa}
\end{figure}

We first examine the dependence of the transient Gaussian ergotropy on the detector--field coupling strength $\kappa^2$. Figure~\ref{fig:dyn_kappa} shows that $\mathcal{G}_U(\tau)$ exhibits an oscillatory decay as the proper time increases. The successive local maxima become larger as $\kappa^2$ increases. For the three displayed coupling strengths, the early peak amplitudes approximately follow a ratio of $1:4:9$, indicating a quadratic dependence on the plotted parameter $\kappa^2$. This is consistent with the leading weak-coupling scaling $\mathcal{G}_U(\tau)\propto\kappa^4$. The oscillation period is approximately $3.6$, while the oscillation envelope decreases progressively. At stronger coupling, the transient response decays more rapidly, and the differences between the curves become less pronounced at later times.

\begin{figure}[!t]
	\centering
	\includegraphics[width=\columnwidth]
	{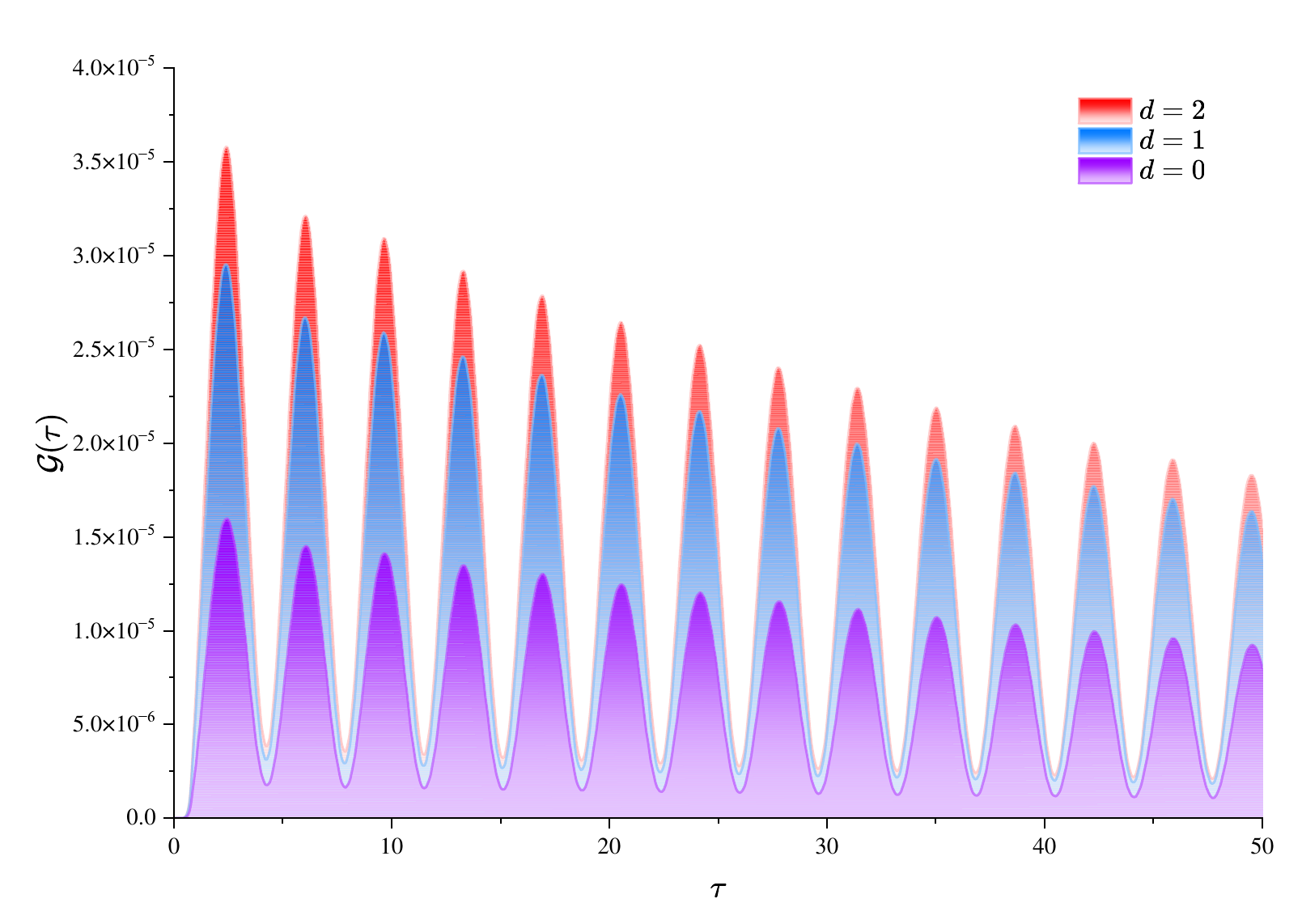}
	\caption{Time evolution of the Gaussian ergotropy for different values of the detector separation $d$, with fixed parameters $\lambda=0.5$, $\omega=1$, $\kappa^2=0.05$, $a=2\pi$, and $N=20000$.}
	\label{fig:dyn_d}
\end{figure}

We next investigate the effect of the detector separation $d$. As shown in Fig.~\ref{fig:dyn_d}, the Gaussian ergotropy exhibits an oscillatory decay for all displayed separations, and its local maxima increase with increasing $d$. The oscillation period remains approximately $3.6$, while the envelopes decay progressively. At fixed acceleration, changing $d$ leaves the local spectral contribution and the Unruh temperature unchanged, but modifies the spatial cross-correlation factor $\gamma_U(\Omega,a,d)$. As the separation increases, the cross-correlation envelope becomes weaker, and the two collective spectral densities progressively approach the same local spectrum. The distance dependence of the transient response consequently becomes less pronounced at sufficiently large separation.

\begin{figure}[!t]
	\centering
	\includegraphics[width=\columnwidth]
	{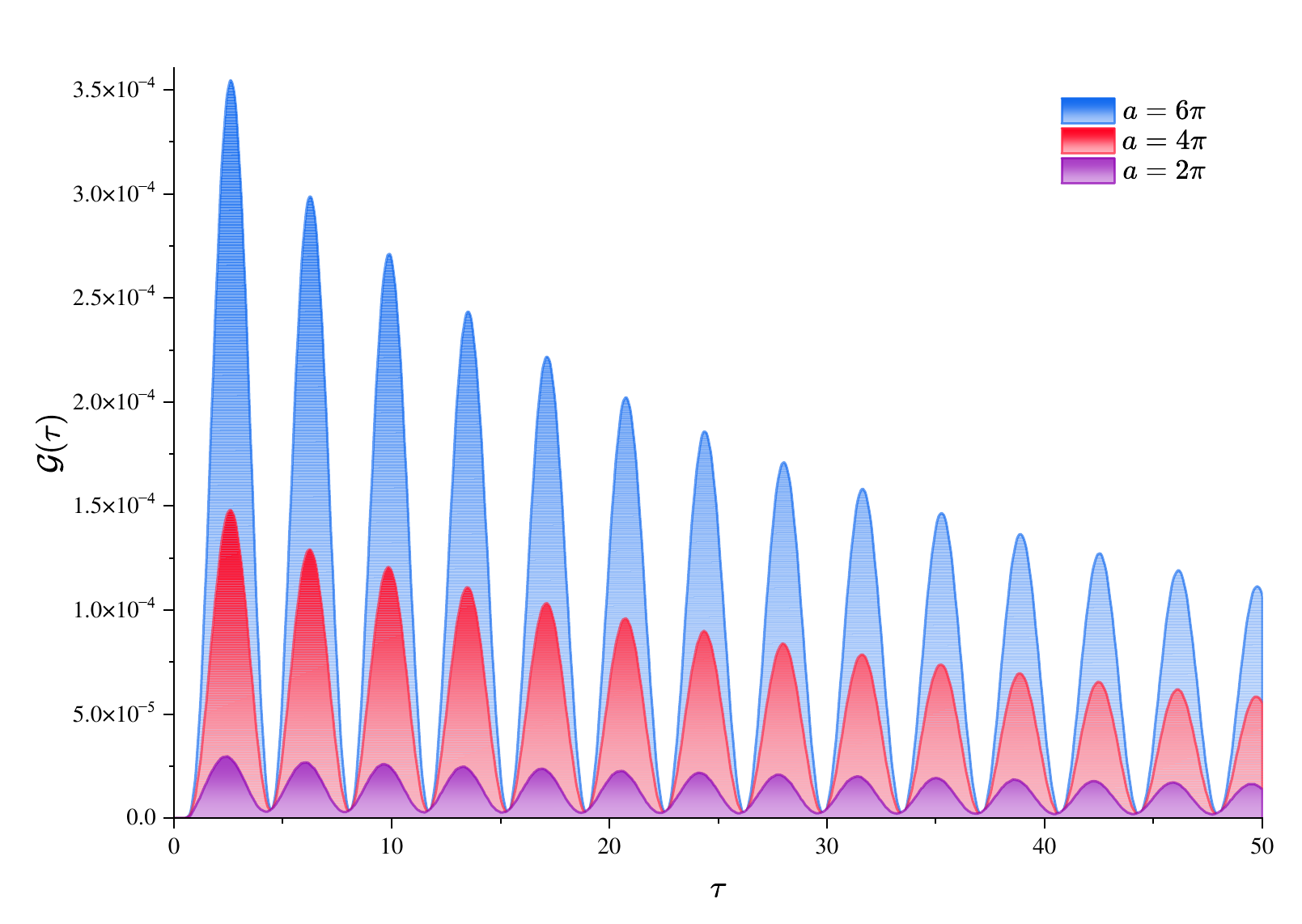}
	\caption{Time evolution of the Gaussian ergotropy for different values of the acceleration $a$, with fixed parameters $\lambda=0.5$, $\omega=1$, $\kappa^2=0.05$, $d=1$, and $N=20000$.}
	\label{fig:dyn_a}
\end{figure}

We further examine the influence of acceleration. Figure~\ref{fig:dyn_a} shows that the transient Gaussian ergotropy increases in magnitude with increasing acceleration, while its oscillation envelope decays more rapidly. The oscillation periods remain approximately unchanged over the displayed range. Notably, the increase in ergotropy indicates that stronger Unruh-induced thermal fluctuations do not necessarily suppress work extraction; instead, they are accompanied by an enhancement of the transient Gaussian ergotropy in the parameter range considered. This behavior reflects the combined influence of the increasing Unruh temperature $T_U=a/(2\pi)$ and the acceleration-dependent spatial correlation factor $\gamma_U(\Omega,a,d)$. Thus, stronger thermal fluctuations can enhance the transient work signal, even though the corresponding oscillations decay more rapidly.

\begin{figure}[!t]
	\centering
	\includegraphics[width=\columnwidth]
	{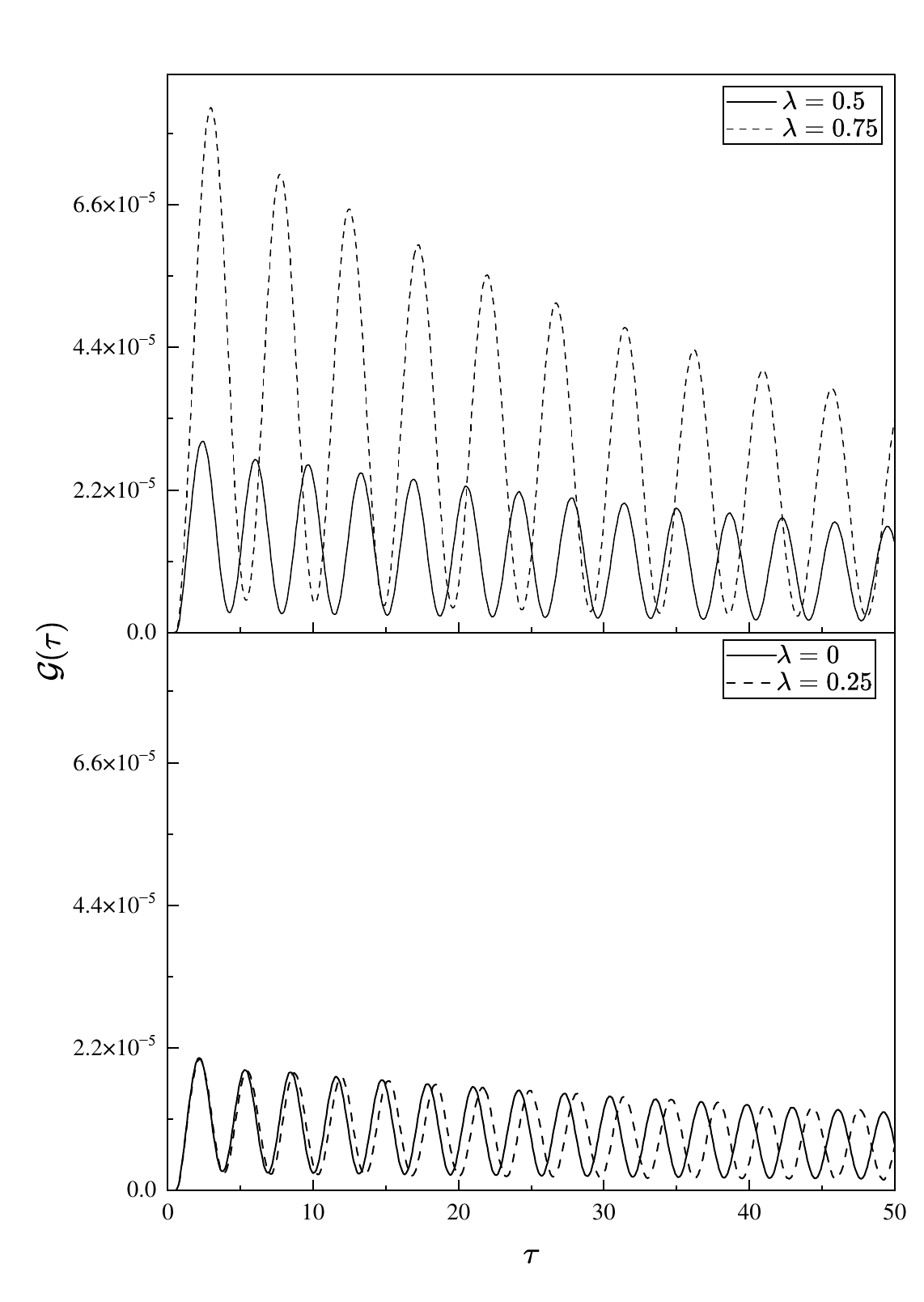}
	\caption{Time evolution of the Gaussian ergotropy for different values of the direct detector coupling strength $\lambda$, with fixed parameters $\omega=1$, $\kappa^2=0.05$, $d=1$, $a=2\pi$, and $N=20000$.}
	\label{fig:dyn_lambda}
\end{figure}

We now turn to the effect of the direct interaction between the detectors. Figure~\ref{fig:dyn_lambda} shows the time evolution for different values of $\lambda$. The Gaussian ergotropy exhibits an oscillatory decay for all displayed values. Increasing $\lambda$ produces higher peaks and more widely separated oscillations, whereas at small $\lambda$ the ergotropy remains relatively small and the curves nearly coincide. This behavior is related to the collective detector frequency
\begin{equation}
	\nu_S=\sqrt{\omega^2-\lambda^2},
\end{equation}
which decreases as $\lambda$ increases. The resulting longer oscillation period is accompanied by enhanced two-mode squeezing of the detector ground state, thereby modifying the response of the detector pair to the thermal field.

\begin{figure}[!t]
	\centering
	\includegraphics[width=\columnwidth]
	{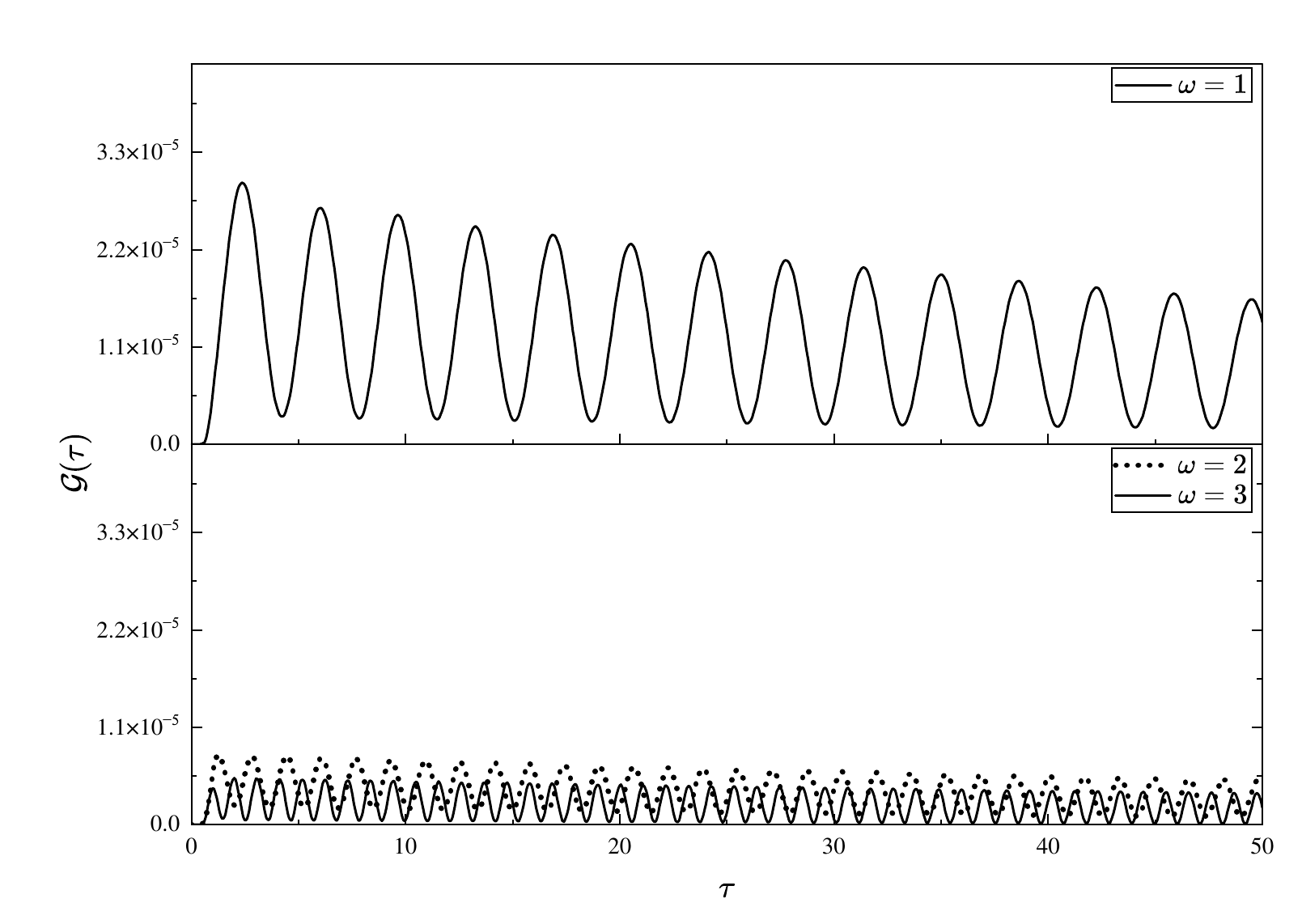}
	\caption{Time evolution of the Gaussian ergotropy for different values of the detector frequency $\omega$, with fixed parameters $\lambda=0.5$, $\kappa^2=0.05$, $d=1$, $a=2\pi$, and $N=20000$.}
	\label{fig:dyn_omega}
\end{figure}

Finally, we investigate the dependence on the detector frequency $\omega$. Figure~\ref{fig:dyn_omega} shows that increasing $\omega$ suppresses the transient peak amplitudes and produces faster oscillations. At fixed $\lambda$, increasing $\omega$ raises the collective frequency $\nu_S$. This reduces the thermal occupation around the detector frequency and also changes the response kernel $B_1$. Together, these changes affect how much Gaussian ergotropy can be extracted from the field fluctuations.

\begin{figure*}[!t]
	\centering
	\includegraphics[width=\textwidth]
	{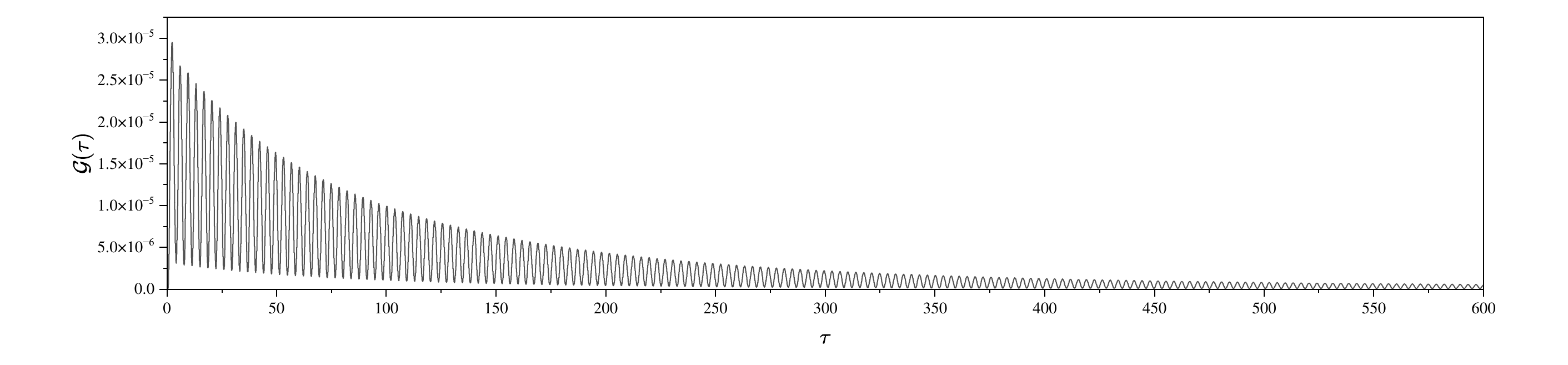}
	\caption{Time evolution of the Gaussian ergotropy in the accelerated vacuum, with fixed parameters $\lambda=0.5$, $\omega=1$, $\kappa^2=0.05$, $d=1$, $a=2\pi$, $\Omega_c=5$, and $N=30000$.}
	\label{fig:dyn_long}
\end{figure*}

We finally examine the long-time evolution of the Gaussian ergotropy. Figure~\ref{fig:dyn_long} shows its time dependence over an extended proper-time interval. The Gaussian ergotropy exhibits persistent oscillations with a gradually decreasing envelope. The oscillation amplitude decreases rapidly at early times and then decays more slowly at later times, while the oscillatory behavior remains visible throughout the displayed interval.

\section{Conclusions}
\label{c}

In this work, we have investigated how the spatial correlations of a bosonic environment are encoded in the Gaussian ergotropy of two harmonic-oscillator detectors. By comparing the accelerated vacuum with an inertial thermal field at the same Unruh temperature, we have shown that identical local spectra do not imply identical collective responses. The leading equilibrium Gaussian ergotropy is governed by the response-weighted spatial correlation $\eta_E$, establishing a direct connection between environmental spatial correlations and extractable work.

We have further shown that a single interaction setting exhibits a unique finite-distance blind point, at which the two environments yield identical leading-order ergotropies despite having different spatial correlations. This ambiguity can be removed by considering two opposite direct-coupling settings, $\lambda=\pm g$. The resulting normalized work asymmetry $\mathcal A_E$ is a monotonic function of $\eta_E$ and can be inverted to reconstruct the weighted spatial correlation. Thus, the two-setting scheme eliminates the blind point and distinguishes environments that cannot be distinguished through their local spectra alone. Calculations using the unexpanded interacting Green function further confirm the validity of this reconstruction at finite coupling. Moreover, when the direct detector coupling vanishes, the additional work accessible through joint Gaussian operations is proportional to $\eta_E^2$. This advantage can persist even when the detector pair remains separable, showing that entanglement is not a necessary resource for additional global work extraction.

Finally, the finite-time dynamics reveal an oscillatory decay of Gaussian ergotropy, whose magnitude and decay behavior depend on the detector--field coupling, separation, acceleration, and detector parameters. Stronger acceleration enhances the transient work while accelerating its decay, whereas increasing the detector frequency suppresses work extraction and speeds up the oscillations. These findings establish Gaussian ergotropy as a useful probe of both the spatial structure of quantum environments and their dynamical influence on detector systems.

	\begin{acknowledgments}
		This work is supported by the Natural Science Foundation of Hainan Province under Grant No. 125RC744; the China Scholarship Council (CSC).
	\end{acknowledgments}
	
    \appendix
	\section{Gaussian conventions, field spectra, and switching preparation\label{A}\label{B}\label{C}\label{D}}
	\subsection{Gaussian-state conventions and quadratic Hamiltonians}
	Throughout this paper, we work in natural units:
	\begin{equation}
		\hbar = c = k_\text{B} = 1.
	\end{equation}
	
	Consider an $n$-mode bosonic system described by $n$ pairs of canonical conjugate coordinates and momenta $(q_i, p_i)$. Since the full Hamiltonian is quadratic in these variables, we assemble them into the column vector $\mathbf{R} = (q_1, q_2, \dots, q_n, p_1, p_2, \dots, p_n)^{\mathrm{T}}$ and introduce the symmetric matrix $H$ via
	\begin{equation}
		\mathcal{H} = \frac{1}{2} \mathbf{R}^{\mathrm{T}} H \mathbf{R}.
	\end{equation}
	Note that $H$ must be positive semidefinite; otherwise the Hamiltonian $\mathcal{H}$ would have a spectrum unbounded from below, which is unphysical.
	
	For each oscillator in the system with canonical coordinate $q$ and momentum $p$, we define the annihilation operator $a$ and its Hermitian conjugate $a^\dag$ via \cite{cohen1977,braunstein2005}
	\begin{equation}
		q = \frac{a + a^\dag}{\sqrt{2}}, \qquad p = - \mathrm{i} \frac{a - a^\dag}{\sqrt{2}}, \qquad [q, p] = \mathrm{i}.
	\end{equation}
	
	Inverting these relations gives $a = (q +  \mathrm{i}p)/\sqrt{2}$, which leads to the identity
	\begin{equation}
		a^\dag a + \frac{1}{2} = \frac{1}{2}\left(q^2 + p^2\right),
	\end{equation}
	Thus, the free Hamiltonian of a harmonic oscillator with frequency $\omega$ is
	\begin{equation}
		\mathcal{H} = \omega\left(a^\dag a + \frac{1}{2}\right) = \frac{\omega}{2}\left(q^2 + p^2\right).
	\end{equation}
	
	For a Gaussian state with vanishing first moments, let us furthermore define the covariance matrix $\sigma$ as \cite{braunstein2005,weedbrook2012}
	\begin{equation}
		\sigma_{ij} = \frac{1}{2}\left\langle R_i R_j + R_j R_i \right\rangle.
	\end{equation}
	It should be noted that a covariance matrix $\sigma$ is symmetric and must satisfy the uncertainty principle \cite{robertson1929,Simon1994QNM,weedbrook2012}
	\[
	\sigma+\frac{\mathrm{i}}{2}\,\Omega_{\mathrm{can}}\geq 0,
	\]
	where, in the ordering of $\mathbf{R}$ above,
	\[
	\Omega_{\mathrm{can}}=
	\begin{pmatrix}
		0 & I_n\\
		-I_n & 0
	\end{pmatrix},
	\]
	with $I_n$ the $n\times n$ identity. in particular
	$\sigma>0$ as $\sigma+\frac{\mathrm{i}}{2}\,\Omega_{\mathrm{can}}\geq0$.
	\subsection{Self-consistency between the switching function and the initial state}
	
	To avoid exciting high-frequency transients induced by abrupt switching on, we adopt a $C^\infty$ S-shaped ramp function.
	
	Define
	\begin{equation}
		\begin{split}
			&f(x) = e^{-1/x}\,\mathbf{1}_{x>0},\\
			&S(x) = \frac{f(x)}{f(x) + f(1-x)},\\
			&o = \frac{\tau - \tau_{\text{on}}}{\tau_s},
		\end{split}
	\end{equation}
	so
	\begin{equation}
		\chi(\tau) =
		\begin{cases}
			0, & \tau \le \tau_{\text{on}}, \\
			S(o), & \tau_{\text{on}} < \tau < \tau_{\text{on}} + \tau_o, \\
			1, & \tau \ge \tau_{\text{on}} + \tau_o.
		\end{cases}
	\end{equation}
	
	The function $S(x)$ has vanishing derivatives of all orders at $x = 0$ and $x = 1$ (all derivatives of $e^{-1/x}$ vanish at $0^+$), hence $\chi \in C^\infty$ \cite{lee2013}. In this work, we set \(\tau_{\text{on}} = 0\) and \(\tau_o = 1\).

	Since $\chi(0) = 0 \implies H_{SB}(0) = H_{CT}(0) = 0 \implies H_{\text{tot}}(0) = H_S + H_B$, the initial state can be legitimately taken as the product state \cite{breuer2002}
	\begin{equation}
		\rho(0) = \rho_S(0) \otimes \rho_B(0),
	\end{equation}
	where $\rho_S(0)$ is the ground state of $\mathcal{H}_S$, and $\rho_B(0)$ is the thermal state at temperature $T_U$.
	\subsection{Fourier transforms of the Wightman functions}
	
	Before proceeding with the calculation, we briefly recall the residue theorem. 
	
	If $f(z)$ is analytic inside a simple positively oriented closed contour $\Gamma$ except for isolated singularities $z_k$, then
	\begin{equation}
		\oint_\Gamma f(z) \,\mathrm{d}z = 2\pi\mathrm{i} \sum_k \mathrm{Res}(f, z_k).
	\end{equation}
	
	If $f(z)$ has a second-order pole at $z_n$ and can be expressed as $f(z) = h(z)/(z-z_n)^2$, where $h(z)$ is analytic at $z_n$, then the residue of $f(z)$ at $z_n$ reads
	\begin{equation}
		\mathrm{Res}(f,z_n) = h'(z_n).
	\end{equation}
	
	From Eq.~\eqref{fly}, it follows that
	\begin{equation}
		\begin{split}
			\widetilde{W}_{AA}(\Omega) &= \widetilde{W}_{BB}(\Omega) = \int_{-\infty}^{\infty} \mathrm{d}\Delta\tau \, e^{i\Omega\Delta\tau} W_{AA}(\Delta\tau).\\
			&=-\frac{a^2}{16\pi^2}\int  \frac{ e^{i\Omega\tau}}{\sinh^2\left[ \frac{a}{2}\left( \tau - i\epsilon \right) \right]} \mathrm{d}\tau.
		\end{split}
	\end{equation}
	
	Here we restrict our attention to the core integral:
	\begin{equation}
		\int  \frac{ e^{i\Omega\tau}}{\sinh^2\left[ \frac{a}{2}\left( \tau - i\epsilon \right) \right]}\mathrm{d}\tau.
	\end{equation}
	
	Performing the substitution \( z = \frac{a}{2}(\tau - \mathrm{i}\epsilon) \). The integration contour $L_\epsilon$ is given by
	\begin{equation}
		L_\epsilon: \quad z = x - \mathrm{i}\frac{a\epsilon}{2}, \qquad x\in\mathbb{R},\qquad x: -\infty \to +\infty,
	\end{equation}
	we obtain
	\begin{equation}
		\int  \frac{ e^{i\Omega\tau}}{\sinh^2\left[ \frac{a}{2}\left( \tau - i\epsilon \right) \right]}\mathrm{d}\tau = \frac{2}{a} \int  \frac{e^{\frac{2\mathrm{i}\Omega z}{a}}}{\sinh^2 z}\mathrm{d}z.
	\end{equation}
	
	Since the zeros of $\sinh z$, given by $z_n = \mathrm{i} n\pi$ with $n\in\mathbb{Z}$, are all second-order poles of $1/\sinh^2 z$, and hence they are second-order poles of the function
	\begin{equation}
		f(z) = \frac{e^{2\mathrm{i}\Omega z/a}}{\sinh^2 z}.
	\end{equation}
	
	Applying the residue theorem, we obtain
	\begin{equation}
		\int_{L_\epsilon} f(z) \,\mathrm{d}z = 2\pi\mathrm{i} \sum_{n=0}^{\infty} \mathrm{Res}(f, z_n).
	\end{equation}
	
	Using the residue formula for second-order poles, we obtain
	\begin{equation}
		\mathrm{Res}(f, z_n) = \frac{2\mathrm{i}\Omega}{a} e^{-2\pi\Omega n / a} \qquad (n = 0, 1, 2, \dots).
	\end{equation}
	
	It is straightforward to see that this is a convergent geometric series:
	\begin{equation}
		\sum_{n=0}^{\infty} \mathrm{Res}(f, z_n) = \frac{2\mathrm{i}\Omega}{a} \sum_{n=0}^{\infty} e^{-2\pi\Omega n / a} = \frac{2\mathrm{i}\Omega}{a} \frac{1}{1-e^{-2\pi\Omega  / a}}.
	\end{equation}
	
	Therefore,
	\begin{equation}
		\int_{L_\epsilon} f(z) \,\mathrm{d}z = 2\pi\mathrm{i} \cdot \frac{2\mathrm{i}\Omega}{a} \frac{1}{1-e^{-2\pi\Omega  / a}} = -\frac{4\pi\Omega}{a} \frac{1}{1-e^{-2\pi\Omega  / a}}.
	\end{equation}
	
	Hence we obtain
	\begin{equation}
		\widetilde{W}_{AA}(\Omega) = \widetilde{W}_{BB}(\Omega) = \frac{\Omega}{2\pi} \frac{1}{1 - e^{-2\pi\Omega/a}}, \qquad \Omega > 0.
	\end{equation}
	
	Following the same procedure, we can compute the Fourier transform of the Wightman function for two distinct detectors:
	\begin{equation}
		\widetilde{W}_{AB}(\Omega) =  \widetilde{W}_{BA}(\Omega) = \frac{1}{2\pi d \sqrt{1 + \left( \frac{ad}{2} \right)^2}} \frac{\sin\left[ \frac{2\Omega}{a} \operatorname{arsinh} \frac{ad}{2} \right]}{1 - e^{-2\pi\Omega/a}}.
	\end{equation}
\subsection{Geometric correlation factor
	\texorpdfstring{$\gamma_E$}{gamma sub E}}
	
	We define the geometric correlation factor $\gamma_E$ as:
	\begin{equation}
		\gamma_E(\Omega, a, d) \equiv \frac{\widetilde{W}_{AB}(\Omega)}{\widetilde{W}_{AA}(\Omega)} =  \frac{\sin\left[ \frac{2\Omega}{a} \operatorname{arsinh} \frac{ad}{2} \right]}{\Omega d \sqrt{1 + \left( \frac{ad}{2} \right)^2}}.
	\end{equation}
	
	We set
	\begin{equation}
		u = \operatorname{arsinh}\left( \frac{a d}{2} \right) \ge 0, \qquad X = \frac{2 \Omega}{a} \, u \ge 0,
	\end{equation}
	then
	\begin{equation}
		\gamma_E = \frac{2u}{\sinh 2u} \frac{\sin X}{X}.
	\end{equation}
	
	For \(d \to 0^+\), we have
	\begin{equation}
		u = \operatorname{arsinh}\left( \frac{ad}{2} \right) = \frac{ad}{2} + O(d^3),
	\end{equation}
	and therefore
	\begin{equation}
		X = \frac{2\Omega}{a} \, u = \Omega d + O(d^3).
	\end{equation}
	It follows that
	\begin{equation}
		\frac{2u}{\sinh 2u} \to 1, \qquad \frac{\sin X}{X} \to 1,
	\end{equation}
	which yields
	\begin{equation}
		\lim_{d\to 0^+} \gamma_E(\Omega, a, d) = 1.
	\end{equation}

	For \(a \to 0^+\), we have
	\begin{equation}
		u = \operatorname{arsinh}\left( \frac{ad}{2} \right) = \frac{ad}{2} + O(a^3).
	\end{equation}
	Substituting into the expression for \(X\) yields
	\begin{equation}
		X = \Omega d + O(a^2).
	\end{equation}
	Meanwhile,
	\begin{equation}
		\frac{2u}{\sinh 2u} \to 1,
	\end{equation}
	and therefore
	\begin{equation}
		\gamma_E = \frac{2u}{\sinh 2u} \frac{\sin X}{X} \to \frac{\sin(\Omega d)}{\Omega d}.
	\end{equation}
	That is,
	\begin{equation}
		\lim_{a\to 0^+} \gamma_E(\Omega, a, d)  = \frac{\sin(\Omega d)}{\Omega d}.
	\end{equation}
	
	From the Taylor expansion of the hyperbolic sine function
	\begin{equation}
		\sinh x = x + \frac{x^3}{6} + \frac{x^5}{120} + \cdots
	\end{equation}
	it follows that $\sinh x > x$ for $x > 0$. Thus,
	\begin{equation}
		0 < \frac{2u}{\sinh 2u} < 1 \qquad (u>0),
	\end{equation}
	and this expression approaches 1 only when $u \to 0$, i.e., when $d \to 0$.
	
	Furthermore, from the Taylor expansion of the sine function
	\begin{equation}
		\sin x = x - \frac{x^3}{6} + \cdots
	\end{equation}
	it follows that $|\sin x| \le |x|$ for all real numbers $x$. Hence, for $X > 0$,
	\begin{equation}
		\left| \frac{\sin X}{X} \right| \le 1,
	\end{equation}
	and it tends to 1 only as $X \to 0$.
	
	Multiplying the two inequalities yields
	\begin{equation}
		|\gamma_E| = \frac{2u}{\sinh 2u} \left| \frac{\sin X}{X} \right| \le \frac{2u}{\sinh 2u} < 1 \qquad (u>0).
	\end{equation}
	Therefore, for all $\Omega>0$, $a>0$, and $d>0$, we have
	\begin{equation}
		|\gamma_E| < 1.
	\end{equation}
	When $d=0$, the value is defined as the limit 1 by continuity, so
	\begin{equation}
		|\gamma_E(\Omega, a, d)| \le 1.
	\end{equation}
	\section{Equilibrium covariance and continuum-bath relaxation}
	\label{app:equilibrium}
	
	In Sec.~\ref{tau}, the detector equilibrium covariance
	is obtained from the Euclidean Green function of the
	coupled detector--bath system. Here we provide the intermediate steps connecting
	the total Hamiltonian of each collective sector
	to the effective Green function and the leading-order
	Gaussian ergotropy.
	We then examine when the resulting Gibbs covariance
	can also describe the long-time limit of the reduced
	detector dynamics.
	\subsection{Bath elimination and equilibrium covariance}
	
	The equilibrium quantities of interest are the detector
	variances $Q_\zeta^{\mathrm{eq}}$ and
	$P_\zeta^{\mathrm{eq}}$. Although the total Hamiltonian
	contains all bath oscillators, these variances can be
	calculated without retaining the bath coordinates
	explicitly. The reason is that the Euclidean action
	in Eq.~\eqref{se} is quadratic, so the bath variables
	can be integrated out exactly.

	We first show how integrating out the bath coordinates
	leads to the effective Green function used in the
	main text. For each collective sector $\zeta$, we
	perform a Matsubara Fourier expansion of the detector
	and bath coordinates:
	\begin{align}
		q_\zeta(u)
		&=
		\sqrt{T_U}\sum_n
		q_{\zeta,n}e^{\mathrm{i}\nu_nu},
		\notag\\
		Q_{j,\zeta}(u)
		&=
		\sqrt{T_U}\sum_n
		Q_{j\zeta,n}e^{\mathrm{i}\nu_nu},
	\end{align}
	where $\nu_n=2\pi nT_U$ are the bosonic
	Matsubara frequencies. Since the coordinates
	are real, their Fourier coefficients satisfy
	$q_{\zeta,-n}=q_{\zeta,n}^*$ and
	$Q_{j\zeta,-n}=Q_{j\zeta,n}^*$.
	
	Substituting these expansions into
		\begin{equation}
		\begin{aligned}
			S_E^{(\zeta)}
			=\int_0^{1/T_U}\mathrm{d}u
			\Bigg[
			&\frac{\dot q_\zeta^2}{2\beta_\zeta}
			+\frac{\alpha_\zeta}{2}q_\zeta^2
			+\frac{\delta\omega_\zeta}{2}q_\zeta^2
			\\
			&+\sum_j
			\left(
			\frac{\dot Q_{j,\zeta}^2}{2\Omega_j}
			+\frac{\Omega_j}{2}Q_{j,\zeta}^2
			+g_{j,\zeta}q_\zeta Q_{j,\zeta}
			\right)
			\Bigg],	
		\end{aligned}
	\end{equation} 
	the contribution of the
	$j$th bath oscillator at frequency $\nu_n$
	is
	\begin{equation}
		\frac{\nu_n^2+\Omega_j^2}{2\Omega_j}
		|Q_{j\zeta,n}|^2
		+
		g_{j,\zeta}
		\operatorname{Re}
		(q_{\zeta,n}^*Q_{j\zeta,n}).
	\end{equation}
	The first term is the free bath contribution,
	while the second describes its coupling to
	the detector coordinate.
	
	To remove $Q_{j\zeta,n}$, we complete the
	square:
	\begin{align}
		&\frac{\nu_n^2+\Omega_j^2}{2\Omega_j}
		|Q_{j\zeta,n}|^2
		+
		g_{j,\zeta}
		\operatorname{Re}
		(q_{\zeta,n}^*Q_{j\zeta,n})
		\notag\\
		&=
		\frac{\nu_n^2+\Omega_j^2}{2\Omega_j}
		\left|
		Q_{j\zeta,n}
		+
		\frac{g_{j,\zeta}\Omega_j}
		{\nu_n^2+\Omega_j^2}
		q_{\zeta,n}
		\right|^2
		\notag\\
		&\quad
		-
		\frac{g_{j,\zeta}^2\Omega_j}
		{2(\nu_n^2+\Omega_j^2)}
		|q_{\zeta,n}|^2.
		\label{app:bath_square}
	\end{align}
	The shifted bath coordinate appears only
	quadratically. Its Gaussian integral contributes
	a factor independent of the detector coordinate,
	which is absorbed into the normalization.
	The remaining term depends only on
	$q_{\zeta,n}$ and therefore modifies the
	detector's effective action.
	
	After integrating out all bath oscillators,
	we obtain
	\begin{align}
		S_{\mathrm{eff}}^{(\zeta)}
		=
		\frac12\sum_n
		\Bigg[
		\alpha_\zeta
		+\frac{\nu_n^2}{\beta_\zeta}
		+\delta\omega_\zeta
		\notag
		-\sum_j
		\frac{g_{j,\zeta}^2\Omega_j}
		{\nu_n^2+\Omega_j^2}
		\Bigg]
		|q_{\zeta,n}|^2,
		\label{app:effective_action}
	\end{align}
	up to a detector-independent constant.
	The last two contributions are the static
	counterterm and the bath-induced correction.
	Using
	\[
	\delta\omega_\zeta
	=
	\sum_j\frac{g_{j,\zeta}^2}{\Omega_j},
	\]
	their sum becomes
	\begin{align}
		\Sigma_\zeta(\nu_n)
		&=
		\sum_jg_{j,\zeta}^2
		\left[
		\frac{1}{\Omega_j}
		-
		\frac{\Omega_j}
		{\nu_n^2+\Omega_j^2}
		\right]
		\notag\\
		&=
		\nu_n^2
		\sum_j
		\frac{g_{j,\zeta}^2}
		{\Omega_j(\nu_n^2+\Omega_j^2)}
		\notag\\
		&=
		\nu_n^2\Lambda_\zeta(\nu_n).
		\label{app:self_energy}
	\end{align}
	Hence the effective action takes the
	Gaussian form
	\begin{equation}
		S_{\mathrm{eff}}^{(\zeta)}
		=
		\frac12\sum_n
		G_\zeta^{-1}(\mathrm{i}\nu_n)
		|q_{\zeta,n}|^2,
	\end{equation}
	where
	\begin{equation}
		G_\zeta^{-1}(\mathrm{i}\nu_n)
		=
		\alpha_\zeta
		+\frac{\nu_n^2}{\beta_\zeta}
		+\Sigma_\zeta(\nu_n).
	\end{equation}
	
	The Green function determines the position
	variance through Eq.~\eqref{qeq}.
	To calculate the momentum variance using
	the stationary virial identity, we also need
	the detector--bath cross-correlation.
	This quantity follows from the same Gaussian
	integration and does not require a separate
	calculation of the bath dynamics.
	
	In Eq.~\eqref{app:bath_square}, the shifted
	Gaussian has the conditional mean
	\begin{equation}
		\left\langle Q_{j\zeta,n}
		\right\rangle_{q_\zeta}
		=
		-\frac{g_{j,\zeta}\Omega_j}
		{\nu_n^2+\Omega_j^2}
		q_{\zeta,n}.
	\end{equation}
	Averaging over the detector fluctuations
	then gives
	\begin{equation}
		\langle q_\zeta Q_{j,\zeta}
		\rangle_{\mathrm{eq}}
		=
		-T_U\sum_n
		\frac{g_{j,\zeta}\Omega_j}
		{\nu_n^2+\Omega_j^2}
		G_\zeta(\mathrm{i}\nu_n).
		\label{app:cross_covariance}
	\end{equation}
	Summing over the bath modes yields
	\begin{align}
		&\sum_jg_{j,\zeta}
		\langle q_\zeta Q_{j,\zeta}
		\rangle_{\mathrm{eq}}
		\notag\\
		&=
		-\delta\omega_\zeta
		Q_\zeta^{\mathrm{eq}}
		+
		T_U\sum_n
		\Sigma_\zeta(\nu_n)
		G_\zeta(\mathrm{i}\nu_n).
		\label{app:cross_sum}
	\end{align}
	The first contribution cancels the
	counterterm in the stationary virial
	identity, Eq.~\eqref{bp}. We therefore
	recover
	\begin{equation}
		P_\zeta^{\mathrm{eq}}
		=
		\frac{T_U}{\beta_\zeta}
		\sum_n
		[\alpha_\zeta+\Sigma_\zeta(\nu_n)]
		G_\zeta(\mathrm{i}\nu_n),
	\end{equation}
	which is Eq.~\eqref{peq}.
	
	We next connect these covariance formulas
	to the weak-coupling work coefficient.
	The leading corrections are obtained by
	expanding the Green function in the bath
	self-energy, which is of order $\kappa^2$.
	The resulting Matsubara sums can be written
	in terms of two kernels,
	\begin{equation}
		B_k(\Omega)
		=
		T_U\sum_n
		\frac{\nu_n^2}
		{(\nu_n^2+\Omega^2)
			(\nu_n^2+\nu_S^2)^k},
		\qquad k=1,2.
		\label{app:Bkernels}
	\end{equation}
	They arise from the same self-energy
	but with different powers of the free
	detector Green function. Indeed, with
	$D_{0,\zeta}=(\nu_n^2+\nu_S^2)/\beta_\zeta$,
	\begin{equation}
		T_U\sum_n
		\frac{\nu_n^2\Lambda_\zeta(\nu_n)}
		{D_{0,\zeta}(\nu_n)^k}
		=
		\beta_\zeta^k
		\sum_j\frac{g_{j,\zeta}^2}{\Omega_j}
		B_k(\Omega_j).
	\end{equation}
	
	The elementary Matsubara identity
	\begin{equation}
		T_U\sum_n
		\frac{1}{\nu_n^2+x^2}
		=
		\frac{1}{2x}
		\coth\frac{x}{2T_U}
	\end{equation}
	gives
	\begin{equation}
		B_1(\Omega)
		=
		\frac{
			\Omega\coth[\Omega/(2T_U)]
			-\nu_Sc_U}
		{2(\Omega^2-\nu_S^2)},
		\qquad
		c_U=\coth\frac{\nu_S}{2T_U}.
	\end{equation}
	The second kernel follows by differentiation,
	\begin{equation}
		B_2(\Omega)
		=
		-\frac{1}{2\nu_S}
		\frac{\partial B_1(\Omega)}
		{\partial\nu_S},
	\end{equation}
	at fixed temperature and bath frequency.
	Both kernels have finite limits at
	$\Omega=\nu_S$, reproducing Eq.~\eqref{bk}.
	
	The role of these kernels becomes clear
	from the first-order covariance corrections:
	\begin{align}
		\delta Q_\zeta
		&=
		-\beta_\zeta^2
		\sum_j\frac{g_{j,\zeta}^2}{\Omega_j}
		B_2(\Omega_j)
		+O(\kappa^4),
		\notag\\
		\delta P_\zeta
		&=
		\frac{\alpha_\zeta}{\beta_\zeta}
		\delta Q_\zeta
		+
		\sum_j\frac{g_{j,\zeta}^2}{\Omega_j}
		B_1(\Omega_j)
		+O(\kappa^4).
		\label{app:covariance_corrections}
	\end{align}
	Thus $B_2$ enters the individual
	covariance corrections, whereas the
	difference relevant to Gaussian work is
	\begin{equation}
		\delta P_\zeta
		-
		\frac{\alpha_\zeta}{\beta_\zeta}
		\delta Q_\zeta
		=
		\sum_j\frac{g_{j,\zeta}^2}{\Omega_j}
		B_1(\Omega_j)
		+O(\kappa^4).
		\label{app:work_imbalance}
	\end{equation}
	The $B_2$ contribution cancels from
	this difference.
	
	To see why this cancellation matters,
	introduce
	$X_\zeta=\alpha_\zeta Q_\zeta^{\mathrm{eq}}$
	and
	$Y_\zeta=\beta_\zeta P_\zeta^{\mathrm{eq}}$.
	The Gaussian work of one collective sector
	can be expressed as
	\begin{equation}
		\mathcal G_\zeta^{\mathrm{eq}}
		=
		\frac{(X_\zeta-Y_\zeta)^2}
		{2(X_\zeta+Y_\zeta
			+2\sqrt{X_\zeta Y_\zeta})}.
		\label{app:work_rationalized}
	\end{equation}
	At zero detector--bath coupling,
	$X_\zeta=Y_\zeta=\nu_Sc_U/2$.
	Equation~\eqref{app:work_imbalance}
	shows that $Y_\zeta-X_\zeta$
	is of order $\kappa^2$.
	Consequently, the leading Gaussian
	work is of order $\kappa^4$.
	
	Moreover, its leading coefficient
	depends only on the $B_1$ integral.
	In the continuum limit,
	\begin{align}
		&\sum_j\frac{g_{j,\zeta}^2}{\Omega_j}
		B_1(\Omega_j)
		\notag\\
		&\longrightarrow
		\frac{\kappa^2}{2\pi^2}
		\int_0^\infty d\Omega\,
		r(\Omega)
		[1 \pm \gamma_E(\Omega)]
		B_1(\Omega).
	\end{align}
	This is the response-weighted spectral
	integral that leads to the work coefficient
	$\aleph_E$ in Eq.~\eqref{NE}.
	The bath integration therefore provides
	the connection between the microscopic
	environmental spectra and the equilibrium
	Gaussian work used in the main text.
	\subsection{Continuum relaxation and the dark mode}
	
	The preceding calculation determines the
	reduced Gibbs covariance of the interacting
	detector--bath Hamiltonian. A separate
	question is whether the detectors actually
	approach this covariance after the interaction
	is switched on. This requires examining
	the real-time response of the coupled
	system.
	
	For a finite number of bath oscillators,
	the total system is closed and its dynamics
	is generally quasiperiodic. The continuum
	reservoir is therefore essential for
	irreversible local relaxation.
	
	To examine this limit, we eliminate the
	bath coordinates from the real-time
	equations of motion. The retarded
	susceptibility of one collective sector
	is
	\begin{equation}
		\chi_\zeta^{-1}(z)
		=
		\alpha_\zeta
		-\frac{z^2}{\beta_\zeta}
		+\delta\omega_\zeta
		-\frac{2}{\pi}
		\int_0^\infty d\Omega\,
		\frac{\Omega J_{E\zeta}(\Omega)}
		{\Omega^2-z^2},
		\label{app:retarded}
	\end{equation}
	where $z$ lies in the upper half-plane.
	
	Under full static compensation,
	the zero-frequency contribution of
	the bath cancels the counterterm,
	so that
	\begin{equation}
		\chi_\zeta^{-1}(0)
		=
		\alpha_\zeta>0.
	\end{equation}
	For positive real frequencies, the
	boundary value satisfies
	\begin{equation}
		\operatorname{Im}
		\chi_\zeta^{-1}(x+\mathrm{i}0^+)
		=
		-J_{E\zeta}(x),
	\end{equation}
	and hence
	\begin{equation}
		\operatorname{Im}
		\chi_\zeta(x+\mathrm{i}0^+)
		=
		\frac{J_{E\zeta}(x)}
		{[\operatorname{Re}
			\chi_\zeta^{-1}(x)]^2
			+J_{E\zeta}(x)^2}.
		\label{app:spectral_response}
	\end{equation}
	
	For nonzero separation $d>0$ and
	nonzero detector--bath coupling,
	both collective spectral densities
	are positive at every $x>0$ in
	the environments considered here.
	The response therefore has no
	isolated positive-frequency pole.
	The zero-frequency condition above,
	together with stability of the
	completed quadratic Hamiltonian,
	also excludes zero-frequency and
	unstable modes.

	Under these conditions, the response contains no
	undamped discrete-mode contribution. Moreover, the
	terms in the detector covariance that depend on
	the initial detector state and the finite switching
	history are Fourier integrals over the regulated
	spectral functions. Since the relevant spectral
	weights are integrable, these terms vanish as
	$\tau\to\infty$ by the Riemann--Lebesgue lemma.
	The loss of dependence on the initial preparation
	is the standard relaxation behavior of a damped
	oscillator coupled to a continuum bath
	\cite{grabert1988,Trushechkin2022}.
	The remaining stationary covariance is therefore
	determined by the bath fluctuations:
	\begin{align}
	Q_\zeta^{\mathrm{eq}}
	&=
	\frac1\pi\int_0^\infty dx\,
	\coth\frac{x}{2T_U}
	\operatorname{Im}
	\chi_\zeta(x+\mathrm{i}0^+),
	\notag\\
	P_\zeta^{\mathrm{eq}}
	&=
	\frac1{\pi\beta_\zeta^2}
	\int_0^\infty dx\,x^2
	\coth\frac{x}{2T_U}
	\operatorname{Im}
	\chi_\zeta(x+\mathrm{i}0^+).
	\label{app:stationary_variances}
	\end{align}
	Analytic continuation to imaginary
	frequency recovers the equilibrium
	Matsubara expressions in
	Eqs.~\eqref{qeq} and~\eqref{peq}.
	This establishes the connection
	between the reduced Gibbs covariance
	and the long-time detector covariance
	in the regular continuum-reservoir
	limit.
	
	There is an important exception at
	zero detector separation.
	When $d=0$, the antisymmetric
	spectral density vanishes for both
	the accelerated-vacuum and matched
	thermal environments:
	\begin{equation}
		J_{E-}(\Omega)=0,
		\qquad E=U,T.
	\end{equation}
	The antisymmetric collective mode
	is then decoupled from the bath.
	Its retarded response retains
	isolated poles at $\pm\nu_S$,
	so it does not undergo bath-induced
	relaxation.
	
\section{Correlation ordering and uniqueness of the blind separation}
\label{cr:proof_order}

The main text shows that the weighted spatial correlations
in the accelerated-vacuum and matched thermal environments
are different and that their work contrast can vanish at
a finite detector separation. Here we establish these
results analytically. We first prove the ordering of the
two correlations. We then show that both decrease strictly
with separation, which allows us to prove the uniqueness
of the finite blind separation and determine the ordering
of the two work minima.

Throughout this section, the temperature $T=T_U>0$,
the detector parameters, and the regulator are fixed,
while the separation $d$ varies.
\subsection{Weighted spatial correlations and their ordering}

We begin by establishing the positivity and ordering
of the weighted spatial correlations. The key step is
to show that the function
\[
f(\Omega)=r(\Omega)B_1(\Omega)
\]
has a strictly positive cosine transform. We therefore
introduce
\begin{align}
	f(\Omega)
	&=
	r(\Omega)B_1(\Omega),
	\notag\\
	h(x)
	&=
	\int_0^\infty d\Omega\,
	f(\Omega)\cos(\Omega x),
	\notag\\
	H(x)
	&=
	\int_0^x h(y)\,dy.
	\label{cr:hH}
\end{align}
Here $B_1(\Omega)$ is the positive kernel defined
in Eq.~\eqref{bk}, and $r(\Omega)$ is the regulator.
We also write
\[
I_0=\int_0^\infty f(\Omega)\,d\Omega.
\]
The regulated integrals satisfy
$\int_0^\infty f(\Omega)d\Omega<\infty$ and
$\int_0^\infty\Omega f(\Omega)d\Omega<\infty$.

To determine the sign of $h(x)$, we use the
Matsubara representation of $B_1$:
\begin{equation}
	B_1(\Omega)
	=
	2T\sum_{n=1}^\infty
	\frac{\nu_n^2}{\nu_n^2+\nu_S^2}
	\frac{1}{\Omega^2+\nu_n^2}.
	\label{cr:positive_mixture}
\end{equation}
Every term in this representation is positive.
Moreover, the regulated sum is integrable, since
\begin{equation}
	2T\sum_{n=1}^\infty
	\frac{\nu_n^2}{\nu_n^2+\nu_S^2}
	\int_0^\infty
	\frac{r(\Omega)}
	{\Omega^2+\nu_n^2}\,d\Omega
	=I_0<\infty.
\end{equation}
This justifies interchanging the frequency integral
and the Matsubara sum.

We now use the properties of the two regulators.
Extend $r(\Omega)$ evenly to the whole frequency
axis and denote its cosine transform by
\[
R(x)=\int_0^\infty r(\Omega)\cos(\Omega x)\,d\Omega.
\]
For the exponential and Gaussian regulators,
respectively, direct integration gives
\begin{align}
	R_{\exp}(x)
	&=
	\frac{p}{p^2+x^2},
	\qquad p=\Omega_c^{-1},
	\notag\\
	R_G(x)
	&=
	\Omega_c
	\exp\left(-\frac{\Omega_c^2x^2}{\pi}\right).
	\label{cr:Rtransforms}
\end{align}
Both transforms are positive, even, integrable,
and strictly decreasing for $x>0$.

The remaining frequency integral can be evaluated
using
\begin{equation}
	\int_0^\infty
	\frac{\cos(\Omega x)}
	{\Omega^2+v^2}\,d\Omega
	=
	\frac{\pi}{2v}e^{-v|x|},
	\qquad v>0.
\end{equation}
Defining $K_v(x)=e^{-v|x|}$, the convolution theorem
for the even extensions gives
\begin{equation}
	h(x)
	=
	T\sum_{n=1}^\infty
	\frac{\nu_n}{\nu_n^2+\nu_S^2}
	(R*K_{\nu_n})(x),
	\label{cr:h_convolution}
\end{equation}
where
\[
(R*K)(x)
=
\int_{-\infty}^{\infty}
R(x-y)K(y)\,dy.
\]
Since both $R$ and $K_{\nu_n}$ are strictly positive,
every convolution in Eq.~\eqref{cr:h_convolution}
is positive. All coefficients in the Matsubara sum
are also positive. Consequently,
\begin{equation}
	h(x)>0,
	\qquad
	H(x)>0
	\quad (x>0).
\end{equation}

We can now express the spatial correlation integrals
in terms of $H$. Integrating the cosine transform
over the finite interval $0\leq y\leq x$ gives
\begin{equation}
	H(x)
	=
	\int_0^\infty d\Omega\,
	f(\Omega)\frac{\sin(\Omega x)}{\Omega}.
\end{equation}
The interchange of the integrals is justified by
the regulated convergence established above.
Using the geometric relations between the two
environments, we obtain
\begin{equation}
	I_{c,T}=\frac{H(d)}{d},
	\qquad
	I_{c,U}=\frac{H(L)}{D},
	\label{cr:cross_H}
\end{equation}
where
\[
L=\frac{2}{a}\operatorname{arsinh}\frac{ad}{2},
\qquad
D=d\sqrt{1+(ad/2)^2}.
\]

For $a,d>0$, these geometric quantities satisfy
\[
0<L<d<D.
\]
Since $H(x)$ is strictly increasing, we have
\[
H(L)<H(d).
\]
Together with $D>d$, this yields
\begin{equation}
	0<I_{c,U}<I_{c,T}.
\end{equation}
In addition, the thermal cross integral can be
written as
\[
I_{c,T}
=
\int_0^\infty d\Omega\,
f(\Omega)\operatorname{sinc}(\Omega d),
\]
where $\operatorname{sinc}(x)=\sin x/x$.
Since $\operatorname{sinc}(\Omega d)<1$ for
$\Omega,d>0$ and $f(\Omega)>0$, it follows that
$I_{c,T}<I_0$.

Using the definition
$\eta_E=I_{c,E}/I_0$, we therefore obtain
\begin{equation}
		0<\eta_U<\eta_T<1,
		\qquad d>0.
\end{equation}
This establishes the correlation ordering
in Eq.~\eqref{cr:ordering}.
\subsection{Monotonicity and uniqueness of the blind separation}

The ordering established above shows that the
accelerated-vacuum correlation is always smaller
than the thermal one. To prove that the blind
separation is unique, we need a further result:
both correlations must decrease strictly with $d$.

We first show that $h(x)$ is strictly decreasing.
The two regulator transforms in Eq.~\eqref{cr:Rtransforms}
are even and strictly decreasing for positive
arguments. For $K_v(x)=e^{-v|x|}$, differentiation
of the convolution gives, for $x>0$,
\begin{align}
	(R*K_v)'(x)
	&=
	\int_{-\infty}^{\infty}
	R(x-y)K_v'(y)\,dy
	\notag\\
	&=
	\int_0^\infty
	K_v'(y)
	[R(x-y)-R(x+y)]\,dy.
	\label{cr:convolution_derivative}
\end{align}
The second equality follows by splitting the
integral into positive and negative $y$ and
using the evenness of $K_v$ and $R$.

For $y>0$, we have
\[
K_v'(y)=-ve^{-vy}<0.
\]
Furthermore, since
$|x-y|<x+y$ and $R$ is even and strictly
decreasing with the absolute value of its
argument,
\[
R(x-y)>R(x+y).
\]
Thus the integrand in the last line of
Eq.~\eqref{cr:convolution_derivative} is
strictly negative, giving
\begin{equation}
	(R*K_v)'(x)<0,
	\qquad x>0.
\end{equation}

The regulator also ensures that
$\int_0^\infty\Omega f(\Omega)d\Omega<\infty$.
Consequently, the differentiated Matsubara
series in Eq.~\eqref{cr:h_convolution} converges
uniformly: the sum of the absolute bounds on
its differentiated terms is bounded by
$\int_0^\infty\Omega f(\Omega)d\Omega$.
We may therefore differentiate the series
term by term. Since all its coefficients
are positive, we conclude that
\begin{equation}
	h'(x)<0,
	\qquad x>0.
\end{equation}

We next show that the normalized function
\begin{equation}
	F(x)=\frac{H(x)}{I_0x}
\end{equation}
is also strictly decreasing. Because $h(x)$
is strictly decreasing,
\[
H(x)=\int_0^x h(y)\,dy>xh(x).
\]
It follows that
\begin{equation}
	F'(x)
	=
	\frac{xh(x)-H(x)}{I_0x^2}<0.
	\label{cr:F_derivative}
\end{equation}

The two weighted correlations can now be
written in the compact form
\begin{equation}
	\eta_T(d)=F(d),
	\eta_U(d)=b(d)F(L(d)),
	b(d)=\frac{L(d)}{D(d)}.
\end{equation}
Since $L'(d)>0$, the monotonicity of
$\eta_U$ also depends on the geometric
factor $b(d)$. To determine its sign,
introduce
\[
u=\operatorname{arsinh}\frac{ad}{2}>0.
\]
Then
\[
b=\frac{2u}{\sinh(2u)}
\]
and
\begin{equation}
	\frac{db}{du}
	=
	\frac{2\sinh(2u)-4u\cosh(2u)}
	{\sinh^2(2u)}<0.
\end{equation}
Indeed, the numerator has the opposite
sign to $2u\cosh(2u)-\sinh(2u)$.
The latter vanishes at $u=0$ and has
derivative $4u\sinh(2u)>0$ for $u>0$.
Therefore $b'(d)<0$. Together with
$F>0$, $F'<0$, and $L'(d)>0$, we obtain
\begin{align}
	\eta_T'(d)
	&=
	F'(d)<0,
	\notag\\
	\eta_U'(d)
	&=
	b'(d)F(L)
	+b(d)F'(L)L'(d)<0.
	\label{cr:strict_derivatives}
\end{align}
Thus both weighted spatial correlations
decrease strictly with the detector separation.

Their endpoint limits follow directly from
the integral representations. At coincidence,
continuity of $h$ gives
\[
\lim_{x\to0^+}\frac{H(x)}{x}
=
h(0)=I_0,
\]
while $b(d)\to1$ as $d\to0^+$. Hence
\[
\lim_{d\to0^+}\eta_U(d)
=
\lim_{d\to0^+}\eta_T(d)=1.
\]
At large separation, the geometric factors
$\operatorname{sinc}(\Omega d)$ and
$b(d)\operatorname{sinc}(\Omega L(d))$
tend to zero for every fixed $\Omega>0$.
Their absolute values are bounded by one.
Since $f$ is integrable, dominated convergence
then yields
\[
\lim_{d\to\infty}\eta_U(d)
=
\lim_{d\to\infty}\eta_T(d)=0.
\]

We can now establish the uniqueness of the
finite blind separation. According to
Eq.~\eqref{cr:contrast}, the work contrast
factorizes as
\begin{equation}
	\aleph_U-\aleph_T
	=
	\aleph_C
	(\eta_U-\eta_T)
	(\eta_U+\eta_T-s).
	\label{cr:contrast_factorized}
\end{equation}
For every finite $d>0$, the ordering proved
above ensures $\eta_U-\eta_T<0$. Thus the
finite blind separation is determined by
\begin{equation}
	\eta_U(d_0)+\eta_T(d_0)=s.
	\label{cr:blind_condition}
\end{equation}

For $0<\lambda<\omega$, we have $0<s<2$.
The sum $\eta_U+\eta_T$ is continuous and
strictly decreasing, with limiting values
$2$ at $d=0$ and $0$ as $d\to\infty$.
It therefore crosses the value $s$ exactly
once. This proves the existence and uniqueness
of the finite blind separation $d_0$.

The crossing is also simple. Differentiating
Eq.~\eqref{cr:contrast_factorized} and using
Eq.~\eqref{cr:blind_condition}, we obtain
\begin{equation}
	\left.
	\partial_d(\aleph_U-\aleph_T)
	\right|_{d=d_0}
	=
	\left.
	\aleph_C
	(\eta_U-\eta_T)
	(\eta_U'+\eta_T')
	\right|_{d=d_0}>0.
	\label{cr:simple_root}
\end{equation}
Both factors in parentheses are strictly
negative, so the derivative is positive.
The two work curves therefore cross with
a nonzero slope difference at $d_0$.
The coincident-detector limit $d=0$ is a
separate boundary case, where
$\eta_U=\eta_T=1$ and the contrast also
vanishes. It is not a finite blind separation.

Finally, we determine the ordering of the
two work minima. From the dependence of
$R_E$ on $\eta_E$, its derivative is
\begin{equation}
	R_E'=(2\eta_E-s)\eta_E'.
\end{equation}
Since each $\eta_E$ decreases continuously
from $1$ to $0$, the corresponding $R_E$
decreases until $\eta_E=s/2$ and increases
thereafter. Both environments therefore
attain their minimum at $\eta_E=s/2$.
The minimum values coincide, as given in
Eq.~\eqref{cr:minimum}, although the
corresponding separations are different.

At the accelerated-vacuum minimum,
$\eta_U=s/2$. Since $\eta_T>\eta_U$,
we have $\eta_U+\eta_T>s$ at that separation.
The strictly decreasing sum
$\eta_U+\eta_T$ reaches $s$ only at a
larger separation. Conversely, at the
thermal minimum, $\eta_T=s/2$ and
$\eta_U<\eta_T$, so that
$\eta_U+\eta_T<s$. The blind separation
must therefore lie between the two minima:
\begin{equation}
		d_{\min,U}<d_0<d_{\min,T}.
\end{equation}

For $\lambda\leq0$, we have $s\leq0$. Since
$0<\eta_U<\eta_T$, their difference is negative,
whereas their sum minus $s$ is strictly positive.
Together with $\aleph_C>0$, Eq.~\eqref{cr:contrast_factorized}
therefore gives
\[
\aleph_U-\aleph_T<0
\qquad (d>0).
\]
Thus, the two work coefficients never become equal
at any finite detector separation, and no finite
blind separation exists for $\lambda\leq0$.
	\section{Finite-coupling corrections to the equilibrium readout}
	\label{cr:sixth}

The main text establishes a relation between the
equilibrium Gaussian ergotropy and the weighted
spatial correlation in the weak-coupling regime.
This relation is obtained from the leading
$O(\kappa^4)$ contribution to the ergotropy.
At finite coupling, however, higher-order terms
also contribute to the work readout. In particular,
these corrections depend on additional frequency
integrals and are not determined solely by the
weighted spatial correlation $\eta_E$.
Consequently, finite coupling can modify both
the separation at which the two environments
give the same work and the correlation inferred
from the two-setting reconstruction.
To quantify these effects, we expand the exact
equilibrium covariance to the next order in
$\kappa^2$. This gives the $O(\kappa^6)$ correction
to the ergotropy, which we then use to calculate
the shift of the blind separation and the leading
finite-coupling bias in the reconstructed
spatial correlation.

	\subsection{Sixth-order correction to the equilibrium Gaussian ergotropy}
	
	We first derive the next-order correction to the
	equilibrium Gaussian ergotropy. The starting point
	is the exact covariance obtained from the Green
	function in Sec.~\ref{tau}. To expand these
	covariances systematically, we introduce
	\[
	t=\kappa^2,\qquad
	z_n=\nu_n^2,\qquad
	D_n=z_n+\nu_S^2.
	\]
	For a fixed environment $E$ and collective sector
	$\zeta$, define, for $n\neq0$,
	\begin{equation}
		\ell_{E\zeta n}
		=
		\frac{1}{2\pi^2}
		\int_0^\infty d\Omega\,
		\frac{r(\Omega)[1 \pm \gamma_E(\Omega)]}
		{z_n+\Omega^2}.
		\label{cr:ell_finite}
	\end{equation}
	The fully compensated self-energy is then
	\[
	\Sigma_{E\zeta}(\nu_n)
	=
	t z_n\ell_{E\zeta n}.
	\]
	In particular, its zero-frequency contribution
	vanishes. We therefore omit the $n=0$ term
	in the following sums.
	
	For later use, introduce three Matsubara sums:
	\begin{align}
		\mathsf A_\zeta
		&=
		T\sum_{n\ne0}
		\frac{z_n\ell_{E\zeta n}}{D_n}
		=
		\frac{I_{E,\zeta}}{2\pi^2},
		\notag\\
		\mathsf B_\zeta
		&=
		T\sum_{n\ne0}
		\frac{z_n^2\ell_{E\zeta n}^2}{D_n^2},
		\notag\\
		\mathsf C_\zeta
		&=
		T\sum_{n\ne0}
		\frac{z_n\ell_{E\zeta n}
			(z_n-\nu_S^2)}{D_n^2}.
		\label{cr:ABC}
	\end{align}
	Here the environment label $E$ is suppressed
	in $\mathsf A_\zeta$, $\mathsf B_\zeta$, and
	$\mathsf C_\zeta$. The first sum reproduces the
	response-weighted spectral integral introduced
	in the main text, whereas the other two retain
	additional information about the frequency
	dependence of the self-energy.
	
	For a fixed regulator, $z_n\ell_{E\zeta n}$
	approaches a finite constant as $|n|\to\infty$.
	The summands in Eq.~\eqref{cr:ABC} therefore
	decay as $n^{-2}$, $n^{-4}$, and $n^{-2}$,
	respectively. All three sums are consequently
	convergent.
	
	We now use these sums to expand the exact
	equilibrium covariances. Define
	\[
	X_\zeta=\alpha_\zeta Q_\zeta^{\mathrm{eq}},
	\qquad
	Y_\zeta=\beta_\zeta P_\zeta^{\mathrm{eq}},
	\qquad
	E_0=\frac{\nu_Sc_U}{2},
	\]
	where $E_0$ is the common uncoupled value
	of $X_\zeta$ and $Y_\zeta$, with
	$c_U=\coth[\nu_S/(2T_U)]$.
	
	The exact covariance formulas and the stationary
	virial identity derived in Sec.~\ref{tau} give
	\begin{align}
		Y_\zeta-X_\zeta
		&=
		t\beta_\zeta T
		\sum_{n\ne0}
		\frac{z_n\ell_{E\zeta n}}
		{D_n+t\beta_\zeta z_n\ell_{E\zeta n}},
		\notag\\
		X_\zeta-E_0
		&=
		-t\nu_S^2\beta_\zeta T
		\sum_{n\ne0}
		\frac{z_n\ell_{E\zeta n}}
		{D_n(D_n+t\beta_\zeta z_n\ell_{E\zeta n})}.
		\label{cr:exact_differences}
	\end{align}
	These expressions show explicitly how the
	bath-induced self-energy modifies the two
	detector variances. In particular, the difference
	between the momentum and position contributions
	is directly determined by the first expression.
	
	Expanding the denominators for sufficiently
	small $t$ at fixed temperature, cutoff, and
	stable detector parameters, we obtain
	\begin{align}
		\delta_\zeta
		\equiv Y_\zeta-X_\zeta
		&=
		t\beta_\zeta\mathsf A_\zeta
		-t^2\beta_\zeta^2\mathsf B_\zeta
		+O(t^3),
		\notag\\
		S_\zeta
		\equiv X_\zeta+Y_\zeta
		&=
		2E_0+t\beta_\zeta\mathsf C_\zeta
		+O(t^2).
		\label{cr:energy_expansions}
	\end{align}
	Thus, $\mathsf A_\zeta$ determines the leading
	difference between the two covariance contributions,
	while $\mathsf B_\zeta$ and $\mathsf C_\zeta$
	enter the next-order correction.
	
	We can now obtain the finite-coupling correction
	to the Gaussian ergotropy. Using the expression
	for one collective sector,
	\begin{equation}
		\mathcal G_\zeta^{\mathrm{eq}}
		=
		\frac{X_\zeta+Y_\zeta}{2}
		-\sqrt{X_\zeta Y_\zeta},
	\end{equation}
	and the identity
	\[
	X_\zeta Y_\zeta
	=
	\frac{S_\zeta^2-\delta_\zeta^2}{4},
	\]
	we rewrite it as
	\begin{align}
		\mathcal G_\zeta^{\mathrm{eq}}
		&=
		\frac{S_\zeta}{2}
		\left[
		1-\sqrt{1-\frac{\delta_\zeta^2}{S_\zeta^2}}
		\right]
		\notag\\
		&=
		\frac{\delta_\zeta^2}{4S_\zeta}
		+O(t^4).
	\end{align}
	Since $\delta_\zeta=O(t)$, the leading ergotropy
	is of order $t^2=\kappa^4$. Retaining the next
	order requires the $O(t^2)$ correction to
	$\delta_\zeta$ and the $O(t)$ correction to
	$S_\zeta$. Substituting
	Eq.~\eqref{cr:energy_expansions}, we find
	\begin{align}
		\mathcal G_\zeta^{\mathrm{eq}}
		=
		\frac{t^2\beta_\zeta^2\mathsf A_\zeta^2}
		{4\nu_Sc_U}
		-t^3\beta_\zeta^3
		\left[
		\frac{\mathsf A_\zeta\mathsf B_\zeta}
		{2\nu_Sc_U}
		+
		\frac{\mathsf A_\zeta^2\mathsf C_\zeta}
		{4\nu_S^2c_U^2}
		\right]
		+O(t^4).
	\end{align}
	
	Summing the two collective sectors gives
	the equilibrium ergotropy of the detector pair:
	\begin{align}
		\mathcal G_E^{\mathrm{eq}}
		&=
		\kappa^4\aleph_E
		+\kappa^6\mathcal M_E
		+O(\kappa^8),
		\notag\\
		\mathcal M_E
		&=
		-\sum_\zeta\beta_\zeta^3
		\left[
		\frac{\mathsf A_\zeta\mathsf B_\zeta}
		{2\nu_Sc_U}
		+
		\frac{\mathsf A_\zeta^2\mathsf C_\zeta}
		{4\nu_S^2c_U^2}
		\right].
		\label{cr:sixth_coefficient}
	\end{align}
	The leading coefficient is
	\begin{equation}
		\aleph_E
		=
		\sum_\zeta
		\frac{\beta_\zeta^2\mathsf A_\zeta^2}
		{4\nu_Sc_U},
	\end{equation}
	which reproduces Eq.~\eqref{NE} upon using
	$\mathsf A_\zeta=I_{E,\zeta}/(2\pi^2)$ and
	$\beta_\zeta=\nu_S^2/\alpha_\zeta$.
	
	The coefficient $\mathcal M_E$ is the first
	finite-coupling correction beyond the leading
	weak-coupling result. Unlike $\aleph_E$, it
	depends on the additional sums
	$\mathsf B_\zeta$ and $\mathsf C_\zeta$.
	\subsection{Shift of the blind separation at finite coupling}
	
	We next use the sixth-order correction to
	determine how the blind separation changes
	when the detector--field coupling is finite.
	At leading order, the blind separation $d_0$
	is determined by the equality of the two
	work coefficients,
	\[
	\Delta\aleph(d_0)=0,
	\qquad
	\Delta\aleph=\aleph_U-\aleph_T.
	\]
	At finite coupling, the corresponding work
	difference also contains the sixth-order
	contribution. Defining
	\[
	\Delta\mathcal M(d)
	=
	\mathcal M_U(d)-\mathcal M_T(d),
	\]
	the equality of the two equilibrium ergotropies
	becomes
	\begin{equation}
		0=
		\Delta\aleph(d)
		+t\Delta\mathcal M(d)
		+O(t^2),
		\label{cr:finite_blind_equation}
	\end{equation}
	after dividing the work difference by $t^2$.
	
	The leading blind separation is a simple root,
	as established in Eq.~\eqref{cr:simple_root}.
	We can therefore expand the finite-coupling
	solution around $d_0$:
	\begin{equation}
		d_0(t)=d_0+t\delta d+O(t^2).
	\end{equation}
	Substituting this expression into
	Eq.~\eqref{cr:finite_blind_equation} and
	retaining terms through first order in $t$
	gives
	\[
	\Delta\aleph'(d_0)\delta d
	+\Delta\mathcal M(d_0)=0.
	\]
	Hence the leading shift is
	\begin{equation}
			\delta d
			=
			-\frac{\Delta\mathcal M(d_0)}
			{\Delta\aleph'(d_0)}.
		\label{cr:blind_shift}
	\end{equation}
	Thus, the sixth-order correction determines
	the leading $O(\kappa^2)$ displacement of
	the blind separation.
	\subsection{Finite-coupling correction to the two-setting reconstruction}
	
	Finally, we examine how finite coupling
	affects the reconstruction of the weighted
	spatial correlation from two separately
	prepared detector settings,
	$\lambda=\pm g$. For a fixed environment $E$,
	write the corresponding ergotropies as
	\begin{equation}
		\mathcal G_\pm(t)
		=
		t^2\aleph_\pm
		+t^3\mathcal M_\pm
		+O(t^4),
	\end{equation}
	where $\aleph_\pm$ and $\mathcal M_\pm$
	are evaluated at $\lambda=\pm g$.
	
	The two-setting readout is defined by the
	normalized difference
	\begin{equation}
		\mathcal A_W(t)
		=
		\frac{\mathcal G_-(t)-\mathcal G_+(t)}
		{\mathcal G_-(t)+\mathcal G_+(t)}.
	\end{equation}
	At leading order, this reduces to
	\begin{equation}
		\mathcal A_E
		=
		\frac{\aleph_--\aleph_+}
		{\aleph_-+\aleph_+}.
	\end{equation}
	Expanding the ratio to first order in $t$
	then gives
	\begin{align}
		\mathcal A_W(t)
		&=
		\mathcal A_E+b_Et+O(t^2),
		\notag\\
		b_E
		&=
		\frac{
			\mathcal M_--\mathcal M_+
			-\mathcal A_E(\mathcal M_-+\mathcal M_+)}
		{\aleph_-+\aleph_+}.
		\label{cr:finite_asymmetry}
	\end{align}
	The coefficient $b_E$ therefore quantifies
	the leading finite-coupling correction to
	the two-setting readout.
	
	To make its effect on the reconstructed
	spatial correlation explicit, recall that
	the fully compensated leading-order relation
	in the main text is
	\begin{equation}
		\mathcal A_E
		=
		\frac{s_g\eta_E}{1+\eta_E^2},
		\qquad
		s_g=\frac{4\omega g}{\omega^2+g^2}.
	\end{equation}
	The reconstruction based on this leading-order
	relation gives
	\begin{equation}
		\widehat{\eta}_E(t)
		=
		\frac{2\mathcal A_W(t)}
		{s_g+\sqrt{s_g^2-4\mathcal A_W(t)^2}}.
	\end{equation}
	Expanding around the true weighted correlation
	$\eta_E$ yields
	\begin{equation}
		\widehat{\eta}_E(t)
		=
		\eta_E
		+
		\frac{(1+\eta_E^2)^2}
		{s_g(1-\eta_E^2)}
		b_Et
		+O(t^2),
		\qquad 0<\eta_E<1.
		\label{cr:reconstruction_bias}
	\end{equation}
	Thus, the leading bias in the reconstructed
	correlation is proportional to $\kappa^2$,
	with its coefficient determined by the
	sixth-order corrections to the two work
	measurements. The inversion becomes
	increasingly sensitive to finite-coupling
	corrections as $\eta_E$ approaches unity,
	where the derivative of the leading-order
	readout relation vanishes.
\section{Partial static compensation}
\label{st:appstatic}

The main text establishes the relation between the equilibrium Gaussian ergotropy and the weighted spatial correlation $\eta_E$ under full static compensation. This compensation removes the static shift induced by the detector--field coupling, allowing us to isolate the contribution of the frequency-dependent bath response. We now examine whether this relation remains valid when the static compensation is incomplete. Specifically, we keep the local static contribution fully compensated but cancel only a fraction $\rho_{\rm ct}$ of the cross contribution. For $\rho_{\rm ct}<1$, a residual static interaction remains in the effective detector potential. This additional contribution modifies the equilibrium covariance and therefore affects the work independently of the frequency-dependent bath response. Our aim is to determine how this residual interaction modifies the relation between the extracted work and the underlying spatial correlation.

We first derive the effective Green function and the
corresponding equilibrium covariance. We then obtain
the leading work coefficient and identify how the
residual static interaction modifies the relation
between the weighted spatial correlation and the
extracted work. Finally, we separate the additional
global work and examine the classical equilibrium
limit.

The coupling to the field induces both local and
cross static contributions to the detector potential.
We retain the full local compensation and introduce
$\rho_{\rm ct}\in[0,1]$ to specify the fraction of the
cross contribution included in the counterterm.

To make this distinction explicit, define
\begin{align}
	\mu_0
	&=
	\frac{1}{2\pi^2}
	\int_0^\infty r(\Omega)\,d\Omega,
	\notag\\
	\mu_{c,E}
	&=
	\frac{1}{2\pi^2}
	\int_0^\infty
	r(\Omega)\gamma_E(\Omega)\,d\Omega.
	\label{st:mu}
\end{align}
Here, $\mu_0$ describes the local static contribution,
whereas $\mu_{c,E}$ describes the cross contribution
and depends on the environment through
$\gamma_E(\Omega)$.

Under full static compensation, the counterterm
coefficient in collective sector $\zeta$ is
$\kappa^2(\mu_0\pm \mu_{c,E})$. With partial
compensation, the corresponding coordinate
coefficient in the total Hamiltonian becomes
\begin{equation}
	\alpha_\zeta+
	\kappa^2(\mu_0\pm \rho_{\rm ct}\mu_{c,E}).
\end{equation}
Thus, $\rho_{\rm ct}=1$ corresponds to full
compensation, while $\rho_{\rm ct}=0$ means that
only the local static contribution is compensated.

After integrating out the bath coordinates, the
bath-induced static shift
$\kappa^2(\mu_0+\zeta\mu_{c,E})$ is subtracted.
The remaining coordinate coefficient is therefore
\begin{equation}
	\alpha_\zeta+\kappa^2h_{E\zeta},
\end{equation}
where
\begin{equation}
	h_{E,+}=-(1-\rho_{\rm ct})\mu_{c,E},
	\qquad
	h_{E,-}=+(1-\rho_{\rm ct})\mu_{c,E}.
	\label{st:residual}
\end{equation}
The residual interaction vanishes when
$\rho_{\rm ct}=1$. Throughout the following
calculation, we require
\[
\alpha_\zeta+\kappa^2h_{E\zeta}>0,
\qquad
\beta_\zeta>0,
\]
so that the quadratic Hamiltonian remains stable.

We now calculate the equilibrium covariance of
this modified detector--field system. Put
\[
t=\kappa^2,\qquad
z_n=\nu_n^2,\qquad
D_n=z_n+\nu_S^2,
\]
and introduce
\begin{equation}
	\ell_{E\zeta n}
	=
	\frac{1}{2\pi^2}
	\int_0^\infty
	\frac{r(\Omega)[1\pm \gamma_E(\Omega)]}
	{z_n+\Omega^2}\,d\Omega.
\end{equation}
At zero Matsubara frequency, the product
$z_n\ell_{E\zeta n}$ is understood by continuity
and vanishes.

Integrating out the bath gives the inverse
coordinate Green function
\begin{align}
	G_{E\zeta}^{-1}(i\nu_n)
	&=
	\frac{D_n}{\beta_\zeta}
	+tR_{\zeta n},
	\notag\\
	R_{\zeta n}
	&=
	z_n\ell_{E\zeta n}+h_{E\zeta}.
	\label{st:kernel}
\end{align}
The first term in $R_{\zeta n}$ is the
frequency-dependent bath contribution that
remains under full static compensation. The
second is the frequency-independent residual
static interaction. In particular,
\[
R_{\zeta 0}=h_{E\zeta}.
\]
This zero-frequency contribution is absent under
full compensation but must be retained when the
cross contribution is only partially compensated.

Let
\[
X_\zeta=\alpha_\zeta Q_\zeta,
\qquad
Y_\zeta=\beta_\zeta P_\zeta,
\]
where $Q_\zeta$ and $P_\zeta$ are the equilibrium
position and momentum variances, respectively.
The coordinate covariance obtained from the
Green function, together with the equilibrium
virial identity, gives
\begin{align}
	X_\zeta
	&=
	\nu_S^2T_U
	\sum_{n\in\mathbb Z}
	\frac{1}{D_n+t\beta_\zeta R_{\zeta n}},
	\notag\\
	Y_\zeta
	&=
	T_U
	\sum_{n\in\mathbb Z}
	\frac{\nu_S^2+t\beta_\zeta R_{\zeta n}}
	{D_n+t\beta_\zeta R_{\zeta n}}.
	\label{st:exactXY}
\end{align}
Both sums converge for the regulated spectrum.

Subtracting the two expressions isolates the
difference between the scaled momentum and
position variances:
\begin{equation}
	Y_\zeta-X_\zeta
	=
	t\beta_\zeta T_U
	\sum_{n\in\mathbb Z}
	\frac{R_{\zeta n}}
	{D_n+t\beta_\zeta R_{\zeta n}}.
	\label{st:anisotropyexact}
\end{equation}
This difference determines the Gaussian
ergotropy of each collective sector. In
particular, the zero-frequency term contributes
\begin{equation}
	\frac{
		t\beta_\zeta T_Uh_{E\zeta}}
	{\nu_S^2+t\beta_\zeta h_{E\zeta}}.
\end{equation}
It is therefore important not to omit the
$n=0$ term when a residual static interaction
is present.

The coordinate variance can also be expressed
relative to its uncoupled equilibrium value.
Using
\[
T_U\sum_n\frac{1}{D_n}
=
\frac{c_U}{2\nu_S},
\qquad
c_U=\coth\frac{\nu_S}{2T_U},
\]
we obtain
\begin{equation}
	X_\zeta-\frac{\nu_Sc_U}{2}
	=
	-t\nu_S^2\beta_\zeta T_U
	\sum_n
	\frac{R_{\zeta n}}
	{D_n(D_n+t\beta_\zeta R_{\zeta n})}.
\end{equation}

We now expand the covariance difference to
leading order in $t=\kappa^2$. This gives
\begin{align}
	Y_\zeta-X_\zeta
	&=
	t\beta_\zeta\widetilde{\mathsf A}_\zeta
	+O(t^2),
	\notag\\
	\widetilde{\mathsf A}_\zeta
	&=
	\frac{I_{E,\zeta}}{2\pi^2}
	+\frac{h_{E\zeta}c_U}{2\nu_S}.
	\label{st:tildeA}
\end{align}
Here, $I_{E,\zeta}$ is the frequency-weighted
spectral integral defined in the main text:
\[
I_{E,\zeta}=I_0 \pm I_{c,E}.
\]
The first term in $\widetilde{\mathsf A}_\zeta$
is the contribution obtained under full static
compensation. The second term is the correction
from the residual static interaction. To see
the origin of the first term, we use
\begin{equation}
	T_U\sum_{n\in\mathbb Z}
	\frac{z_n}
	{D_n(z_n+\Omega^2)}
	=
	B_1(\Omega),
\end{equation}
and then perform the frequency integral in
$\ell_{E\zeta n}$.

We can now calculate the Gaussian ergotropy.
For one collective sector, it is
\begin{equation}
	\mathcal G_\zeta
	=
	\frac{X_\zeta+Y_\zeta}{2}
	-\sqrt{X_\zeta Y_\zeta}.
\end{equation}
Equivalently, it can be written as
\begin{equation}
	\mathcal G_\zeta
	=
	\frac{(Y_\zeta-X_\zeta)^2}
	{2\left(
		X_\zeta+Y_\zeta+2\sqrt{X_\zeta Y_\zeta}
		\right)}.
\end{equation}
At zero coupling, the two scaled variances
coincide:
\[
X_\zeta=Y_\zeta
=\frac{\nu_Sc_U}{2}.
\]
The uncoupled detector therefore has zero
Gaussian ergotropy. The leading nonzero
contribution comes from the square of the
coupling-induced variance difference. Substituting
Eq.~\eqref{st:tildeA}, we obtain
\begin{equation}
	\mathcal G_\zeta
	=
	\frac{t^2\beta_\zeta^2}
	{4\nu_Sc_U}
	\widetilde{\mathsf A}_\zeta^2
	+O(t^3).
	\label{st:workA}
\end{equation}

To express this result in terms of the spatial
correlation defined in the main text, write
\[
J_{c,E}=2\pi^2\mu_{c,E},
\qquad
\eta_E=\frac{I_{c,E}}{I_0}.
\]
Using Eq.~\eqref{st:residual}, the coefficient
in Eq.~\eqref{st:tildeA} becomes
\begin{equation}
	\widetilde{\mathsf A}_\zeta
	=
	\frac{I_0}{2\pi^2}(1+\zeta\xi_E),
\end{equation}
where
\begin{equation}
		\xi_E
		=
		\eta_E
		-(1-\rho_{\rm ct})
		\frac{c_UJ_{c,E}}{2\nu_S I_0}.
	\label{st:xi}
\end{equation}
Thus, the residual static interaction changes
the parameter entering the work readout from
$\eta_E$ to $\xi_E$. When $\rho_{\rm ct}=1$,
the residual term vanishes and
$\xi_E=\eta_E$, recovering the fully compensated
result.

Summing Eq.~\eqref{st:workA} over the two
collective sectors gives
\begin{align}
	&\aleph_E^{(\rho_{\rm ct})}\notag
	= \\ 
	&\frac{I_0^2}{16\pi^4\nu_Sc_U}
	\left[
	(\omega-\lambda)^2(1+\xi_E)^2
	\right.
	\left.
	+(\omega+\lambda)^2(1-\xi_E)^2
	\right].
\end{align}
The part independent of the spatial correlation
is the uncorrelated work coefficient
\begin{equation}
	\aleph_C
	=
	\frac{I_0^2(\omega^2+\lambda^2)}
	{8\pi^4\nu_Sc_U}.
\end{equation}
Dividing by $\aleph_C$ and defining
\[
s=\frac{4\omega\lambda}{\omega^2+\lambda^2},
\]
we obtain
\begin{equation}
		\frac{\aleph_E^{(\rho_{\rm ct})}}{\aleph_C}
		=
		1+\xi_E^2-s\xi_E.
	\label{st:criterion}
\end{equation}
This is the relation used in the main text.
It shows explicitly that incomplete static
compensation changes the work readout through
the replacement of $\eta_E$ by $\xi_E$.

We can also identify how the residual static
interaction changes the additional global work.
At $\lambda=0$, the leading work coefficient
takes the form
\begin{equation}
	\left.\aleph_E^{(\rho_{\rm ct})}\right|_{\lambda=0}
	=
	\left.\aleph_C\right|_{\lambda=0}
	(1+\xi_E^2).
\end{equation}
The first term is the local-work contribution,
whereas the second is the additional contribution
obtained by operating jointly on the detector pair.
Thus, the additional global-work coefficient is
\begin{equation}
		\left.
		\left(
		\aleph_E^{(\rho_{\rm ct})}-\aleph_C
		\right)
		\right|_{\lambda=0}
		=
		\left.\aleph_C\right|_{\lambda=0}\xi_E^2.
	\label{st:appgap}
\end{equation}
The residual static interaction therefore
changes both the correlation parameter entering
the work readout and the additional work
available from the joint operation.

We now ask whether the work induced by incomplete static compensation requires quantum bath fluctuations or can also arise from the residual static interaction in the classical limit. To answer this question, we take the classical equilibrium limit of the detector--bath system and integrate out the bath coordinates. The resulting reduced equilibrium distribution of the detector pair is
\begin{equation}
	\rho_{S,\mathrm{cl}}
	\propto
	\exp\left[
	-\frac{1}{2T_U}
	\sum_\zeta
	\left(
	(\alpha_\zeta+th_{E\zeta})q_\zeta^2
	+\beta_\zeta p_\zeta^2
	\right)
	\right].
\end{equation}
The corresponding classical variances are
\begin{equation}
	Q_{\zeta,\mathrm{cl}}
	=
	\frac{T_U}{\alpha_\zeta+th_{E\zeta}},
	\qquad
	P_{\zeta,\mathrm{cl}}
	=
	\frac{T_U}{\beta_\zeta}.
\end{equation}

Evaluating the same covariance-based work
expression relative to the original detector
Hamiltonian $\mathcal H_S$, we find
\begin{align}
	\mathcal G_{\zeta,\mathrm{cl}}
	&=
	\frac{T_U}{2}
	\left[
	1+
	\frac{\alpha_\zeta}
	{\alpha_\zeta+th_{E\zeta}}
	-2\sqrt{
		\frac{\alpha_\zeta}
		{\alpha_\zeta+th_{E\zeta}}
	}
	\right]
	\notag\\
	&=
	\frac{T_Ut^2h_{E\zeta}^2}
	{8\alpha_\zeta^2}
	+O(t^3).
	\label{st:classical}
\end{align}
When the static contribution is fully compensated, $h_{E\zeta}=0$, the classical equilibrium variances reduce to $Q_{\zeta,\mathrm{cl}}=T_U/\alpha_\zeta$ and $P_{\zeta,\mathrm{cl}}=T_U/\beta_\zeta$. These are precisely the variances of the classical Gibbs state of the original detector Hamiltonian $\mathcal H_S$. The state is therefore passive with respect to $\mathcal H_S$, and its extractable work vanishes. In contrast, when $h_{E\zeta}\neq0$, the residual static interaction changes the effective stiffness from $\alpha_\zeta$ to $\alpha_\zeta+t h_{E\zeta}$, while the momentum variance remains unchanged. The resulting classical equilibrium state is no longer the Gibbs state of $\mathcal H_S$. Consequently, it has nonzero ergotropy when the work is evaluated relative to the original detector Hamiltonian, as shown in Eq.~\eqref{st:classical}. This contribution arises from the residual static interaction itself and does not require quantum bath fluctuations. Thus, under incomplete static compensation, the extracted work contains a contribution from the change in the static detector potential, in addition to any contribution associated with quantum bath fluctuations.
	\section{Numerical methods and convergence checks}
	\label{num:validation}
	\label{F}
	
	This appendix describes the numerical methods used
	in the transient and equilibrium calculations and
	provides the corresponding convergence checks.
		\subsection{Finite-coupling equilibrium covariance and numerical verification}
		
		The finite-coupling equilibrium work is calculated
		from the unexpanded Matsubara Green function,
		without performing a weak-coupling expansion
		in $\kappa^2$. The frequency integrals entering
		the self-energy are evaluated numerically.
		The zero-frequency contribution is treated
		separately, while the large-frequency tail
		of the Matsubara sum is evaluated using its
		asymptotic expansion and Hurwitz zeta functions.
		The resulting equilibrium covariances are
		then used to calculate the Gaussian ergotropy.
		
		To independently verify the Matsubara results,
		we also calculate the equilibrium covariance
		by diagonalizing the full Hamiltonian of a
		discretized detector--bath system. In contrast
		to the Matsubara approach, which integrates
		out the bath degrees of freedom, this method
		retains the bath oscillators explicitly and
		evaluates the detector covariance directly
		from the interacting Gibbs state.
		
		For each collective sector $\zeta=\pm$,
		the total quadratic Hamiltonian, including
		the static counterterm, can be written as
		\begin{equation}
			\mathcal H_{\mathrm{tot},\zeta}
			=
			\frac12\mathbf p_\zeta^T B_\zeta
			\mathbf p_\zeta
			+
			\frac12\mathbf q_\zeta^T K_\zeta
			\mathbf q_\zeta,
		\end{equation}
		where
		\begin{equation}
			B_\zeta
			=
			\operatorname{diag}
			(\beta_\zeta,\Omega_1,\ldots,\Omega_N),
		\end{equation}
		and the nonzero entries of $K_\zeta$ are
		\begin{align}
			(K_\zeta)_{00}
			&=
			\alpha_\zeta+\delta\omega_\zeta,
			\notag\\
			(K_\zeta)_{jj}
			&=
			\Omega_j,
			\notag\\
			(K_\zeta)_{0j}
			&=
			(K_\zeta)_{j0}
			=
			g_{j,\zeta}.
		\end{align}
		Here,
		$\mathbf q_\zeta=(q_\zeta,Q_{1,\zeta},
		\ldots,Q_{N,\zeta})^T$ and
		$\mathbf p_\zeta=(p_\zeta,P_{1,\zeta},
		\ldots,P_{N,\zeta})^T$.
		The index $0$ denotes the detector,
		while $j=1,\ldots,N$ labels the bath
		oscillators.
		
		We introduce the canonical transformation
		\begin{equation}
			\mathbf q_\zeta
			=
			B_\zeta^{1/2}\mathbf x_\zeta,
			\qquad
			\mathbf p_\zeta
			=
			B_\zeta^{-1/2}\boldsymbol\pi_\zeta,
		\end{equation}
		which brings the Hamiltonian into the form
		\begin{equation}
			\mathcal H_{\mathrm{tot},\zeta}
			=
			\frac12\boldsymbol\pi_\zeta^T
			\boldsymbol\pi_\zeta
			+
			\frac12\mathbf x_\zeta^T
			B_\zeta^{1/2}K_\zeta B_\zeta^{1/2}
			\mathbf x_\zeta.
		\end{equation}
		The normal-mode frequencies are obtained
		from the orthogonal diagonalization
		\begin{equation}
			B_\zeta^{1/2}K_\zeta B_\zeta^{1/2}
			=
			O_\zeta
			\operatorname{diag}(f_k^2)
			O_\zeta^T,
			\label{num:normalmodes}
		\end{equation}
		where $f_k>0$ are the normal-mode
		frequencies.
		
		The transformed Hamiltonian describes
		independent harmonic oscillators.
		Using their thermal variances at
		temperature $T_U$, we obtain the
		equilibrium detector covariances
		\begin{align}
			Q_\zeta
			&=
			\frac{\beta_\zeta}{2}
			\sum_k (O_\zeta)_{0k}^2
			\frac{\coth(f_k/2T_U)}{f_k},
			\notag\\
			P_\zeta
			&=
			\frac{1}{2\beta_\zeta}
			\sum_k (O_\zeta)_{0k}^2
			f_k\coth(f_k/2T_U).
			\label{num:gibbsnormal}
		\end{align}
		These expressions give the detector
		covariance of the reduced interacting
		Gibbs state, including the effects of
		the detector--bath coupling and the
		static counterterm. No expansion in
		the coupling strength is required.
		
		The Gaussian ergotropy is evaluated
		with respect to the original detector
		Hamiltonian
		\begin{equation}
			\mathcal H_{S,\zeta}
			=
			\frac12
			\left(
			\alpha_\zeta q_\zeta^2
			+
			\beta_\zeta p_\zeta^2
			\right),
		\end{equation}
		consistent with the work definition
		used in the main text. For numerical
		stability, we use
		\begin{equation}
			\begin{aligned}
				\mathcal G_\zeta
				&=
				\frac{(X_\zeta-Y_\zeta)^2}
				{2\left(
					X_\zeta+Y_\zeta+
					2\sqrt{X_\zeta Y_\zeta}
					\right)},
				\\
				X_\zeta
				&=
				\alpha_\zeta Q_\zeta,
				\qquad
				Y_\zeta
				=
				\beta_\zeta P_\zeta.
			\end{aligned}
			\label{num:stablework}
		\end{equation}
		This expression is algebraically
		equivalent to the Gaussian ergotropy
		formula used in the main text, but
		avoids subtracting nearly equal
		energies when the extractable work
		is small.

		To verify the finite-coupling equilibrium
		calculation, we compare two independent methods:
		the continuum Matsubara Green-function approach,
		in which the bath degrees of freedom are
		integrated out, and the finite-bath normal-mode
		approach, in which the full detector--bath
		Hamiltonian is diagonalized directly.
		Neither method involves a weak-coupling expansion.
		
		For both environments, $E=U,T$, at
		$d=1$ and $d=d_0$, with $\kappa^2=0.05$,
		the relative difference in the equilibrium
		Gaussian ergotropy obtained from the two
		methods is below $1.9\times10^{-8}$.
		This agreement provides an independent
		numerical check of the finite-coupling
		equilibrium results.
	\subsection{Numerical convergence of the dynamical calculations}
	
	We first check the convergence of the original
	finite-time calculations with respect to the
	discretization of the bath spectrum. The continuum
	frequency interval is fixed at
	\[
	\Omega\in[10^{-5},50],
	\]
	and the spectral integral is approximated by an
	$N$-point Gauss--Legendre quadrature. To isolate
	the effect of the quadrature resolution, all other
	parameters and the propagation protocol are kept
	fixed.
	
	The parameters used in this comparison are
	\begin{center}
		\small
		\begin{tabular}{|cccccccc|}
			\hline
			$\omega$ & $\lambda$ & $\kappa^2$ & $d$ & $a$
			& $\Omega_c$ & $\Omega_{\min}$ & $\Omega_{\max}$ \\
			\hline
			$1$ & $0.5$ & $0.05$ & $1$ & $2\pi$
			& $5$ & $10^{-5}$ & $50$ \\
			\hline
		\end{tabular}
	\end{center}
	
	We calculate the transient ergotropy for
	$N=5000$, $10000$, and $20000$, and use the
	$N=40000$ result as the numerical reference.
	The differences are summarized in the following
	table.
	
	\begin{center}
		\small
        \setlength{\tabcolsep}{3pt}
        \renewcommand{\arraystretch}{1.15}
        \resizebox{\columnwidth}{!}{%
		\begin{tabular}{rccccc}
			\hline
			$N$
			& $\Delta$
			& rel.\ [\%]
			& $\delta G_{\rm pk}$ [\%]
			& $\delta T$ [\%]
			& $\delta G(50)$ [\%] \\
			\hline
			$5000$
			& $1.55\times10^{-15}$
			& $5.26\times10^{-9}$
			& $3.76\times10^{-10}$
			& $0.00$
			& $0.00$ \\
			$10000$
			& $1.33\times10^{-15}$
			& $4.51\times10^{-9}$
			& $1.13\times10^{-9}$
			& $0.00$
			& $0.00$ \\
			$20000$
			& $1.55\times10^{-15}$
			& $5.26\times10^{-9}$
			& $1.13\times10^{-9}$
			& $0.00$
			& $1.61\times10^{-9}$ \\
			\hline
		\end{tabular}%
        }
	\end{center}
	
	Here, $\Delta$ is the maximum absolute
	difference between the entire ergotropy curve
	and the $N=40000$ reference curve:
	\[
	\Delta
	=
	\max_{0\leq\tau\leq50}
	\left|
	\mathcal G_N(\tau)
	-
	\mathcal G_{40000}(\tau)
	\right|.
	\]
	The relative deviation is defined by
	\[
	\mathrm{rel.}
	=
	\frac{\Delta}{G_{\rm pk,ref}}\times100\%,
	\]
	where $G_{\rm pk,ref}$ is the main peak height of
	the reference curve. The quantities
	$\delta G_{\rm pk}$, $\delta T$, and $\delta G(50)$
	denote, respectively, the relative deviations
	of the main peak height, the separation between
	successive peaks, and the ergotropy at the endpoint
	$\tau=50$, each evaluated relative to its
	corresponding reference value.
	
	The maximum absolute difference is approximately
	$10^{-15}$ for all three resolutions, indicating
	that the grid-refinement differences have reached
	the floating-point precision level of the
	calculation. The relative deviations of the peak
	height and endpoint value are also extremely small.
	The entries displayed as $0.00$ for $\delta T$
	and $\delta G(50)$ are zero at the precision
	reported in the table. These results show that
	the transient ergotropy curve is insensitive to
	further refinement of the quadrature grid for
	the parameters considered here.

	We next examine the dependence of the numerical
	results on the frequency interval. Taking
	\[
	\Omega\in[10^{-6},80]
	\]
	as the reference interval, we vary one endpoint
	at a time while keeping the other fixed. The
	maximum relative deviations, normalized by the
	reference peak amplitude, are listed below.
	
	\begin{center}
		\small
		\begin{tabular}{lc}
			\hline
			Frequency bound & Relative error [\%] \\
			\hline
			$\Omega_{\max}=30$
			& $9.150\times10^{-3}$ \\
			$\Omega_{\max}=50$
			& $1.314\times10^{-4}$ \\
			$\Omega_{\max}=80$
			& $0.000$ \\
			$\Omega_{\min}=10^{-4}$
			& $2.330\times10^{-2}$ \\
			$\Omega_{\min}=10^{-5}$
			& $2.118\times10^{-3}$ \\
			$\Omega_{\min}=10^{-6}$
			& $0.000$ \\
			\hline
		\end{tabular}
	\end{center}
	
	The upper-cutoff test is performed with
	$\Omega_{\min}=10^{-6}$ fixed, whereas the
	lower-cutoff test is performed with
	$\Omega_{\max}=80$ fixed. For the frequency
	interval $\Omega\in[10^{-5},50]$ adopted in
	this work, the corresponding deviations are
	$1.314\times10^{-4}\%$ for the upper cutoff
	and $2.118\times10^{-3}\%$ for the lower cutoff.
	Both are sufficiently small compared with the
	reference peak amplitude. These results support
	the use of $\Omega\in[10^{-5},50]$ in the
	transient calculations.
	
	We further examine how the transient work
	depends on the spectral cutoff $\Omega_c$.
	For each value of $\Omega_c$, we set
	$\Omega_{\max}=10\Omega_c$ to ensure sufficient
	coverage of the regulated spectral tail. The
	first peak amplitude, the decay rate of the
	oscillation envelope, and the ergotropy at
	$\tau=50$ are compared for
	$\Omega_c=3,5,8$.
	
	\begin{center}
		\small
		\begin{tabular}{cccc}
			\hline
			$\Omega_c$ & $G_{\rm pk}$
			& Envelope rate & $G(50)$ \\
			\hline
			$3$ & $2.691\times10^{-5}$
			& $-1.104\times10^{-2}$
			& $1.438\times10^{-5}$ \\
			$5$ & $2.954\times10^{-5}$
			& $-1.186\times10^{-2}$
			& $1.378\times10^{-5}$ \\
			$8$ & $3.152\times10^{-5}$
			& $-1.215\times10^{-2}$
			& $1.263\times10^{-5}$ \\
			\hline
		\end{tabular}
	\end{center}
	
	Here, $G_{\rm pk}$ denotes the height of the
	first ergotropy peak. The envelope rate is
	obtained by a least-squares fit of
	\[
	\log A_m=r\tau_m+b
	\]
	to the sequence of main peak amplitudes
	$A_m$, where $r<0$ is the fitted decay rate.
	The quantity $G(50)$ denotes the ergotropy
	at $\tau=50$.
	
	As $\Omega_c$ increases from $3$ to $8$,
	the first peak amplitude increases from
	$2.691\times10^{-5}$ to $3.152\times10^{-5}$,
	while the envelope decay rate varies from
	$-1.104\times10^{-2}$ to $-1.215\times10^{-2}$.
	The ergotropy at $\tau=50$ decreases over
	the same range. Thus, changing the spectral
	cutoff affects the quantitative work amplitude
	and its late-time value, while the characteristic
	decay rate remains within a relatively narrow
	range. The damped oscillatory behavior is
	preserved throughout the parameter range
	considered here.

    \bibliographystyle{apsrev4-2}
    \bibliography{ref}

\end{document}